\documentclass[lettersize,journal]{IEEEtran}
\usepackage{amsmath,amsfonts}
\usepackage{algorithmic}
\usepackage{algorithm}
\usepackage{array}
\usepackage[caption=false,font=normalsize,labelfont=sf,textfont=sf]{subfig}
\usepackage{textcomp}
\usepackage{stfloats}
\usepackage{verbatim}
\usepackage{cite}
\usepackage{url}
\usepackage{booktabs}
\usepackage{graphicx}
\usepackage{fontawesome5}
\usepackage[many]{tcolorbox}
\usepackage{import}
\usepackage{xspace}
\usepackage{multirow}
\usepackage{amssymb}
\usepackage{lscape} 
\usepackage{longtable} 
\usepackage{hyperref} 
\usepackage{tikz} 
\usepackage{enumitem} 
\usepackage{pgfplots}
\usepackage{pgfplotstable}
\pgfplotsset{compat=1.18} 
\usepackage[table]{xcolor} 
\usepackage{soul} 
\usepackage{float}
\usepackage[flushleft]{threeparttable}
\usepackage{xurl} 

\makeatletter
\@namedef{ver@lineno.sty}{9999/12/31}
\@namedef{opt@lineno.sty}{}
\makeatother
\usepackage{flushend}

\usepackage{macros}
\usepackage{diagbox}
\usepackage{tikz}

\usepackage{xparse}

\usepackage{makecell}
\usepackage{mathtools}

\IEEEtriggeratref{346} 

\hypersetup{
    colorlinks,
    linkcolor={black!80!black},
    citecolor={blue!80!blue},
    urlcolor={blue!80!black}
}

\usepackage{orcidlink}

\begin{document}

\title{LLM in the Middle: A Systematic Review of Threats and \\Mitigations to Real-World LLM-based Systems}

\author{Vitor Hugo Galhardo Moia\,\orcidlink{0000-0003-0396-2873}, 
        Igor Jochem Sanz\,\orcidlink{0000-0002-1122-0784}, 
        Gabriel Antonio Fontes Rebello\,\orcidlink{0000-0003-3344-0734}, \\
        Rodrigo Duarte de Meneses\,\orcidlink{0009-0008-7026-6863},
        Briland Hitaj\,\orcidlink{0000-0001-5925-3027}, and
        Ulf Lindqvist\,\orcidlink{0009-0002-5941-0947}
        
\thanks{Vitor Hugo Galhardo Moia, Igor Jochem Sanz, Gabriel Antonio Fontes Rebello, and Rodrigo Duarte de Meneses are with Instituto de Pesquisas Eldorado, Av. Alan Turing, 275 - Cidade Universit{\'a}ria, Campinas - SP, 13083-898, Brazil (e-mail:~\href{mailto:vitor.moia@eldorado.org.br}{vitor.moia@eldorado.org.br};~\href{mailto:igor.sanz@eldorado.org.br}{igor.sanz@eldorado.org.br}; \href{mailto:gabriel.rebello@eldorado.org.br}{gabriel.rebello@eldorado.org.br};
\href{mailto:rodrigo.meneses@eldorado.org.br}{rodrigo.meneses@eldorado.org.br}}
\thanks{Briland Hitaj and Ulf Lindqvist are with the Computer Science Lab, SRI International, 333 Ravenswood Ave, Menlo Park, CA 94025, USA (e-mail:~\href{mailto:briland.hitaj@sri.com}{briland.hitaj@sri.com};~\href{mailto:ulf.lindqvist@sri.com}{ulf.lindqvist@sri.com}}
}



\maketitle

\begin{abstract}

The success and wide adoption of generative AI (GenAI), particularly large language models (LLMs), has attracted the attention of cybercriminals seeking to abuse models, steal sensitive data, or disrupt services.
Moreover, providing security to LLM-based systems is a great challenge, as both traditional threats to software applications and threats targeting LLMs and their integration must be mitigated.
In this survey, we shed light on security and privacy concerns of such LLM-based systems by performing a systematic review and comprehensive categorization of threats and defensive strategies considering the entire software and LLM life cycles.
We analyze real-world scenarios with distinct characteristics of LLM usage, spanning from development to operation.
In addition, threats are classified according to their severity level and to which scenarios they pertain, facilitating the identification of the most relevant threats. Recommended defense strategies are systematically categorized and mapped to the corresponding life cycle phase and possible attack strategies they attenuate.
This work paves the way for consumers and vendors to understand and efficiently mitigate risks during integration of LLMs in their respective solutions or organizations. It also enables the research community to benefit from the discussion of open challenges and edge cases that may hinder the secure and privacy-preserving adoption of LLM-based systems.

\end{abstract}

\begin{IEEEkeywords}
Cybersecurity, Threats, Risks, Mitigations, Defenses, Threat Modeling, Use Cases, Generative AI, Large Language Models, LLM, Real-world Deployment.
\end{IEEEkeywords}
\section{Introduction}

The widespread use of large language models (LLMs) has attracted the attention of many to this new artificial intelligence (AI) technology~\cite{neumann2024future}. The ability of LLMs to engage in human-like conversations and provide answers to complex questions has changed the way users search for content online, with many adopting it into their daily lives. 
LLMs have become the source of information for many users, helping them to write, summarize, learn, and develop software, among other tasks. 
This evolution has extended to companies, as shown by an IBM study~\cite{ibm-study-2023} in which nearly 80\% of UK business leaders confirmed that they have already deployed LLMs or have plans to do so in their organizations, with the goal of enhancing customer experience, advancing modernization, and improving operational efficiency.
But even though the benefits of this new technology are many,
adopting it widely brings old and new security concerns to software developers.

First, LLMs are an attractive target for adversaries due to the sensitive information they contain.
Privacy issues arise from the memorization of training data by LLMs, as demonstrated in empirical studies~\cite{hartmann2023sok,staab2023beyond,aditya2024evaluating}.
The origin of LLM information is largely the public Internet, but in some cases, it also contains proprietary data, obtained from an organization and used during model fine-tuning.
Thus, the model may unintentionally generate content during its responses that contain sensitive data (e.g., personal information, application programming interface (API) keys, proprietary data), illicit content (e.g., detailed instructions on how to build a bomb or generate malware), or copyrighted data.

Second, developing and deploying LLMs is a complex and never-ending task.
LLM-based systems inherit threats from common software applications and also present new threats specific to the LLMs and their integration.
We define an LLM system as a set of software components responsible for leveraging an LLM to fulfill its goal, as a chat-bot or agent, or integrated into applications to perform specific tasks, and containing some of the following elements: user interface, APIs, databases, input and output processing modules, among others, including a foundation or fine-tuned LLM.
The components may vary according to the purpose and requirements of the solution.
Thus, vulnerabilities coming from the elements that compose the system should also be considered during the security assessment of an LLM-based system.

Software vulnerabilities may exist in the development pipeline 
due to supply chain attacks aiming to expose businesses' secrets (e.g., source code, APIs, training data) via compromised plugins used by developers, 
or taking advantage of improper input validation in web API tools that could lead to remote code execution.
We have seen vulnerabilities in commercial LLMs, such as ChatGPT \cite{march20ChatGPTVuln}, having problems with their open source software dependencies, leaking sensitive data from active users of the platform (e.g., first and last name, email address, payment address, credit card number, etc.).
Many Common Vulnerability and Exposure (CVE) records are being disclosed for all types of software used during the development and deployment of LLMs, including CVE-2024-8309, CVE-2024-28088, CVE-2024-3924, and CVE-2024-10044.
There are also initiatives that foster vulnerability discovery, such as bug bounty programs specific for AI-related software~\cite{huntr-bug-bounty}.
Known (classical) vulnerabilities~\cite{snykJanVuln2025, Kovacs-news-2025, Flesch2025supervised, chatgpt-interface-vuln-2023, lindqvist2002map, tinnel2022importance} also continue to be a problem for LLM-based systems.

Particular threats to LLMs are also seen in recent news~\cite{Cuthbertson-news-aiworm, Chen2024news, Russinovich-jailbreakattack, news-Wickens-shadowlogic2024, news-Kassianik-TAP2023, Kovacs-news-2025-2}.
From zero-click AI worms~\cite{cohen-aiworm}, to jailbreaks~\cite{Chen2024news, Russinovich-jailbreakattack, Shimony-news-jailbreak2025, news-Kassianik-TAP2023} and backdoor attacks~\cite{Wickens-shadowlogic2024}, LLMs are a constant target of new forms of attacks aimed at obtaining sensitive information, implanting malware, and influencing model behavior.
Concerned with the risks and vulnerabilities in LLM development, the Open Worldwide Application Security Project (OWASP) released a Top 10 list specific for this domain~\cite{owasp_llm_top10}.
Many works in the literature have also been presenting different forms of attacks to LLMs \cite{nist_adversarial_ml_attacks_2023, 
nist_adversarial_ml_attacks_2023, liu2024demystifying, shen2024anything, yao2024asurvey, altun2024securing, hui2024pleak, greshake2023not},
but gaps still exist regarding LLM security.
With so many threats discovered so far, it is difficult to keep track and identify those applicable to a specific scenario with unique design characteristics.
Given the limited resources available, it is paramount that we understand where to focus efforts while protecting LLM-based systems from the most critical threats.

In this survey, we address some gaps in the field by designing and analyzing distinctive LLM use case scenarios under a security perspective, highlighting different design choices that may affect the security and privacy of systems. We perform a systematic review and characterization of threats and defenses considering the development and deployment of LLM-based systems, presenting different forms of executing attacks, defining a severity level for each threat, and classifying possible defense strategies according to the threat category they mitigate and to which software development life cycle phase they apply. By considering different LLM scenarios with specific design choices, we present an analysis of threats and defensive strategies to these scenarios.

This work seeks to answer the following research questions:

\begin{itemize}

    \item[RQ1] \textbf{What are the main use cases and design choices of LLM systems from a security perspective?}

    \item[RQ2] \textbf{What are the major threats to real-world LLM-based systems and how can they be mitigated?}
    
    \item[RQ3] \textbf{How do different use cases and design choices affect the security and privacy of LLM-based systems?}
    
\end{itemize}

By addressing these research questions, we hope to provide guidance for future development and deployment of secure real-world LLM-based systems in different scenarios and use of this technology.
It is important to note that while this work 
focuses on security and privacy risks to LLMs, these concepts also apply to the entire GenAI field.

In this paper, we first present in Section~\ref{sec:methodology} the methodology adopted for the systematic review we performed, followed by a summary of the literature in Section~\ref{sec:related-work} to 
compare our contributions to related surveys.
In Section~\ref{sec:llm_scenarios}, we show the phases of the software and LLM life cycles considered in this work and define different LLM scenarios in which one can apply this technology.
Section~\ref{sec:attacks_to_LLM} presents the results of a review and a characterization of threats identified in LLM-based systems, followed by a severity level analysis of each threat in Section~\ref{sec:threats_severity_level}.
In Section~\ref{sec:mitigation_strategies} we categorize possible defensive strategies to be adopted by LLM developers and identify the software and LLM life cycle phase to which they can be applied.
Based on the previously defined scenarios, in Section~\ref{sec:threats_LLM_use_cases} we analyze the application of a threat modeling methodology to some LLM scenarios, highlighting threats and defenses.
Section~\ref{sec:discussion} discusses results, open challenges, and limitations, and Section~\ref{sec:conclusion} concludes the paper. Table~\ref{tab:Abbreviations} presents the full list of abbreviations used in the paper. 

\begin{table}[H]
\centering
\scriptsize
\caption{Glossary of Technical Terms and Survey Acronyms}
\begin{tabular}{ll}
\toprule

\multicolumn{2}{c}{\textbf{Technical Terms Definitions}} \\
\midrule
API & Application Programming Interface \\
ASR & Automated Speech Recognition \\
CSP & Cloud Service Provider \\
CVE & Common Vulnerabilities and Exposures \\
CVSS & Common Vulnerability Scoring System \\
DevSecOps & Development, Security, and Operations \\
GenAI & Generative AI \\
GPS & Global Positioning System \\
LLM & Large Language Model \\
LLMOps & Large Language Model Operations \\
MITM & Man-in-the-Middle \\
MLBOM & Machine Learning Bill of Materials \\
RAG & Retrieval-Augmented Generation \\
RCE & Remote Code Execution \\
SBOM & Software Bill of Materials \\
SQL & Structured Query Language \\
SSDLC & Secure Software Development Lifecycle \\
TTP & Tactics, Techniques and Procedures \\
\midrule
\multicolumn{2}{c}{\textbf{Survey Acronyms Definitions}} \\
\midrule

A[id] & Availability (Threat) \\
AG & Agent (Use Case) \\
AP & Integrated App (Use Case) \\
AR & Access to Resources (Design Choice) \\
C[id] & Confidentiality (Threat) \\
CB & Chat-bot (Use Case) \\
CL & Continuous Learning (Design Choice) \\
D & Development \\
DI & Dev. and Deploy Infrastructure (Design Choice) \\
DP & Data Provenance (Design Choice) \\
FM & Foundation Model (Use Case) \\
FT & Fine-Tuning (Use Case) \\
I[id] & Integrity (Threat) \\
IO & Prompt Input Origin (Design Choice) \\
M[id] & Mitigation \\
O & Operation \\
RG & RAG (Use Case) \\
SI & Shared Infrastructure (Design Choice) \\
SL & SW Libraries and Dependencies (Design Choice) \\
ST & Stage \\
UC & Use Case \\

\bottomrule
\end{tabular}
\label{tab:Abbreviations}
\end{table}
\section{Methodology}
\label{sec:methodology}

\begin{figure}[hbt]
    \centering
    \includegraphics[scale=.6]{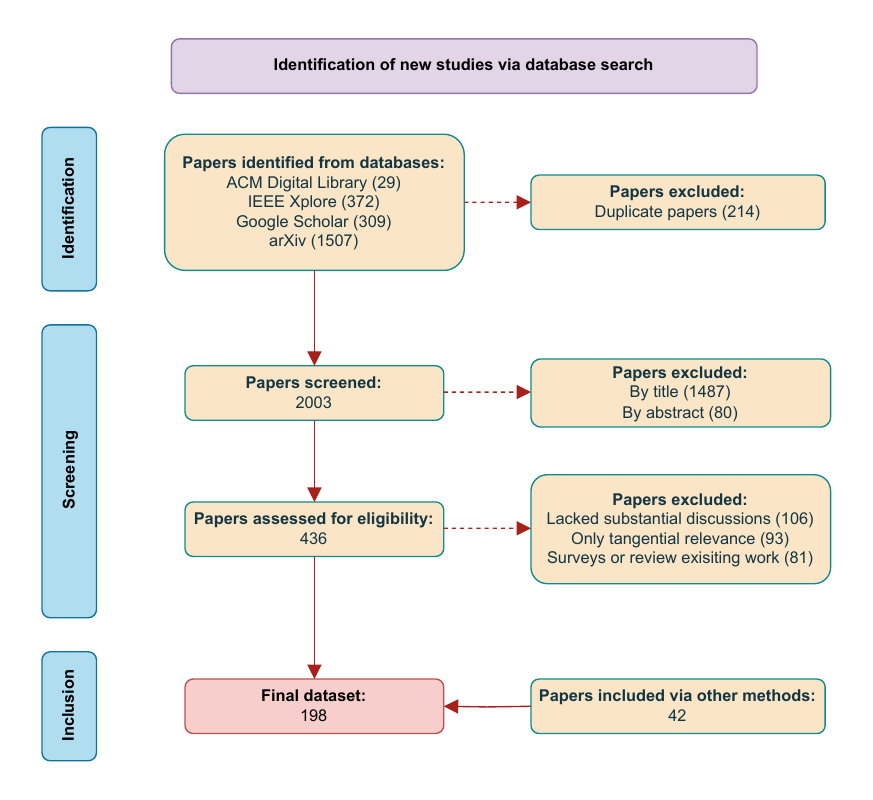}
    \caption{Steps and results from the application of the PRISMA guidelines -- the final list of papers comprises a total of \Nfinallist{} references.
    }
    \label{fig:prisma_flow_diagram}
\end{figure}

This study follows a systematic approach to identify and select relevant work related to threats and defenses of LLM-based systems. 
The methodology is structured according to the Preferred Reporting Items for Systematic Reviews and Meta-Analyses (PRISMA) guidelines \cite{haddaway2020prisma}, a well-documented framework that helps make systematic reviews meaningful, transparent, and reproducible. 
Figure~\ref{fig:prisma_flow_diagram} presents an overview of the selection process in the form of a PRISMA flow diagram. Next, we detail the steps performed in each of the three phases
illustrated in Figure~\ref{fig:prisma_flow_diagram}.

\subsection{Identification Phase}

The first step to identify relevant work is to create a precise search query using Boolean operators, wildcards (*), and keywords based on research goals. The search query constructed for this work is as follows:

\vspace{.3cm}

\fbox{\begin{minipage}{7.6cm}
\textbf{Search query:} (``LLM*" $\vee$ ``Language Model*" $\vee$ ``Generative AI" $\vee$ ``ChatGPT" $\vee$ ``GPT*" $\vee$ ``Chatbot*")  $\wedge$ (``Cybersecurity" $\vee$ ``Security" $\vee$ ``Attack*" $\vee$ ``Defense*" $\vee$ ``Protection" $\vee$ ``Risk*" $\vee$ ``Jailbreak*" $\vee$ ``Threat*" $\vee$ ``Privacy" $\vee$ ``Vulnerabilit*" $\vee$ ``Mitigation*" $\vee$ ``Prompt Injection*" $\vee$ ``Red Team*" $\vee$ ``Side-Channel" $\vee$ ``Membership Inference" $\vee$ ``Backdoor*")
\end{minipage}}

\vspace{.4cm}

We conducted a comprehensive search using the aforementioned query in ACM Digital Library, IEEE Xplore, Google Scholar, and arXiv. 
The search yielded a total of \Ninitialsearch{} references, of which \Ndupsremoved{} were identified as duplicates and removed. 
Hence, \Nfirstscreening{} papers continued to the screening phase.

\subsection{Screening Phase}

We performed a two-stage screening process to select the most relevant studies for 
this review, as described below.

\subsubsection{Title/Abstract Screening}

The first screening stage involved a preliminary review of the titles and abstracts of all references with the goal of removing publications that were not explicitly related to the security or privacy of LLMs. 
Studies that were deemed out of scope based on their title and abstract were excluded. This step reduced the analyzed list to \Nfulltextscreening{} references, excluding \Nremovalbytitle{} papers based on their titles and \Nspecialistremoval{} references based on their abstracts.

\subsubsection{Full-text Screening}
We reviewed the full-text versions of the remaining \Nfulltextscreening{} studies with two goals: (i)~ensure they are indeed high-quality articles relevant to the research objectives of this work; and (ii)~classify the papers according to their scope. A subjective analysis conducted by the authors filtered \Nlackedsubst{} works that lacked substantial discussion and \Ntangrelevance{} that presented only tangential relevance. The remaining references were classified into two types: (a) papers that propose specific attacks and mitigation for LLM systems, which are the object of study of this Systematization of Knowledge (SoK); and (b) references that provide reviews of LLM security and privacy literature, similar to this work. 
In Section~\ref{sec:related-work} we analyze  the \NsurveystypeB{} works classified as (b) and separate the relevant works for discussion. After completion of the eligibility assessment, \Nfulltextelegible{} studies of type (a) remained.

\subsection{Inclusion Phase}

The final step of PRISMA involves investigating and retrieving relevant works cited by the references selected in the screening phase. Throughout this process, we discovered \Nadditionalrefs{} additional type (a) references that should be added to the list. The final list contains \Nfinallist{} references that propose either attack techniques (presented and discussed in Section~\ref{sec:attacks_to_LLM}), defenses strategies (presented and detailed in Section~\ref{sec:mitigation_strategies}), or both.
\section{Related Work}
\label{sec:related-work}

Several studies review the literature on the security of GenAI,
and in particular LLMs. 
In this section, we present and compare some of these works with respect to the research methodology, scope, coverage, systematic analysis, and limitations. Namely, we consider the \NsurveystypeB{} references classified as type (b) during our selection process and position our contribution in relation to it.

The studies discussed in this section are further categorized into three groups:
(\ref{sec:related_work_group_risks})~publications that analyze the potential impact of LLMs regarding security, safety and ethics; 
(\ref{sec:related_work_group_attacks})~papers that provide taxonomies of attack and defense techniques in the context of LLMs; and
(\ref{sec:related_work_group_security})~works that addresses the security of LLM-based systems.

\subsection{Risks and Impacts of Generative AI} 
\label{sec:related_work_group_risks}
Works in this category provide directions on which aspects of cybersecurity can change in the presence of LLMs. The authors of \cite{slattery2024airisk} focus on creating a comprehensive and categorized database of AI risks extracted from various sources, dividing risks between causal taxonomy (i.e., characterizing by entity, intentionality, and timing factors) and domain taxonomy (separating by technological fields). 
As a timing factor, they split the risks considering two phases of the AI life cycle: pre-deployment and post-deployment. 
We follow this approach in the attack classification described in Section~\ref{sec:threats_LLM_use_cases}. 
The authors of \cite{barrett2023identifying} explore the potential for attacks, defenses, and safety issues in GenAI, proposing short- and long-term goals for the research community. The work of \cite{chua2024aisafety} also provides a systematic investigation of safety risks in LLMs focusing on practical implementation of safety measures in the development stages of LLMs, such as training data, model training, prompting, model alignment, and scaling. Similarly, industry-oriented work \cite{fed24} systematizes the security risks of Generative AI. 
Although they discuss several countermeasures to mitigate risks in the LLM development life cycle, these works do not address scenarios and issues that appear in LLM systems. References~\cite{derner2023safeguards} and \cite{gupta2023chatgpt} discuss the security risks associated with LLMs, focusing specifically on ChatGPT, and how LLM safeguards can be bypassed. References~\cite{golda2024privacy} and \cite{zhang2024llms} provide a comprehensive overview of privacy and security challenges in GenAI, covering technical, ethical, and regulatory aspects related to users and institutions. 

The impact of LLMs on cybersecurity in general, as well as safety and ethics concerns, is out of the scope of this work. 
Although we borrow concepts such as LLM life cycle phases from the aforementioned works to discuss security measures, our work differs from them 
by focusing
on attacks against LLM systems rather than the consequences of misuse.

\begin{table*}[t]
\centering
\scriptsize
\caption{
Comparison of related surveys. The filled circles (\fullcirc[0.6ex]) indicate a true correspondence of a research aspect to the work, half-filled circles (\halfcirc[0.6ex]) indicate a limited correspondence, blank indicates an absence of the aspect, and a cross ($\times$) indicates the specified limitation is present. 
}
\begin{tabular}{lp{0cm}p{0cm}p{0cm}p{0cm}p{0cm}p{0cm}p{0cm}p{0cm}p{0cm}p{0cm}p{0cm}p{0cm}p{0cm}p{0cm}p{0cm}p{0cm}p{0cm}p{0cm}p{0cm}p{0cm}p{0cm}p{0cm}p{-1cm}}

\toprule

  & \multicolumn{3}{c}{\textbf{Methodol.}} & \multicolumn{2}{c}{\textbf{Scope}} & \multicolumn{5}{c}{\textbf{Target}} & \multicolumn{5}{c}{\textbf{Attack Coverage}} & \multicolumn{6}{c}{\textbf{SoK Analysis}} & \multicolumn{2}{c}{\textbf{Limitation}}\\

\cmidrule(lr{1ex}){2-4}
\cmidrule(lr{1ex}){5-6}
\cmidrule(lr{1ex}){7-11}
\cmidrule(lr{0.5ex}){12-16}
\cmidrule(lr{0.5ex}){17-22}
\cmidrule(lr{0.5ex}){23-24}

\multicolumn{1}{c}{\multirow{-9.5}{*}{\textbf{Related Work}}}  & \rotatebox[origin=l]{90}{\textbf{Survey}}  & \rotatebox[origin=l]{90}{\textbf{Systematic Review}} & \rotatebox[origin=l]{90}{\textbf{Taxonomy}} & \rotatebox[origin=l]{90}{\textbf{Threats}} & \rotatebox[origin=l]{90}{\textbf{Mitigations}} & \rotatebox[origin=l]{90}{\textbf{LLM }} & \rotatebox[origin=l]{90}{\textbf{LLM-based System}} & \rotatebox[origin=l]{90}{\textbf{LLM-based Agents}} & \rotatebox[origin=l]{90}{\textbf{Database}} & \rotatebox[origin=l]{90}{\textbf{Supply-Chain}} & \rotatebox[origin=l]{90}{\textbf{Jailbreak}} & \rotatebox[origin=l]{90}{\textbf{Privacy}} & \rotatebox[origin=l]{90}{\textbf{Poisoning}} & \rotatebox[origin=l]{90}{\textbf{Disruption}} & \rotatebox[origin=l]{90}{\textbf{LLM App. Flaws}} & \rotatebox[origin=l]{90}{\textbf{LLM Lifecycle}} & \rotatebox[origin=l]{90}{\textbf{Design Choices}} & \rotatebox[origin=l]{90}{\textbf{Attack/Defense Map}} & \rotatebox[origin=l]{90}{\textbf{Use Cases}} & \rotatebox[origin=l]{90}{\textbf{Threat Model}} & \rotatebox[origin=l]{90}{\textbf{Risk Assessment}} & \rotatebox[origin=l]{90}{\textbf{Attack-Specific}} & \rotatebox[origin=l]{90}{\textbf{Model-Specific}} \\

\midrule

Neel et al. (2024) \cite{neel2023privacy}  & \fullcirc[0.6ex] & & \multicolumn{1}{c|}{ }  & \fullcirc[0.6ex] & \multicolumn{1}{c|}{ \fullcirc[0.6ex]}  & \fullcirc[0.6ex] & & & & \multicolumn{1}{c|}{ }  & & \fullcirc[0.6ex] & & & \multicolumn{1}{c|}{ }  & & & & & & \multicolumn{1}{c|}{ }  & $\times$  &  \\
Ahmed and Jothi (2024) \cite{surur2024jailbreak}  & \fullcirc[0.6ex] & & \multicolumn{1}{c|}{ \halfcirc[0.6ex]}  & \fullcirc[0.6ex] & \multicolumn{1}{c|}{ \fullcirc[0.6ex]}  & \fullcirc[0.6ex] & & & & \multicolumn{1}{c|}{ }  & \fullcirc[0.6ex] & & & & \multicolumn{1}{c|}{ }  & & & & & & \multicolumn{1}{c|}{ }  & $\times$  &  \\
Guo et al. (2022) \cite{guo2022threats}  & \fullcirc[0.6ex] & & \multicolumn{1}{c|}{ \fullcirc[0.6ex]}  & \fullcirc[0.6ex] & \multicolumn{1}{c|}{ }  & \fullcirc[0.6ex] & & & & \multicolumn{1}{c|}{ }  & & \fullcirc[0.6ex] & \fullcirc[0.6ex] & & \multicolumn{1}{c|}{ }  & \fullcirc[0.6ex] & & & & & \multicolumn{1}{c|}{ }  & &  \\
Huang et al. (2024) \cite{huang2023identifying}  & \fullcirc[0.6ex] & \halfcirc[0.6ex] & \multicolumn{1}{c|}{ \fullcirc[0.6ex]}  & \fullcirc[0.6ex] & \multicolumn{1}{c|}{ \fullcirc[0.6ex]}  & \fullcirc[0.6ex] & & & & \multicolumn{1}{c|}{ }  & & \fullcirc[0.6ex] & & & \multicolumn{1}{c|}{ }  & \fullcirc[0.6ex] & \fullcirc[0.6ex] & & & & \multicolumn{1}{c|}{ }  & $\times$  &  \\
Yi et al. (2024) \cite{yi2024jailbreak}  & \fullcirc[0.6ex] & & \multicolumn{1}{c|}{ \fullcirc[0.6ex]}  & \fullcirc[0.6ex] & \multicolumn{1}{c|}{ \fullcirc[0.6ex]}  & \fullcirc[0.6ex] & & & & \multicolumn{1}{c|}{ }  & \fullcirc[0.6ex] & & & & \multicolumn{1}{c|}{ }  & & & & \fullcirc[0.6ex] & & \multicolumn{1}{c|}{ \halfcirc[0.6ex]}  & $\times$  &  \\
Rababah et al. (2024) \cite{rabalah2024sok}  & \fullcirc[0.6ex] & \fullcirc[0.6ex] & \multicolumn{1}{c|}{ }  & \fullcirc[0.6ex] & \multicolumn{1}{c|}{ \fullcirc[0.6ex]}  & \fullcirc[0.6ex] & & & & \multicolumn{1}{c|}{ }  & \fullcirc[0.6ex] & \halfcirc[0.6ex] & & & \multicolumn{1}{c|}{ }  & & & & \fullcirc[0.6ex] & & \multicolumn{1}{c|}{ \halfcirc[0.6ex]}  & $\times$  &  \\
Huang et al. (2024) \cite{huang2024harmful}  & \fullcirc[0.6ex] & & \multicolumn{1}{c|}{ \fullcirc[0.6ex]}  & \fullcirc[0.6ex] & \multicolumn{1}{c|}{ \fullcirc[0.6ex]}  & \fullcirc[0.6ex] & & & & \multicolumn{1}{c|}{ }  & \halfcirc[0.6ex] & & \fullcirc[0.6ex] & & \multicolumn{1}{c|}{ }  & \halfcirc[0.6ex] & & & & & \multicolumn{1}{c|}{ }  & $\times$  &  \\
Miranda et al. (2025) \cite{miranda2024preserving}  & \fullcirc[0.6ex] & & \multicolumn{1}{c|}{ \fullcirc[0.6ex]}  & \fullcirc[0.6ex] & \multicolumn{1}{c|}{ \fullcirc[0.6ex]}  & \fullcirc[0.6ex] & & & & \multicolumn{1}{c|}{ }  & & \fullcirc[0.6ex] & & & \multicolumn{1}{c|}{ }  & \halfcirc[0.6ex] & & & & & \multicolumn{1}{c|}{ }  & $\times$  &  \\
Peng et al. (2024) \cite{peng2024jailbreaking}  & \fullcirc[0.6ex] & & \multicolumn{1}{c|}{ \fullcirc[0.6ex]}  & \fullcirc[0.6ex] & \multicolumn{1}{c|}{ \fullcirc[0.6ex]}  & \fullcirc[0.6ex] & & \halfcirc[0.6ex] & & \multicolumn{1}{c|}{ }  & \fullcirc[0.6ex] & & \fullcirc[0.6ex] & & \multicolumn{1}{c|}{ }  & & & & & & \multicolumn{1}{c|}{ }  & $\times$  &  \\
Chowdhury et al. (2024) \cite{chowdhury2024breaking}  & \fullcirc[0.6ex] & & \multicolumn{1}{c|}{ \fullcirc[0.6ex]}  & \fullcirc[0.6ex] & \multicolumn{1}{c|}{ \fullcirc[0.6ex]}  & \fullcirc[0.6ex] & & & & \multicolumn{1}{c|}{ }  & \fullcirc[0.6ex] & \halfcirc[0.6ex] & \fullcirc[0.6ex] & & \multicolumn{1}{c|}{ }  & & & & & & \multicolumn{1}{c|}{ }  & &  \\
Liu et al. (2024) \cite{liu2024formalizing}  & \fullcirc[0.6ex] & & \multicolumn{1}{c|}{ \fullcirc[0.6ex]}  & \fullcirc[0.6ex] & \multicolumn{1}{c|}{ \fullcirc[0.6ex]}  & \fullcirc[0.6ex] & & & & \multicolumn{1}{c|}{ }  & \fullcirc[0.6ex] & & & & \multicolumn{1}{c|}{ \fullcirc[0.6ex]}  & & & & & \fullcirc[0.6ex] & \multicolumn{1}{c|}{ }  & $\times$  &  \\
Chen et al. (2024) \cite{chen2024security}  & \fullcirc[0.6ex] & \fullcirc[0.6ex] & \multicolumn{1}{c|}{ \fullcirc[0.6ex]}  & \fullcirc[0.6ex] & \multicolumn{1}{c|}{ \fullcirc[0.6ex]}  & \fullcirc[0.6ex] & & & & \multicolumn{1}{c|}{ }  & & & \fullcirc[0.6ex] & & \multicolumn{1}{c|}{ }  & \halfcirc[0.6ex] & & & \fullcirc[0.6ex] & & \multicolumn{1}{c|}{ }  & & $\times$   \\
Ferrag et al. (2025) \cite{amine2024generative}  & \fullcirc[0.6ex] & & \multicolumn{1}{c|}{ }  & \halfcirc[0.6ex] & \multicolumn{1}{c|}{ \fullcirc[0.6ex]}  & \fullcirc[0.6ex] & \fullcirc[0.6ex] & & & \multicolumn{1}{c|}{ }  & \halfcirc[0.6ex] & & \fullcirc[0.6ex] & \fullcirc[0.6ex] & \multicolumn{1}{c|}{ \fullcirc[0.6ex]}  & \halfcirc[0.6ex] & & & \fullcirc[0.6ex] & & \multicolumn{1}{c|}{ }  & &  \\
Yan et al. (2025) \cite{yan2024protecting}  & \fullcirc[0.6ex] & \fullcirc[0.6ex] & \multicolumn{1}{c|}{ }  & \fullcirc[0.6ex] & \multicolumn{1}{c|}{ \fullcirc[0.6ex]}  & \fullcirc[0.6ex] & & & & \multicolumn{1}{c|}{ }  & & \fullcirc[0.6ex] & \fullcirc[0.6ex] & & \multicolumn{1}{c|}{ }  & \fullcirc[0.6ex] & & \fullcirc[0.6ex] & & \halfcirc[0.6ex] & \multicolumn{1}{c|}{ }  & $\times$  & $\times$   \\
Abdali et al. (2024) \cite{abdali2024securing}  & \fullcirc[0.6ex] & & \multicolumn{1}{c|}{ \fullcirc[0.6ex]}  & \fullcirc[0.6ex] & \multicolumn{1}{c|}{ \fullcirc[0.6ex]}  & \fullcirc[0.6ex] & & & & \multicolumn{1}{c|}{ }  & \fullcirc[0.6ex] & \fullcirc[0.6ex] & \fullcirc[0.6ex] & & \multicolumn{1}{c|}{ }  & \halfcirc[0.6ex] & & \halfcirc[0.6ex] & & & \multicolumn{1}{c|}{ }  & &  \\
Cui et al. (2024) \cite{cui2024recent}  & \fullcirc[0.6ex] & & \multicolumn{1}{c|}{ }  & \fullcirc[0.6ex] & \multicolumn{1}{c|}{ \fullcirc[0.6ex]}  & \fullcirc[0.6ex] & & & \halfcirc[0.6ex] & \multicolumn{1}{c|}{ \halfcirc[0.6ex]}  & \fullcirc[0.6ex] & \fullcirc[0.6ex] & \fullcirc[0.6ex] & \fullcirc[0.6ex] & \multicolumn{1}{c|}{ \halfcirc[0.6ex]}  & \fullcirc[0.6ex] & & \fullcirc[0.6ex] & & & \multicolumn{1}{c|}{ \halfcirc[0.6ex]}  & &  \\
Das et al. (2025) \cite{das2024security}  & \fullcirc[0.6ex] & \halfcirc[0.6ex] & \multicolumn{1}{c|}{ \fullcirc[0.6ex]}  & \fullcirc[0.6ex] & \multicolumn{1}{c|}{ \fullcirc[0.6ex]}  & \fullcirc[0.6ex] & & & & \multicolumn{1}{c|}{ }  & \fullcirc[0.6ex] & \fullcirc[0.6ex] & \fullcirc[0.6ex] & & \multicolumn{1}{c|}{ }  & \fullcirc[0.6ex] & \halfcirc[0.6ex] & \fullcirc[0.6ex] & & & \multicolumn{1}{c|}{ }  & &  \\
He et al. (2024) \cite{he2024the}  & \fullcirc[0.6ex] & & \multicolumn{1}{c|}{ \halfcirc[0.6ex]}  & \fullcirc[0.6ex] & \multicolumn{1}{c|}{ \fullcirc[0.6ex]}  & \fullcirc[0.6ex] & \fullcirc[0.6ex] & \fullcirc[0.6ex] & & \multicolumn{1}{c|}{ }  & \fullcirc[0.6ex] & \fullcirc[0.6ex] & \fullcirc[0.6ex] & & \multicolumn{1}{c|}{ }  & & & \fullcirc[0.6ex] & \fullcirc[0.6ex] & & \multicolumn{1}{c|}{ }  & &  \\
Wang et al. (2024) \cite{wang2018unique}  & \fullcirc[0.6ex] & & \multicolumn{1}{c|}{ \fullcirc[0.6ex]}  & \fullcirc[0.6ex] & \multicolumn{1}{c|}{ \fullcirc[0.6ex]}  & \fullcirc[0.6ex] & & \fullcirc[0.6ex] & \fullcirc[0.6ex] & \multicolumn{1}{c|}{ }  & \fullcirc[0.6ex] & \fullcirc[0.6ex] & \fullcirc[0.6ex] & & \multicolumn{1}{c|}{ }  & \fullcirc[0.6ex] & & \fullcirc[0.6ex] & \fullcirc[0.6ex] & \fullcirc[0.6ex] & \multicolumn{1}{c|}{ }  & &  \\
Gan et al. (2024) \cite{gan2024navigating}  & \fullcirc[0.6ex] & & \multicolumn{1}{c|}{ \fullcirc[0.6ex]}  & \fullcirc[0.6ex] & \multicolumn{1}{c|}{ \fullcirc[0.6ex]}  & \fullcirc[0.6ex] & \fullcirc[0.6ex] & \fullcirc[0.6ex] & & \multicolumn{1}{c|}{ }  & \fullcirc[0.6ex] & \fullcirc[0.6ex] & \fullcirc[0.6ex] & & \multicolumn{1}{c|}{ }  & & \fullcirc[0.6ex] & & \fullcirc[0.6ex] & \fullcirc[0.6ex] & \multicolumn{1}{c|}{ }  & &  \\
Yao et al. (2024) \cite{yao2024asurvey}  & \fullcirc[0.6ex] & \fullcirc[0.6ex] & \multicolumn{1}{c|}{ \fullcirc[0.6ex]}  & \fullcirc[0.6ex] & \multicolumn{1}{c|}{ \fullcirc[0.6ex]}  & \fullcirc[0.6ex] & & & & \multicolumn{1}{c|}{ \fullcirc[0.6ex]}  & \fullcirc[0.6ex] & \fullcirc[0.6ex] & \fullcirc[0.6ex] & \halfcirc[0.6ex] & \multicolumn{1}{c|}{ \fullcirc[0.6ex]}  & \halfcirc[0.6ex] & & \fullcirc[0.6ex] & & & \multicolumn{1}{c|}{ }  & &  \\
Cui et al. (2024) \cite{cui2024risk}  & \fullcirc[0.6ex] & & \multicolumn{1}{c|}{ \fullcirc[0.6ex]}  & \fullcirc[0.6ex] & \multicolumn{1}{c|}{ \fullcirc[0.6ex]}  & \fullcirc[0.6ex] & \fullcirc[0.6ex] & & & \multicolumn{1}{c|}{ \fullcirc[0.6ex]}  & \fullcirc[0.6ex] & \fullcirc[0.6ex] & \fullcirc[0.6ex] & \fullcirc[0.6ex] & \multicolumn{1}{c|}{ \fullcirc[0.6ex]}  & \halfcirc[0.6ex] & & & & \halfcirc[0.6ex] & \multicolumn{1}{c|}{ \halfcirc[0.6ex]}  & &  \\
Huang et al. (2024) \cite{huang2024lifting}  & \fullcirc[0.6ex] & \fullcirc[0.6ex] & \multicolumn{1}{c|}{ \fullcirc[0.6ex]}  & \fullcirc[0.6ex] & \multicolumn{1}{c|}{ \fullcirc[0.6ex]}  & \fullcirc[0.6ex] & \fullcirc[0.6ex] & & \fullcirc[0.6ex] & \multicolumn{1}{c|}{ \fullcirc[0.6ex]}  & \fullcirc[0.6ex] & \fullcirc[0.6ex] & \fullcirc[0.6ex] & \fullcirc[0.6ex] & \multicolumn{1}{c|}{ \fullcirc[0.6ex]}  & \fullcirc[0.6ex] & \halfcirc[0.6ex] & & \fullcirc[0.6ex] & & \multicolumn{1}{c|}{ }  & $\times$  &  \\
\midrule \textbf{This work} & \fullcirc[0.6ex] & \fullcirc[0.6ex] & \multicolumn{1}{c|}{ \fullcirc[0.6ex]}  & \fullcirc[0.6ex] & \multicolumn{1}{c|}{ \fullcirc[0.6ex]}  & \fullcirc[0.6ex] & \fullcirc[0.6ex] & \fullcirc[0.6ex] & \fullcirc[0.6ex] & \multicolumn{1}{c|}{ \fullcirc[0.6ex]}  & \fullcirc[0.6ex] & \fullcirc[0.6ex] & \fullcirc[0.6ex] & \fullcirc[0.6ex] & \multicolumn{1}{c|}{ \fullcirc[0.6ex]}  & \fullcirc[0.6ex] & \fullcirc[0.6ex] & \fullcirc[0.6ex] & \fullcirc[0.6ex] & \fullcirc[0.6ex] & \multicolumn{1}{c|}{ \fullcirc[0.6ex]}  & &  \\

\bottomrule

\end{tabular}

\label{tab:related_work}
\end{table*}

\subsection{LLM Attacks and Defenses}
\label{sec:related_work_group_attacks}

While systematizing LLM-related risks may provide an overview of the landscape of offensive techniques, some studies focus specifically on attacks to LLMs and their protection mechanisms. 
Research on privacy attacks conducted in \cite{rigaki2023survey, huang2023identifying, neel2023privacy, miranda2024preserving, yan2024protecting}, and~\cite{hartmann2023sok} analyzes attack methods, their impacts, and corresponding mitigation strategies for securing LLM development, effectively proposing taxonomies for attack techniques and defensive measures. The authors of \cite{huang2024harmful} focus on categorizing and evaluating harmful fine-tuning attacks against LLMs, while jailbreak attacks are systematized in depth by several authors following multiple approaches and taxonomies~\cite{shen2024anything,xu2024comprehensive,yi2024jailbreak,rabalah2024sok,maria2024what,benjamin2024systematically,liu2023jailbreaking,cai2024rethinking,schulhoff2023ignore,liu2024formalizing,peng2024jailbreaking,surur2024jailbreak,wang2024from,
yi2024jailbreak, rabalah2024sok
}. 
The work in \cite{shen2024anything} surveys real jailbreak prompts collected from online forums and evaluates their effectiveness against state-of-the-art models, while \cite{xu2024comprehensive} analyzes nine jailbreak attack techniques and benchmarks their performance against their corresponding defensive countermeasures. 
The works in \cite{maria2024what} and \cite{benjamin2024systematically} focus on understanding how models can be jailbroken. 
In \cite{liu2023jailbreaking,cao2023red}, the authors empirically explore the effectiveness of jailbreak attacks against \textit{ChatGPT}, while \cite{cantini2024are} explores LLM biases and robustness in the face of jailbreak prompts. In \cite{cai2024rethinking}, the authors argue that current jailbreak evaluation processes present limitations and propose new ways to evaluate such threat.
Concerning prompt injection attacks, \cite{schulhoff2023ignore} studies real prompt injection attacks from a crowd-sourced collection approach and proposes a taxonomy for 29 prompt attack techniques. 
Research on \cite{liu2024formalizing} advances the state-of-the-art taxonomy of prompt injection attacks by developing a formal characterization framework, generating a benchmark for evaluating prompt injection attacks and defenses.

All aforementioned works share the limitation of focusing on one attack class, either privacy, fine-tuning, jailbreak, or prompt injection, with the last two attack classes often being used interchangeably. On the other hand, \cite{chen2024security} explores unique threats and mitigations of LLMs specialized in coding tasks. With a broader attack landscape, \cite{cui2024recent} and \cite{guo2022threats} explore recent research on LLM threats and vulnerabilities, discussing current challenges and open problems, while \cite{chowdhury2024breaking} and \cite{abdali2024securing} provide comprehensive surveys of several attack classes, including jailbreaks, prompt injections, and data poisoning. 
Considering safety issues, \cite{dong2024attacks} examines methods to attack LLMs focusing on conversational safety, exploring factors such as toxicity, discrimination, privacy, and misinformation.
Our work not only aims to extend the coverage of attack classes presented in these papers but also includes threats against LLM systems and agents with specific architectures.

Industry-oriented efforts also play a role in creating taxonomies for LLM attacks. 
MITRE \cite{mitre} created a knowledge base to catalog known adversarial tactics, techniques, and procedures (TTPs) relevant to AI systems, and the National Institute of Standards and Technology (NIST)~\cite{nist_adversarial_ml_attacks_2023} developed a taxonomy and terminology in the field of adversarial machine learning. 
Although both initiatives are crucial to help standardize LLM attack techniques, they present the same drawbacks as academic works by offering only limited coverage of attacks against LLM systems.
Finally,~\cite{AISafetyReport2025}, an international effort to provide guidelines on developing LLM systems safely, recommends threat modeling as a supportive process but does not explore it in depth. We address these gaps in this paper.

\subsection{Security of LLM-based Systems} 
\label{sec:related_work_group_security}

Works of this category are the most similar to ours in the sense that they not only discuss threats against the model, but also consider the system to which the LLM integrates. The work from~\cite{altun2024securing} presents an overview of AI attacks and defense strategies and proposes a security checklist to be followed by AI developers when building an AI-based application, although it does not cover LLM-specific security aspects. 
The work in \cite{das2024security} reviews the security and privacy issues of LLM systems, with special attention to the architectural components of LLMs, analyzing two examples of scenarios in which LLM vulnerabilities are exploited. The authors of \cite{yao2024asurvey} proposed a taxonomy of LLM threats and defense techniques, including attacks that can be executed using LLMs and threats targeting the LLM-integrated application, such as Remote Code Execution (RCE), side-channel, and supply-chain attacks, although
the 
respective mitigations are outside the scope of their
paper. 
The research from \cite{liu2024demystifying} proposes LLMSmith, a static analysis tool to scan the source code of LLM systems to detect RCE vulnerabilities and classify real threats detected in the wild. 
In turn, \cite{brokman2024insights} provides a comparative study of open-source LLM vulnerability scanners, assessing their performance and capabilities. 
Both \cite{liu2024demystifying} and \cite{brokman2024insights} address real LLM system security issues, but do not explore threat modeling and use case scenarios in which the detected security issues could be exploited. Reference~\cite{amine2024generative} explores six different scenarios in which attacks can occur. 
The survey in~\cite{cui2024risk} proposes a comprehensive risk taxonomy in LLM systems and catalogs benchmarks for LLM safety and security evaluations. 
It also categorizes risks according to the corresponding modules of the architecture of the LLM system and proposes mitigation techniques, thereby providing an initial effort on threat modeling, although from a high-level perspective and limited to a specific type of application. 
The study in \cite{tete2024threat} explores the security of LLM-integrated applications related to six types of attacks and proposes a threat model framework based on the STRIDE~\cite{kohnfelder1999threats} and DREAD~\cite{dread} frameworks, 
providing an interesting study case of a custom-built LLM-powered application to demonstrate the proposed framework. 
The surveys in~\cite{he2024the} and \cite{gan2024navigating} review attacks against LLM agents, while \cite{wang2018unique} also addresses threats of LLM-related systems, such as the Retrieval-Augmented Generation (RAG) database. Finally, the survey of \cite{huang2024lifting}, instead, reviews the risks of the LLM supply chain.

Table~\ref{tab:related_work} presents a detailed characterization of the most relevant literature surveys that focus on threats to LLMs or LLM systems and compares different research aspects to our work.
To the best of our knowledge, no prior work covers LLM threats and defensive strategies considering multiple 
LLM scenarios, with different use case and design choices that may affect security and privacy. Therefore, we analyze not only the LLM development life cycle, but also the integration between LLMs and common software development phases, inspired by real-world implementations.
We assign severity score levels to the analyzed threats and map them to the LLM scenarios they apply. Defensive strategies are grouped and associated with the LLM development life cycle phases in which they should be implemented, making mitigation easier and more precise. Finally, we provide a threat modeling evaluation based on the STRIDE framework
for different LLM scenarios, showing the security implications of different system design choices and recommending defensive strategies to attenuate threats.
\section{Characterization of LLM Scenarios}
\label{sec:llm_scenarios}

\begin{figure*}[ht]
\includegraphics[scale=0.5]{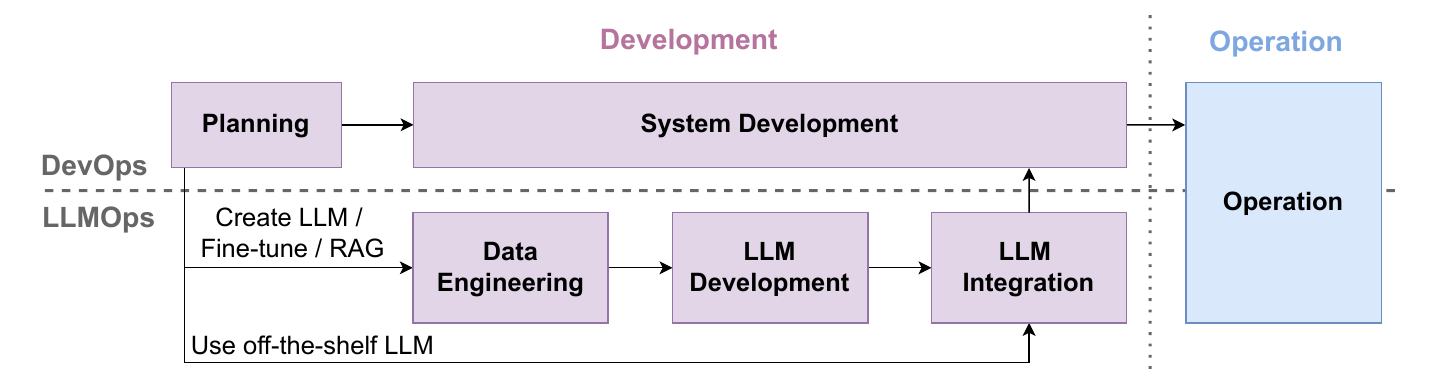}
\centering
\caption{Adopted life cycle model of an LLM system. The phases can be classified along two axes according to their nature: system-related phases (DevOps) vs. LLM-related phases (LLMOps), and Development vs. Operation phases.}
\label{fig:llm_lifecycle}
\end{figure*}

Developing an LLM-based system is a multifaceted process that requires careful planning, execution, and continuous improvement.
Furthermore, the system security is intrinsically linked to securing the process itself, highlighting the need for a thorough threat assessment at every stage of the development and operation pipeline.
Hence, it is imperative to understand the life cycle of an LLM-based system to elucidate how the threats examined in this paper may disrupt such a system. 

The LLM scenarios discussed in this paper are characterized by their life cycle stage (ST), use case (UC), and decisions for the existing design choices (DP, DI, SL, SI, IO, AR, CL), further explained in this section. To uniquely represent a combination of these variables, we propose a canonical representation of an LLM scenario defined by a string, 

{\small 
\[
 \text{LLM Scenario} \coloneqq \text{ST:} \! * \! / \text{UC:} \! * \!  / \text{DP:} \! * \! / \text{DI:} \! * \! / \text{SL:} \! * \! / \text{SI:} \! * \! / \text{IO:} \! * \! / \text{AR:} \! * \! / \text{CL:} *
\]
}

\noindent whereas the asterisks represent the possible configuration values from stage, use case, and the 7 design choices. Finally, we provide four examples of real scenarios created by this framework.

\subsection{Life Cycle of LLM-based Systems}
\label{sec:lifecycle}

The life cycle of LLM-based systems has been conceptualized in various ways, leading to references that describe in detail the numerous actions involved 
from the creation of the model to its integration into a system and its operation~\cite{pahune_llmops, AISafetyReport2025}. 
For clarity, this paper adopts a simplified model tailored to the classification of threats that aggregates such actions into phases. 
Figure~\ref{fig:llm_lifecycle} depicts such a model, 
with the vertical axis separating the \textit{Development} stage, which encompasses all pre-production processes for both the LLM and the system (or software application that interacts with the LLM), from the \textit{Operation} stage, which contains the processes occurring post-deployment. 
Similarly, the horizontal axis distinguishes the processes related to system development and deployment (\textit{DevOps}) from the phases specific to the development and integration of LLMs (\textit{LLMOps}).

The phases involved in the system development are:

\begin{itemize}
    \item \textbf{Planning Phase}. In this initial phase, the development team delineates system requirements, schedules, milestones, and the overarching architecture of the system, including the selection of the appropriate LLM. 
    A critical decision during this phase is whether to develop a proprietary model through training or fine-tuning, which necessitates the collection and preparation of relevant data, or to integrate a general-purpose, off-the-shelf LLM directly into the system. 
    Such a decision defines which path will be taken in the LLMOps cycle.
    \item \textbf{System Development Phase}. This phase covers most of the software development life cycle, including coding, building, and testing. 
    Note that independently of the planning phase decision regarding which LLM to use (create a new one, customize an existing one, or develop a RAG application), the system with the integrated LLM must be tested prior to deployment.
\end{itemize}

The LLMOps process must be undertaken when a new model is to be developed, when a general-purpose model needs to acquire specific knowledge to enhance response quality in the application context, or even when an off-the-shelf LLM is just integrated into a system, requiring fewer steps in this case.
For the particular case of providing specific knowledge to an LLM, 
this can be achieved through two primary approaches: model fine-tuning and RAG. 
For model fine-tuning, particular data is collected to retrain the model, enabling it to provide more specific and accurate responses. 
In contrast, the RAG approach seeks to construct a database of pertinent information, which the LLM consults prior to reasoning, thereby yielding more precise answers. 
Regardless of the approach, the LLMOps cycle comprises three phases:

\begin{itemize}
    \item \textbf{Data Engineering Phase}. In this initial phase, relevant data must be collected, analyzed, and prepared
    for the creation of the LLM or 
    to assist in fine-tuning an existing LLM 
    to align with the primary objectives of the solution.
    The goal is to acquire high-quality and diverse data to address a broad spectrum of cases. 
    For RAG applications, this phase involves collecting and curating data to establish an efficient database.
    \item \textbf{LLM Development Phase}. This phase entails the 
    creation of a foundation model or the
    selection and fine-tuning of an existing one using the data prepared in the previous phase. 
    For RAG applications, the database is built, and the RAG process is established through the selection of an efficient embedding and retrieval algorithm. 
    Subsequently, the LLM is evaluated by developers to mitigate any unexpected behavior.
    \item\textbf{LLM Integration Phase}. After the developed LLM is fully tested, it is integrated into the software, culminating in the release and deployment of the system. 
\end{itemize}

When the LLM is successfully integrated and tested, the software can be considered ready for operation: 
\begin{itemize}
    \item \textbf{Operation Phase}. Upon deployment, the system transitions into the production phase, in which both the developed software and the LLM operate together to provide the desired service. 
    Continuous monitoring is implemented to collect user feedback and identify new data that may assist in aligning the LLM with the desired objectives.
    Although for simplicity, this iterative cycle is not depicted in the Figure~\ref{fig:llm_lifecycle}, this feedback-induced process can be conceptualized as small iterative cycles of the LLMOps collect-train-integrate process~\cite{AISafetyReport2025}. 
    This ensures that the model is perpetually fine-tuned based on observed data.
    This phase can also receive data related to system failures and any other anomalies.
\end{itemize}

We emphasize that the life cycle refers to software companies that need to adapt LLMs to align with the specific requirements of an application. 
AI companies that develop general-purpose LLMs such as ChatGPT, DeepSeek, and Llama may follow a shorter version of the LLMOps process in which the model itself constitutes the final product, bypassing the integration phase. Moreover, the development of foundation models involves extensive data collection, substantially more than that required for fine-tuning or RAG, and involves training the model from scratch, which demands considerably greater effort.

Using the proposed life cycle, we define high-level \textit{use cases} for LLM-based systems to better characterize the unique threats they are subject to. 
We categorize the use cases using the two main stages of the life cycle: Development and Operation. 
Shaped by a combination of design choices, the use cases are derived into more granular cases that differ in variety, impact, and likelihood of threats. 
We refer to each unique variation of design choices within a use case as a \textit{scenario}. The comprehensive list of all use cases discussed in this paper and their potential design choices is presented in Table~\ref{tab:LLM_use_cases}. 
In the following sections, we characterize the use cases and the aspects that may vary within them.

\begin{table*}[ht]
\centering
\scriptsize
\caption{Map of possible design choices for each development and operation uses cases of LLM Systems. Circles indicate that a design choice can have such configuration for a scenario, while crosses indicate that it is not a valid configuration by definition. When a design category is not suitable for either (D) or (O) use cases, it is marked as not applicable.}

\begin{tabular}{llcccccc}

\toprule

\multicolumn{2}{c}{} & \multicolumn{3}{c}{\textbf{Development Use Cases (D)}} & \multicolumn{3}{c}{\textbf{Operation Use Cases (O)}} \\
\cmidrule(lr){3-5}
\cmidrule(lr){6-8}
\multicolumn{2}{c}{\multirow{-1.6}{*}{\textbf{Design Choice}}} & \textbf{Foundation Model} & \textbf{Fine-Tuning} & \textbf{RAG} & \textbf{Chat-bot} & \textbf{Integr. App} & \textbf{Agent} \\
\multicolumn{2}{c}{\multirow{-2.6}{*}{}} & \textbf{(FM)} & \textbf{(FT)} & \textbf{(RG)} & \textbf{(CB)} & \textbf{(AP)} & \textbf{(AG)} \\

\midrule

\cellcolor[HTML]{FFFFFF} & \textbf{Public (U)} & \fullcirc[0.6ex] & \fullcirc[0.6ex] & \fullcirc[0.6ex] & \multicolumn{3}{c}{} \\

\cellcolor[HTML]{FFFFFF} & \textbf{Private (R)} & \fullcirc[0.6ex] & \fullcirc[0.6ex] & \fullcirc[0.6ex] & \multicolumn{3}{c}{} \\

\multirow{-3}{*}{\cellcolor[HTML]{FFFFFF}\textbf{Data Provenance (DP)}} & \textbf{Hybrid (H)} & \fullcirc[0.6ex] & \fullcirc[0.6ex] & \fullcirc[0.6ex] & \multicolumn{3}{c}{\multirow{-3}{*}{\textbf{Not applicable}}} \\ \midrule

\cellcolor[HTML]{FFFFFF} & \textbf{On-Premises (P)} & \fullcirc[0.6ex] & \fullcirc[0.6ex] & \fullcirc[0.6ex] & \fullcirc[0.6ex] & \fullcirc[0.6ex] & \fullcirc[0.6ex] \\

\cellcolor[HTML]{FFFFFF} & \textbf{On-Cloud (C)} & \fullcirc[0.6ex] & \fullcirc[0.6ex] & \fullcirc[0.6ex] & \fullcirc[0.6ex] & \fullcirc[0.6ex] & \fullcirc[0.6ex] \\

\cellcolor[HTML]{FFFFFF} & \textbf{On-Device (D)} & $\times$ & $\times$ & \fullcirc[0.6ex] & \fullcirc[0.6ex] & \fullcirc[0.6ex] & \fullcirc[0.6ex] \\

\multirow{-4}{*}{\cellcolor[HTML]{FFFFFF}\textbf{Dev. and Deploy. Infrastructure (DI)}} & \textbf{Hybrid (H)} & $\times$ & $\times$ & \fullcirc[0.6ex] & \fullcirc[0.6ex] & \fullcirc[0.6ex] & \fullcirc[0.6ex] \\ \midrule
\cellcolor[HTML]{FFFFFF} & \textbf{Proprietary (P)} & \fullcirc[0.6ex] & \fullcirc[0.6ex] & \fullcirc[0.6ex] & \fullcirc[0.6ex] & \fullcirc[0.6ex] & \fullcirc[0.6ex] \\

\cellcolor[HTML]{FFFFFF} & \textbf{Open-source (O)} & \fullcirc[0.6ex] & \fullcirc[0.6ex] & \fullcirc[0.6ex] & \fullcirc[0.6ex] & \fullcirc[0.6ex] & \fullcirc[0.6ex] \\

\multirow{-3}{*}{\cellcolor[HTML]{FFFFFF}\textbf{SW Libraries and Dependencies (SL)}} & \textbf{Hybrid (H)} & \fullcirc[0.6ex] & \fullcirc[0.6ex] & \fullcirc[0.6ex] & \fullcirc[0.6ex] & \fullcirc[0.6ex] & \fullcirc[0.6ex] \\ \midrule
\cellcolor[HTML]{FFFFFF} & \textbf{Yes (Y)} & \fullcirc[0.6ex] & \fullcirc[0.6ex] & \fullcirc[0.6ex] & \fullcirc[0.6ex] & \fullcirc[0.6ex] & \fullcirc[0.6ex] \\

\multirow{-2}{*}{\cellcolor[HTML]{FFFFFF}\textbf{Shared Infrastructure (SI)}} & \textbf{No (N)} & \fullcirc[0.6ex] & \fullcirc[0.6ex] & \fullcirc[0.6ex] & \fullcirc[0.6ex] & \fullcirc[0.6ex] & \fullcirc[0.6ex] \\ \midrule
\cellcolor[HTML]{FFFFFF} & \textbf{Text Field (T)} & \multicolumn{3}{c}{} & \fullcirc[0.6ex] & $\times$ & \fullcirc[0.6ex] \\

\cellcolor[HTML]{FFFFFF} & \textbf{App (A)} & \multicolumn{3}{c}{} & $\times$ & \fullcirc[0.6ex] & \fullcirc[0.6ex] \\

\cellcolor[HTML]{FFFFFF} & \textbf{Voice (V)} & \multicolumn{3}{c}{} & \fullcirc[0.6ex] & $\times$ & \fullcirc[0.6ex] \\

\multirow{-4}{*}{\cellcolor[HTML]{FFFFFF}\textbf{Prompt Input Origin (IO)}} & \textbf{Hybrid (H)} & \multicolumn{3}{c}{\multirow{-4}{*}{\textbf{Not applicable}}} & \fullcirc[0.6ex] & $\times$ & \fullcirc[0.6ex] \\ \midrule
\cellcolor[HTML]{FFFFFF} & \textbf{No (N)} & \multicolumn{3}{c}{} & \fullcirc[0.6ex] & \fullcirc[0.6ex] & \fullcirc[0.6ex] \\

\cellcolor[HTML]{FFFFFF} & \textbf{Tools (T)} & \multicolumn{3}{c}{} & $\times$ & \fullcirc[0.6ex] & \fullcirc[0.6ex] \\

\cellcolor[HTML]{FFFFFF} & \textbf{DBs (D)} & \multicolumn{3}{c}{} & \fullcirc[0.6ex] & \fullcirc[0.6ex] & \fullcirc[0.6ex] \\

\cellcolor[HTML]{FFFFFF} & \textbf{HW/SW Sensors (S)} & \multicolumn{3}{c}{} & $\times$ & $\times$ & \fullcirc[0.6ex] \\

\cellcolor[HTML]{FFFFFF} & \textbf{Internet (I)} & \multicolumn{3}{c}{} & $\times$ & \fullcirc[0.6ex] & \fullcirc[0.6ex] \\

\multirow{-6}{*}{\cellcolor[HTML]{FFFFFF}\textbf{Access to Resources (AR)}} & \textbf{Hybrid (H)} & \multicolumn{3}{c}{\multirow{-6}{*}{\textbf{Not applicable}}} & $\times$ & \fullcirc[0.6ex] & \fullcirc[0.6ex] \\ \midrule

\cellcolor[HTML]{FFFFFF} & \textbf{No (N)} & \multicolumn{3}{c}{} & \fullcirc[0.6ex] & \fullcirc[0.6ex] & \fullcirc[0.6ex] \\
\cellcolor[HTML]{FFFFFF} & \textbf{User Feedback (U)} & \multicolumn{3}{c}{} & \fullcirc[0.6ex] & \fullcirc[0.6ex] & \fullcirc[0.6ex] \\
\multirow{-3}{*}{\cellcolor[HTML]{FFFFFF}\textbf{Continuous Learning (CL)}} & \textbf{Federated (F)} & \multicolumn{3}{c}{\multirow{-3}{*}{\textbf{Not applicable}}} & $\times$ & \fullcirc[0.6ex] & \fullcirc[0.6ex] \\

\bottomrule

\end{tabular}
\label{tab:LLM_use_cases}
\end{table*}

\subsection{Development Use Cases}
\label{sec:development_use_cases}

LLMs need large amounts of text data to be created from scratch.
Some of the databases used to train LLaMA \cite{touvron2023llama}, GPT-3, and other popular LLMs include C4 \cite{raffel2020exploring}, CommonCrawl \cite{CommonCrawlDB}, WebText2 \cite{brown2020language}, Github, Wikipedia, and others.
Since this is an expensive process in terms of both human and computational effort, most applications reuse pre-trained models adapted for particular needs.
Pre-existing foundation models, or off-the-shelf LLMs, can be obtained from public repositories on the Internet, such as \textit{Hugging Face}\footnote{Available at: \url{https://huggingface.co/}}, and further trained to address a particular task of interest using techniques like fine-tuning~\cite{du2024privacy} or RAG~\cite{lewis2020retrieval}.
The three common use cases related to development phases of LLMs are summarized below:

\begin{itemize}

    \item \textbf{Foundation Model Creation}: The process of creating a general-purpose LLM, trained on large volumes of data, usually collected from public sources, such as blogs, news, social media, web repositories, etc.

    \item \textbf{Fine-Tuning}: The process of taking a foundation model and specializing it for particular tasks using additional labeled data from public or private sources (or both).

    \item \textbf{Retrieval-Augmented Generation (RAG) Preparation (Offline)}: The process of constructing an additional knowledge database and a retrieval system that augments prompt requests with contextually relevant information to improve the accuracy and relevance of LLM responses.
    RAG is a way to specialize the LLM system without model retraining, which is usually more efficient than fine-tuning with respect to computational effort.

\end{itemize}

\subsection{Operation Use Cases}
\label{sec:operation_use_cases}

After creating, training and, if necessary, specializing an LLM, the next step is to deploy it. At this point, the purpose of the model and how users will access and interact with it must be clear. The different use cases related to the LLM operation considered in this work are presented below:

\begin{itemize}

    \item \textbf{Chat-bot:}
    In chat-bots, users can interact directly with the LLM system using a text field with instructions (prompts) such as questions and requests, mimicking human-like conversations. Chat-bots are arguably the core drivers of GenAI popularization in recent years, with the most notorious example being ChatGPT~\cite{chatgpt2022}. 
    The key characteristic of this use case security-wise is that the user is in full control of what the LLM receives. 
    
    \item \textbf{LLM-Integrated Application:}
    In LLM-integrated applications, users do not interact directly with the LLM, but via an application that queries an integrated model to solve specific problems\cite{weber2024large, greshake2023not, liu2023prompt, evertz2024whispers}.
    In this process, the application acts as an intermediary responsible for converting the user input into an efficient query to the LLM, obtaining its response, processing it, and finally presenting the result to the user.
    Some examples of this use case include code-generating tasks performed with LLM assistant plugins, LLM-powered summarization of user reviews in marketplaces, LLM-assisted web searches, copilots, and others.

    \item \textbf{LLM-based Agent:} 
    The last use case considers an LLM system that leverages an LLM to reason, take decisions and act~\cite{yao2023react}.
    For a particular task, the LLM-based agent prompts an LLM, which is a central component, to perform multi-step reasoning and decide over predefined actions, such as invoking tools, calling APIs, querying databases or interacting with user or the environment.
    Agents are usually composed of a central system, a memory that stores past context, and a module to call and interact with external tools.
    LLM agents also have the ability to interact with each other and take decisions autonomously\cite{li2024personal}.
    An example of agent-based application is Microsoft Copilot \cite{copilot-microsoft2023}, a tool that assists Windows users with many different tasks, such as drafting documents, summarizing emails, enhancing work efficiency, etc.
    We can also employ LLM agents as Personal Assistants \cite{li2024personal}, or even for interactive environment simulation, in which LLM agents can make inferences about themselves, other agents, and the environment~\cite{park2023generative}.
    Frameworks to support the creation, deployment, and management of such agents include Auto-GPT \cite{autogpt}, LangChain \cite{langchain}, and AutoDroid \cite{wen2023empowering} (for mobile).

\end{itemize}

\subsection{LLM System Design Choices}
\label{sec:llm_scenarios_design_choices}

When developing LLM systems, more specifically, in the planning phase, several aspects can affect security, as design choices may expose the LLM to different threats. The major aspects to consider are:

\begin{itemize}

    \item \textbf{Data Provenance (DP)}:
    Concerns the type or source of data used for training the model for any of the cases presented in Section~\ref{sec:development_use_cases}. For \textbf{public} data sources, the Internet is the default option, as it contains almost all publicly disclosed information ever registered. Most LLMs available today obtained training data from the Internet \cite{liu2024understanding, huang2024harmful}.
    For \textbf{private} data, data sources can include any private data source, such as network data, databases, or other sources of information with restricted access (e.g., sensitive information within the premises of a private company).
    However, since LLMs require massive amounts of data to train the models, is generally infeasible to use only private data due to its scarceness.
    Thus, a common practice is to adopt a \textbf{hybrid} approach that mixes both public and private data; public data is used to create a foundation model and private data is used to specialize the model to address specific problems within a private context.

    \item \textbf{Dev. and Deployment Infrastructure (DI):}
    This decision concerns where the developing and deployment will occur. During the development phase, the choices for installing, configuring, and using tools, storing the collected data for training/fine-tuning/RAG, creating and manipulating the model, and dealing with any other sensitive information are restricted to the company's premises or third-party Cloud Service Providers (CSP).
    During the model operation, one must decide where the LLM system will be deployed, with choices varying from \textbf{user's device} \cite{yin2024llm, chen2024llm, llm_on_device_2024, xu2024device}, \textbf{company premises}, and \textbf{CSP}.
    A hybrid choice is also possible for both scenarios. 
    For instance, a common practice is to have an on-device application interface for user interaction while heavy data is processed on-premises/cloud.
    
    \item \textbf{Software Libraries and Dependencies (SL)}:
    Concerns the use of external libraries and dependencies. Considering the complex environment and features expected by an LLM-based system, using only \textbf{proprietary} (in-house developed) libraries for developing an LLM-based system can be infeasible due to the amount of effort and human resources needed to maintain such tools.
    Thus, many developers either use \textbf{Open-Source Software} (OSS) or adopt a \textbf{mix} of OSS and proprietary software to deal with complex tasks.
    Security-wise, the use of OSS demands a deep analysis of available choices to avoid supply chain attacks or other maintenance risks.
    
    \item \textbf{Shared Infrastructure (SI)}:
    Concerns whether the same model is shared among different users. LLM deployment in external servers (on-premises/cloud) usually follows the shared infrastructure approach, in which each user has access to a particular instance of the same model -- a practice known as LLM-as-a-Service (LLMaaS).
    Although less common, on-device deployments can also follow a similar architecture that deploys a unique LLM to be shared among applications~\cite{yin2024llm}.
    In both types of deployment (on-device/external servers), \textbf{sharing the infrastructure} exposes LLM-based systems to additional threats if mitigation strategies are not implemented, especially due to the possibility of unintended data leaks. In a shared infrastructure, the storage of context information per-user or per-app should be ~\cite{yin2024llm}. \textbf{Isolated instances} of LLMs can be a safer choice, but it can incur in extra financial costs or be constrained due to device resources.

    \item \textbf{Prompt Input Origin (IO)}:
    Prompt is the way users communicate with LLMs.
    There are different forms of how the LLM can receive a prompt.
    In chat-bot applications, the usual way is via a \textbf{text field}, with users providing a text with their wishes to the LLM system directly.
    Another way is using template prompts adapted at runtime and supplied to the LLM via \textbf{Apps}. that integrate and use the model to solve complex problems \cite{greshake2023not}.
    A third option could be by \textbf{voice} \cite{llmvoicebot}, using a multi modal model \cite{shen2024voice} (out of scope of this survey) or using an LLM coupled with text-to-speech and speech-to-text models.
    For instance, GPT-3.5 and GPT-4 use three models to handle users' voice: one to transcribe the audio to text, the LLM to process the text as a normal prompt, and a third model to transform the outputted text to audio \cite{gpt4o2024}.
    It is also possible that an LLM system allows more than one of these methods, using one at a time.
    
    \item \textbf{Access to Resources (AR)}:
    One interesting feature of LLMs is their ability to interact with their environment.
    Giving their vast knowledge base, they can be empowered to solve problems beyond the capabilities of natural language processing, and overcome some of their intrinsic limitations of having access to up-to-date information.
    All of this is achieved by the use of external resources, such as \textbf{tools}, \textbf{databases} (e.g., RAG), \textbf{hardware and software sensors}, and the \textbf{internet}.
    LLMs can be given the ability to interact with functions written by developers, APIs (including external ones), or other resources, using libraries such as LangChain \cite{langchain, toolcalling-langchain}, for instance.
    Some authors have shown that LLMs are also able to teach themselves how to use some tools, requiring just some demonstrations \cite{schick2023toolformer}.
    Besides calling functions to perform some specific operations, LLMs can also access databases to overcome issues related to performance and outdated information \cite{he2024the, li2024personal}.
    To this end, RAG \cite{lewis2020retrieval} can be used, giving the LLM access to a database of specialized knowledge without the costs of retraining (or fine-tuning) it.
    LLM-based agents can access device information via sensors --- hardware (accelerometers, gyroscopes, GPS, etc.) and/or software (app usage, call records, typing), e.g., personal assistant \cite{li2024personal}.
    For last, LLMs can also have the ability to access Internet data, obtaining data from a websites to perform some action (e.g., text summarization), process users' emails, or fetch data from websites and repositories as part of its functionality \cite{chatgpt-plugins2023, greshake2023not}.
    Note that an LLM system can have access to more than one resource type.

    \item \textbf{Continuous Learning (CL)}:
    After creating an LLM, we need to consider how we are going to keep it up-to-date, and even detecting and correcting mistakes.
    Some possibilities are using fine-tuning and RAG (Section~\ref{sec:development_use_cases}), but we can also collect data from end-users during the system operation and use it to refine the model.
    One possible approach is using \textbf{user feedback}, that different from reinforced learning approaches (used in the alignment process), aims to obtain the end-user suggestion about the use of the LLM system to improve it, being a simple thumbs up/down about the LLM response, or using other feedback mechanisms \cite{openai-feedback, anthropic-feedback}.
    \textbf{Federated Learning} is another possibility to improve the performance of LLMs on specific tasks, using data from multiple entities that, without directly sharing (and exposing) sensitive data of users, can contribute to the training process.
    This is a well-known approach for ML algorithms, but can also be applied to LLMs \cite{kuang2024federatedscope, fan2023fate}.

\end{itemize}

Although there are many interesting possibilities here, there are also threats that we need to be aware of depending on the characteristics we choose when deploying LLMs in real world.
Each of these choices will affect the security of the whole system, so understanding the threats associated with possible scenarios can help mitigate potential risks and allow LLM developers to make better choices from a security perspective.

\subsection{Examples of LLM Scenarios and Possible Design Choices}
\label{sec:ex_llm_scenarios}

Some of the scenarios presented in Table~\ref{tab:LLM_use_cases} are detailed in this section.
Here, we are not interested in how complex it is to build an application, training an LLM, the pros and cons of using a CSP, among other aspects.
We focus on showing different forms of building and using LLMs in real-world scenarios and elucidating the threats associated with these scenarios and possible defensive strategies.
Besides, we do not aim to present all possible combinations of scenarios and design choices but at least to cover the most important ones from a security perspective.

\subsubsection{Development of an LLM for Chat-bot Application (ST:D/UC:FM/DP:U/DI:P/SL:H/SI:N/IO/AR/CL)}
\label{sec:llm_scenario_1}

\begin{figure}[ht]
\includegraphics[width=\columnwidth]{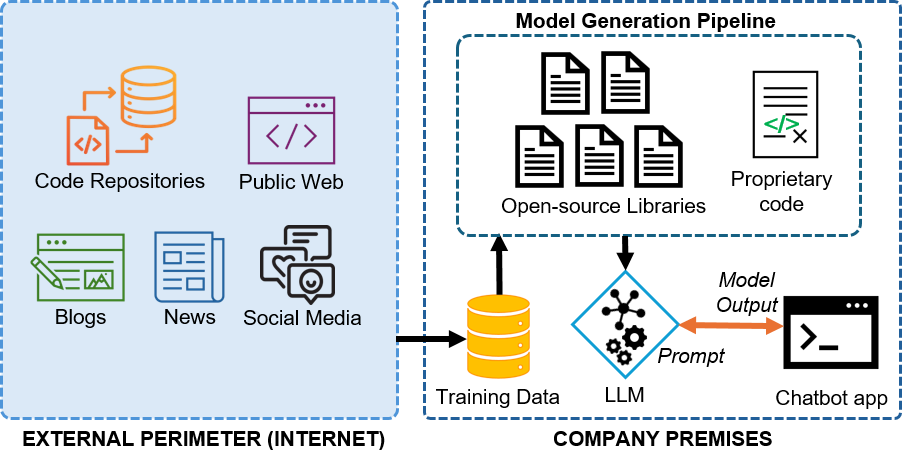}
\centering
\caption{Scenario 1: LLM development process (on company premises) for a chat-bot application.}
\label{fig:uc_on_premises_chatobot_model_creation}
\end{figure}

The first example of LLM scenario is depicted in Figure~\ref{fig:uc_on_premises_chatobot_model_creation}.
In this scenario, a company chooses to build an LLM-based Chat-bot application, using a model created from scratch.
The data used for training the model is collected entirely from the Internet, from sources such as code repositories, social media, news websites, blogs, forums, and other public sources.
This data is pre-processed and stored in a database on company premises.

Next, the company develops all necessary code considering the whole software and LLM life cycle, such as model development, data analysis, training, and evaluation, combining open source libraries and proprietary code, and using a unique infrastructure to perform all activities.
The company also creates the application interface, responsible for taking user inputs and presenting model output.

\subsubsection{Chat-bot Application on User's Device (ST:O/ UC:CB/DP:U/DI:D/SL:O/SI:N/IO:T/AR:N/CL:N)}
\label{sec:llm_scenario_2}

\begin{figure}[ht]
\includegraphics[width=0.5\columnwidth]{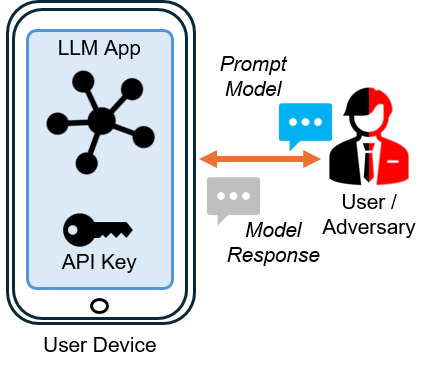}
\centering
\caption{Scenario 2: A chat-bot application operating on users' device with an off-the-shelf LLM.}
\label{fig:uc_on_device_chatobot_no_internet}
\end{figure}

Figure~\ref{fig:uc_on_device_chatobot_no_internet} presents an example of using LLMs deployed on the user's device.
This scenario is respect to a chat-bot application developed using only open-source code and an off-the-Shelf LLM trained on public data.
The user can interact via prompts with the LLM, without the delay or need for an Internet connection.
Besides faster responses, all data exchanged with the application remain on the device (a desirable privacy feature).
Note that an API key may be used to restrict LLM access and functionalities.

Although this is not a common scenario for LLMs, as we advance in creating more specialized and compact models (e.g., using quantization, caching, and other techniques to reduce processing power and memory during LLM inference \cite{nist_adversarial_ml_attacks_2023, llm_on_device_2024}), we expect to see more examples of models deployed on smartphones and IoT \cite{yin2024llm, xu2024device}.
Google AICore is practical example of such deployment \cite{aicore_google}, acting like a system service, allowing apps to incorporate it.

\subsubsection{LLM-Integrated App. on-Cloud (ST:O/ UC:AP/DP:H/DI:H/SL:H/SI:Y/IO:A/AR:H/CL:U)}
\label{sec:llm_scenario_3}


LLMs can also be integrated by applications to perform specific tasks, e.g., solving a mathematical problem \cite{liu2024demystifying}.
The user accesses an application and provides a problem.
The application creates a prompt to the LLM to solve it.
The model returns a code able to solve the problem to the app, which executes it to get the answer.
Finally, the application processes the result and displays the solution to the user.

\begin{figure}[ht]
\includegraphics[width=\columnwidth]{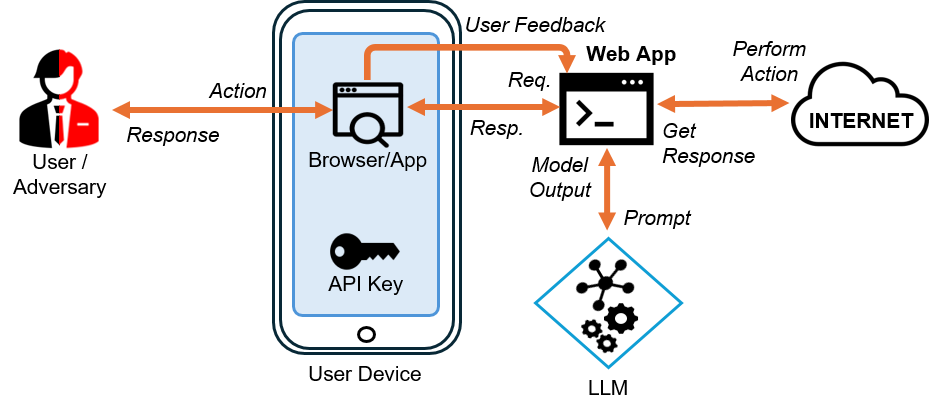}
\centering
\caption{Scenario 3: An LLM-integrated application deployed on-cloud and with Internet access.}
\label{fig:uc_on_cloud_llm_app_with_internet}
\end{figure}

In this scenario, illustrated in Figure~\ref{fig:uc_on_cloud_llm_app_with_internet}, 
we consider an application developed using a mix of open-source and proprietary code, a foundation model obtained from public sources, trained on public data and fine-tuned to solve specific problems using private information.
The LLM system runs on a shared infrastructure of a CSP, in a hybrid deployment configuration, with the app interface on user device, and the LLM and other resources on the external server.
The LLM has the ability to execute tools in the environment and to access the internet as an additional information source.
Upon user's requests, the application can fetch content from the internet to help in solving a problem.
Moreover, users can evaluate the application's effectiveness in solving problems via a feedback feature available.

\subsubsection{LLM-based Agent for User Assistance (ST:O/ UC:AG/DP:H/DI:H/SL:H/SI:Y/IO:H/AR:H/CL:F)}
\label{sec:llm_scenario_4}

\begin{figure}[ht]
\includegraphics[width=\columnwidth]{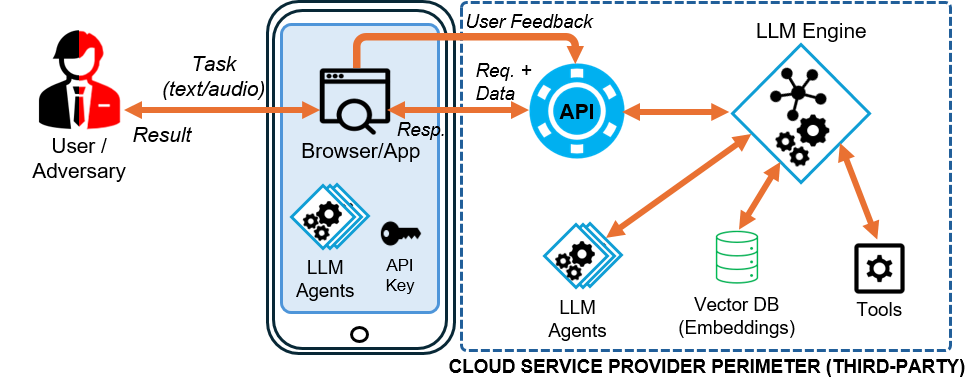}
\centering
\caption{Scenario 4: LLM-based Agent system deployed on user device and the cloud, with access to resources and tools.}
\label{fig:uc_on_cloud_device_llm_agent}
\end{figure}

The last example scenario presented here is the use of LLM-based agents, such as for user assistance.
Figure~\ref{fig:uc_on_cloud_device_llm_agent} depicts this scenario.
Upon receiving a task, the agent plans how to solve it, access its memory (past context data), execute tools, collect necessary environment data (from hardware or software sensors), and, in a autonomous manner, completes the task.
This scenario considers a hybrid deployment, with some agents deployed on users' device to receive tasks (via voice commands or text), collect data, and send all information to a central system (another LLM) deployed on cloud (shared infrastructure) for further processing.
The LLM-based agent system is developed using proprietary and open-source code, trained on public and private data, and adopts federated learning for continuous learning.

\section{Characterization of Threats to LLMs}
\label{sec:attacks_to_LLM}

Given the widespread adoption of LLMs, threats to this technology have become increasingly common.
In this section, we review different manners of abusing or damaging LLM systems and present a new form of classifying and characterizing the threats identified in our literature review from Section~\ref{sec:methodology}.

Our approach centers on the Confidentiality-Integrity-Availability (CIA) triad model, similar to NIST's  classification~\cite{nist_adversarial_ml_attacks_2023}, but we did not include the \textit{misuse} category as NIST does because we understand this is a subcategory of the Integrity aspect.\footnote{Although jailbreak attacks may not change the model binary (a strong indication of integrity violation), we believe they can affect how the model will respond to users. A successful attack ``breaks'' the security mechanisms and guardrails, making the model behavior unpredictable and different from the one programmed by their developers.}
We also explore the threats characteristics (e.g., the adversary knowledge, strategies to perform an attack to LLM systems, goals and targets, use cases, form of system interaction during the attack) and present other threats associated with solutions using this technology.

\begin{table*}
\centering
\setlength\tabcolsep{5pt}
\renewcommand{\citepunct}{]\,\allowbreak[}
\scriptsize
\caption{List of Threats to LLM Development and their characteristics.}

\begin{tabular}{cp{3.9cm}cccC{0.9cm}C{1.2cm}C{1.3cm}C{1.4cm}} 
\toprule

\multirow{2}{*}{\textbf{ID}} & \centering \multirow{2}{*}{\textbf{Threat}} & \multirow{2}{*}{\textbf{Goal}} & \multirow{2}{*}{\textbf{Class}} & \multirow{2}{*}{\textbf{Target}} & \textbf{Attack Strategy} & \textbf{Adversary Knowledge} & \textbf{LLM Interaction} & \textbf{Affected Use~Case} \\ 
\midrule

I01 & Training Data Poisoning \cite{fed24, fu2024poisonbench} & LLM Misuse & Data Infection & Data & \aSC\,\aPO\,\aIN & Black-box & Indirectly & FM, FT, RG \\
I02 & Model Poisoning \cite{fed24, wen2023survey, bhatt2024talk, cao2024stealthy, chen2024multiturn, cheng2024transferring, cheng2024syntactic, du2023uor, raghuram2024study, huang2024catastrophic, hung2024attention, jiao2024can, li2024large, li2024badedit, wangbadagent, yang2024sos, yang2024watch, yao2024poisonprompt, zhang2024instruction, zhao2024weaktostrong, zhao2024universal, zhao2023prompt, zhao2024exploring, li2024modeleditingbased, huang2025virus}  & LLM Misuse & Data Infection & LLM & \aSC\,\aPO\,\aIN & White-box & Indirectly & FM, FT, RG \\
I03 & Fine-Tuning Poisoning \cite{huang2024harmful} & LLM Misuse & Data Infection & Data & \aPO\,\aIN & Black-box & Indirectly & FT, RG \\
I04 & RAG Poisoning \cite{zhang2024hijackrag, cheng2024trojanrag, di2024pirates, kwartler2024backdoored, chen2024blackbox} & LLM Misuse & Data Infection & Data & \aDP\,\aPO\,\aIN & Black-box & Indirectly & CB, AP, AG \\
I05 & Model Dependencies Exploitation \cite{nist_adversarial_ml_attacks_2023} & LLM Misuse & Third-Party SW Compromise & LLM System & \aSC & Black-box & Indirectly & FM, FT, RG \\
I06 & Fake Plugins \cite{greshake2023not} & LLM Misuse & Third-Party SW Compromise & LLM System & \aSC & Black-box & Indirectly & FM, FT, RG \\
\bottomrule

\end{tabular}
\label{tab:threats_to_LLM_development}
\end{table*}

\begin{table*}
\setlength\tabcolsep{4pt}
\renewcommand{\citepunct}{]\,\allowbreak[}
\centering
\scriptsize
\caption{List of Threats to LLM Operation and their characteristics.}
\begin{tabular}{cp{5.2cm}cccC{0.9cm}C{1.2cm}C{1.3cm}C{1.3cm}} 
\toprule

\multirow{2}{*}{\textbf{ID}} & \centering \multirow{2}{*}{\textbf{Threat}} & \multirow{2}{*}{\textbf{Goal}} & \multirow{2}{*}{\textbf{Class}} & \multirow{2}{*}{\textbf{Target}} & \textbf{Attack Strategy} & \textbf{Adversary Knowledge} & \textbf{LLM Interaction} & \textbf{Affected Use~Case} \\ 
\midrule

A01 & Disrupting Search Results \cite{greshake2023not} & Service Disruption & Bad LLM Response & LLM & \aIP\,\aSC\,\aMW & Black-box & Indirectly & CB, AP, AG \\
A02 & Inhibiting Capabilities \cite{greshake2023not} & Service Disruption & Bad LLM Response & LLM & \aIP\,\aSC\,\aMW & Black-box & Indirectly & CB, AP, AG \\
A03 & Output Hijacking \cite{liu2024demystifying, agnew2024hallucinating, pan2023on} & Service Disruption & Bad LLM Response & LLM System & \aSC\,\aIN\,\aSM & Black-box & Indirectly & CB, AP, AG \\
A04 & Bit-Flip Attack \cite{das2024attentionbreaker, yao2020deephammer, rakin2019bit} & Service Disruption & Bad LLM Response & Infrastructure & \aIN\,\aSM\,\aSD & White-box & Indirectly & CB, AP, AG \\
A05 & Sponge Examples \cite{shumailov2021sponge} & Service Disruption & Resource Drain & LLM & \aDP\,\aIP\,\aMW & Black-box & Multi-turn & CB \\
A06 & Time Consuming \cite{greshake2023not, kumar2025overthink} & Service Disruption & Resource Drain & LLM & \aIP\,\aSC\,\aMW & Black-box & Indirectly & CB, AP, AG \\
A07 & Token Wasting \cite{liu2023prompt, gao2024denialofservice}  & Service Disruption & Resource Drain & LLM System & \aKV\,\aCS\,\aSM & Black-box & Indirectly & CB, AP, AG \\
C01 & Embedding Inversion \cite{morris2023text, liu2024mitigating} & Data Stealing & Inference & LLM System & \aKV\,\aCS\,\aSM & White-box & Indirectly & CB, AP, AG \\
C02 & Gradient Inversion \cite{petrov2024dager, huang2021evaluating} & Data Stealing & Inference & LLM System & \aIN\,\aKV\,\aSM & White-box & Indirectly & CB, AP, AG \\
C03 & Model Fingerprinting \cite{nazari2024llm} & Data Stealing & Inference & Infrastructure & \aIN\,\aSD & Black-box & Indirectly & CB, AP, AG \\
C04 & Token-length Inf. \cite{weiss2024your} & Data Stealing & Inference & Infrastructure & \aIN\,\aKV\,\aSD & Black-box & Indirectly & CB, AP, AG \\
C05 & GPU Information Leak \cite{maia2022can} & Data Stealing & Inference & Infrastructure & \aIN\,\aSD & Black-box & Indirectly & CB, AP, AG \\
C06 & Shared-cache Hit Inf. \cite{zheng2024inputsnatch, song2024early} & Data Stealing & Inference & Infrastructure & \aDP\,\aSD & Black-box & Multi-turn & CB, AP, AG \\
C07 & User Data Exfiltration \cite{bhusal2024exfiltration, samoilenko2023new} & Data Stealing & Inference & LLM & \aIP & Black-box & One-shot & CB, AP, AG \\
C08 & Membership Inf. \cite{meeus2024did, shokri2017membership, nist_adversarial_ml_attacks_2023, amit2024sok, chang2024contextaware, chen2024a, galli2024noisy, mattern2023membership, mireshghallah2022quantifying, rrv2024semantic, zhang2023code} & Data Stealing & Inference & LLM & \aDP & Black-box & Multi-turn & CB \\
C09 & Distribution Inf. \cite{rigaki2023survey, suri2021formalizing, hartmann2023sok, nist_adversarial_ml_attacks_2023} & Data Stealing & Inference & LLM & \aDP & Black-box & Multi-turn & CB \\
C10 & User Inference \cite{kandpal2023user} & Data Stealing & Inference & LLM & \aDP & Black-box & Multi-turn & CB \\
C11 & Attribute Inference \cite{hartmann2023sok, li2023privacy, staab2023beyond} & Data Stealing & Inference & LLM & \aDP & Black-box & Multi-turn & CB \\
C12 & Memorized Data Extraction \cite{rigaki2023survey, carlini2021extracting} & Data Stealing & Extraction & LLM & \aDP & Black-box & Multi-turn & CB \\
C13 & Model Replication \cite{rigaki2023survey, owasp_llm_top10, birch2023model} & Data Stealing & Extraction & LLM & \aDP & Black-box & Multi-turn & CB \\
C14 & Model Reverse Engineering \cite{zhou2024investigating, chen2024llm, hajipour2024codelmsec} & Data Stealing & Extraction & LLM & \aRE & Black-box & Indirectly & CB, AP, AG \\
C15 & API Key Stealing \cite{liu2024demystifying} & Data Stealing & Extraction & LLM System & \aIP\,\aCS\,\aMW & Black-box & Indirectly & CB, AP, AG \\
C16 & Prompt Leaking \cite{perez2022ignore, agarwal2024investigating, agarwalprompt} & Data Stealing & Extraction & LLM System & \aIP\,\aCS\,\aMW & Black-box & Indirectly & CB, AP, AG \\
C17 & System Dependencies Exploitation \cite{chatgpt-interface-vuln-2023} & Data Stealing & Extraction & LLM System & \aSC\,\aKV & Black-box & Indirectly & CB, AP, AG \\
I07 & Byzantine Attack \cite{wen2023survey} & LLM Misuse & Data Infection & LLM & \aPO & Black-box & Indirectly & CB, AP, AG \\
I08 & Feedback Poisoning \cite{chen2024dark, altun2024securing} & LLM Misuse & Data Infection & LLM & \aDP\,\aPO & Black-box & Indirectly & CB, AP, AG \\
I09 & Instruction Manipulation \cite{wei2024jailbroken, shang2024intentobfuscator, schulhoff2023ignore, huang2024obscureprompt, liu2023jailbreaking, jiang2023prompt, rao2023tricking, liu2023prompt, sun2024scaling, debenedetti2024agentdojo} & LLM Misuse & LLM Prompting & LLM & \aDP\,\aIP & Black-box & One-shot & CB, AP, AG \\
I10 & Obfuscation \cite{rao2023tricking, kang2024exploiting, Figueroa2024news, li2024exploiting, berezin2025the, yong2023low, liu2023jailbreaking, li2024a, siu2025speak, schulhoff2023ignore} & LLM Misuse & LLM Prompting & LLM & \aDP\,\aIP & Black-box & One-shot & CB, AP, AG \\
I11 & Pretending \cite{liu2023jailbreaking, shen2024anything, rao2023tricking, Huang2024news, shanahan2023role, chu2024comprehensive, deng2023attack, deng2024masterkey, CatoReport2025} & LLM Misuse & LLM Prompting & LLM & \aDP\,\aIP & Black-box & One-shot & CB, AP, AG \\
I12 & Noise-based Attack \cite{yan2024flipattack} & LLM Misuse & LLM Prompting & LLM & \aDP\,\aIP & Black-box & One-shot & CB, AP, AG \\
I13 & Recursive Prompt Hacking \cite{schulhoff2023ignore} & LLM Misuse & LLM Prompting & LLM & \aDP\,\aIP & Black-box & One-shot & CB, AP, AG \\
I14 & RCE \cite{liu2023jailbreaking, liu2024demystifying, beckerich2023ratgpt} & LLM Misuse & LLM Prompting & LLM & \aIP & Black-box & One-shot & AP, AG \\
I15 & Prompt-to-SQL injections \cite{pedro2023prompt, liu2024demystifying} & LLM Misuse & LLM Prompting & LLM & \aDP\,\aIP & Black-box & One-shot & AP, AG \\
I16 & Copied Prompt Injection \cite{samoilenko2023new} & LLM Misuse & LLM Prompting & LLM & \aIP & Black-box & One-shot & CB \\
I17 & Fuzzing \cite{yao2024fuzzllm, yu2023gptfuzzer, gong2024papillon} & LLM Misuse & LLM Prompting & LLM & \aDP & Black-box & Multi-turn & CB, AP, AG \\
I18 & Context Manipulation \cite{russinovich2025jailbreaking, Chen2024news, russinovich2024great, schulhoff2023ignore, rao2023tricking, bianchi2024large, botacin2023gpthreats, liu2023jailbreaking, kang2024exploiting, Figueroa2024news} & LLM Misuse & LLM Prompting & LLM & \aDP & Black-box & Multi-turn & CB \\
I19 & LLM Jailbreak Helper \cite{xu2024an, siu2025speak, das2024humanreadable, liu2023goaloriented} & LLM Misuse & LLM Prompting & LLM & \aDP & Black-box & Multi-turn & CB \\
I20 & LLM Jailbreak Helper w/feedback \cite{jiang2025decomposition, lee2025xjailbreak, lin2024figure, draguns2024when, ding2024a, li2024jailpo, liu2024autodan, chao2023jailbreaking, yu2024llmstinger, mehrotra2023tree} & LLM Misuse & LLM Prompting & LLM & \aDP & Black-box & Multi-turn & CB \\
I21 & Function Calling Jailbreak \cite{wu2024the} & LLM Misuse & Data Infection & LLM & \aIP & Black-box & Indirectly & CB, AP, AG \\
I22 & Multi-stage Exploit \cite{greshake2023not} & LLM Misuse & LLM Prompting & LLM & \aIP & Black-box & Indirectly & AP, AG \\
I23 & Agent Infection \cite{lee2024prompt, cohen-aiworm} & LLM Misuse & LLM Prompting & LLM & \aIP & Black-box & Indirectly & AG \\
I24 & Output metadata Manipulation \cite{andriushchenko2024jailbreaking, li2024lockpicking, du2023analyzing} & LLM Misuse & LLM Prompting & LLM & \aDP & White-box & Multi-turn & CB, AP, AG \\
I25 & Chat Template Injection \cite{zhao2024sql, jiang2024chatbug} & LLM Misuse & LLM Prompting & LLM & \aDP & White-box & Multi-turn & CB \\
I26 & Optimization-based Attack \cite{zou2023universal, xu2024continuous, hu2024droj, hu2024efficient, li2024fastergcg, jia2024improved, zhang2024targetdriven, shi2024optimization} & LLM Misuse & LLM Prompting & LLM & \aDP & White-box & Multi-turn & CB \\
I27 & Optimization-based Attack w/ LLM Helper \cite{zhang2024enja, lin2024towards, zhao2024weaktostrong, liu2024feint} & LLM Misuse & LLM Prompting & LLM & \aDP & White-box & Multi-turn & CB \\
I28 & Jailbreak Bit-Flip Attack \cite{coalson2024prisonbreak} & LLM Misuse & HW Manipulation & LLM & \aIN\,\aKV\,\aSM & White-box & Indirectly & CB, AP, AG \\

\bottomrule

\end{tabular}
\label{tab:threats_to_LLM_operation}
\end{table*}

Tables~\ref{tab:threats_to_LLM_development} and \ref{tab:threats_to_LLM_operation} contain the list of threats to LLM systems and their main characteristics, highlighting important information to understand the attack requirements, goals, and vulnerable scenarios.
In Table~\ref{tab:threats_to_LLM_development}, we present LLM threats that occur in development phases, mostly related to poisoning and supply chain, while in Table~\ref{tab:threats_to_LLM_operation}, we explore the threats to LLMs in their operation.
Each identified threat is marked to one aspect of the CIA triad, denoted by the letter of its ID.
This means that the main type of damage a threat may cause to a system are: 
(C) stealing sensitive and valuable data; (I) manipulating or jailbreaking the guardrails and other security mechanisms designed to prevent model misuse; or (A) disrupting services and functionalities or wasting resources.
Note that whenever we refer to a threat, risk, or possible attack to LLM systems in this paper, we will associate them with the threats listed in Tables~\ref{tab:threats_to_LLM_development} and \ref{tab:threats_to_LLM_operation}.

We further expand adversaries' goals into the CIA aspects and define subgroups based on the similarities the attacking forms share, which we define as classes, as presented in the following for each goal.

\begin{itemize}

\item \textbf{Confidentiality:} Adversaries aim to obtain:
sensitive and valuable data, such as 
data used to train an LLM (C16); 
parameters or properties enabling them to create a \textit{shadow model} (C13) (i.e., generating a model that mimics the behavior of the original model for specific tasks, based on inputs and responses collected through queries \cite{fed24}); 
data learned by the LLM such as businesses secrets (Intellectual Property) (C12); 
API keys that grant access to a system (C15); or inputs provided by users that may contain sensitive data (C16).
These attacks can be further classified into \textit{extraction} or \textit{inference}, based on the form the information is obtained from the target -- either directly or estimated based on data obtained, respectively.

\item \textbf{Integrity:} Adversaries aim to tamper with the LLM system, by 
poisoning the data used to train (or update) the model (I01, I03-04), 
changing model parameters and inserting backdoor (especially in those pre-trained models shared in public repositories) (I02),
manipulating system dependencies (I05), 
or even 
jailbreaking the security mechanisms and guardrails of the LLM to manipulate its behavior (I09-28).
By changing the data or how the model will respond, the adversary can disseminate disinformation, hate speeches, discrimination, and other offensive/biased content, distribute malware, obtain sensitive data from the LLM (restricted content, such as ``how to build a bomb''), cause malfunctioning, etc.
These attacks focus on compromising the software supply chain \textit{(Third-Party SW Compromise)}, poisoning training data or auxiliary files used by the LLM system \textit{(Data Infection)}, or sending malicious prompts to the system to perform jailbreak \textit{(LLM Prompting)}.
Another form of achieving jailbreak is via model manipulation during runtime (\textit{HW Manipulation}).

\item \textbf{Availability:} Adversaries aim to 
take down the LLM system, 
making the model provide useless results (A01-04),
causing instability in the service (A05-06), 
or
causing monetary losses to individuals or organizations (consuming a paid service in the name of the target victim)(A07).
Attacks on availability are motivated by financial gain (via extortion), brand damage, or as a form of hacktivism.
These attacks can focus on consuming all resources from the infrastructure or users \textit{(Resource Drain)} or making the LLM system produce useless results as response to legitimate users \textit{(Bad LLM Response)}.

\end{itemize}

A threat aims to cause damage on a specific target.
In our analysis, we set as target the main element in which defenders should focus their attention to build defenses.
In case of LLM systems, these targets can be the:

\begin{enumerate}
    \item \textbf{LLM:} Causing denial of service (including performance degradation), data/behavior manipulation, and/or leakage of sensitive data.
    \item \textbf{LLM System:} Causing denial of service (inability to access the model or financial loss), remote code execution on the target system, and/or leakage of sensitive data (such as users API keys). 
    The system includes the application interface and other associated services.
    \item \textbf{Data:} Aiming to manipulate (poison) or steal sensitive and confidential data used in the training or model specialization process (fine-tuning or RAG).
    \item \textbf{Infrastructure:} Having leakage of proprietary and sensitive data (source code or other intellectual property content), inability to provide services, brand damage, and/or financial loss.
\end{enumerate}

Other important aspects of our characterization involve the attack strategy (Section \ref{sec:att_methods}), how the adversary interact with the LLM system (Section \ref{sec:att_interaction}), and also the LLM use case (Sections \ref{sec:development_use_cases} and \ref{sec:operation_use_cases}), since different scenarios will be vulnerable to a particular set of threats.
In the next subsections, we detail the first two aspects, and in Section~\ref{sec:threats_LLM_use_cases}, we provide an analysis of the third.

\subsection{Attack Strategies}
\label{sec:att_methods}

\begin{figure*}[ht]
\includegraphics[width=\textwidth]{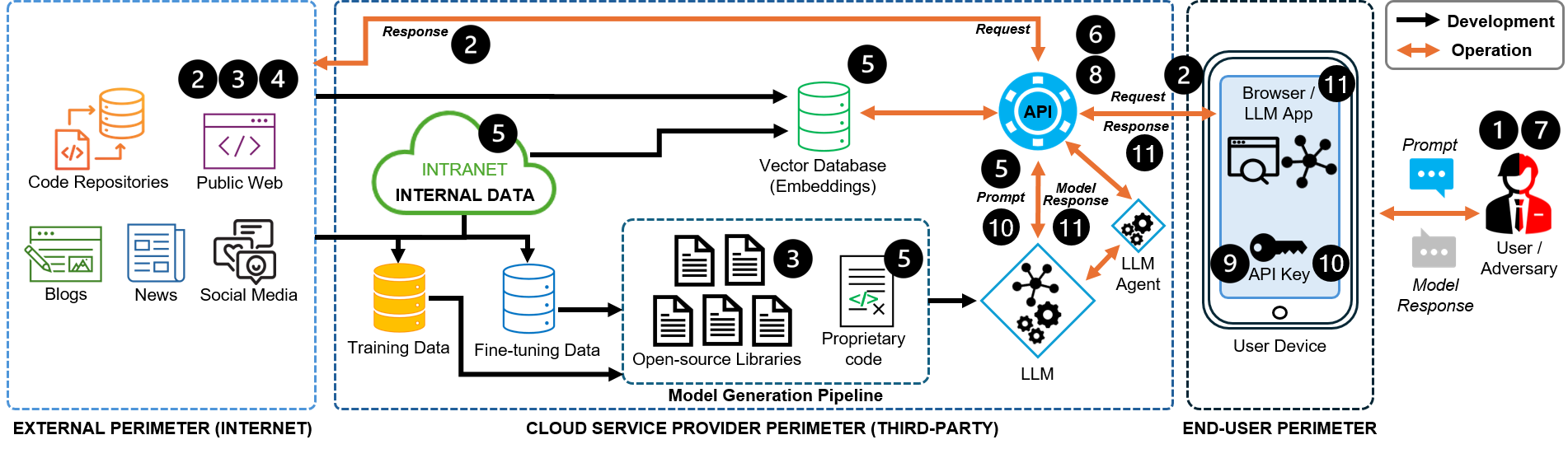}
\centering
\caption{Possible attack strategies on a general view of an LLM-based system architecture, from development to operation.}
\label{fig:LLM_Model_Architecture_and_Threats}
\end{figure*}

There are many forms of attacking LLM systems. 
Adversaries can attack the model directly, instructing it to perform a malicious or non-allowed action, via direct prompts, or indirectly, by having the model processing a malicious content obtained from an external source.
LLM systems can also be exploited using traditional attacks, targeting the authentication process, deployment infrastructure, among other elements and application services.
These different attack forms are referred here as attack strategies.
Next, we present the strategies an adversary can adopt to attack an LLM system.
In our threat characterization, we consider that adversaries will use at least one of these strategies to cause harm, although we understand that some threats can be executed using different strategies.
We also map the attack strategies to the components of a general LLM system architecture to which they can be applied, encompassing parts of the development and deployment processes, as depicted in Figure~\ref{fig:LLM_Model_Architecture_and_Threats}.

\newcommand*\circled[1]{\tikz[baseline=(char.base)]{\node[shape=circle,fill=black,draw,inner sep=1.5pt] (char) {#1};}}

\begin{enumerate}[label=\protect\circledone{\color{white}\small \arabic*}]

    \item \textbf{Direct Prompt:} 
    LLMs, especially chat-bots, allow users to submit questions (via a prompt) directly to the model. 
    Adversaries can exploit this feature to perform jailbreak attacks, obtain sensitive data, or cause service disruption at some level.

    \item \textbf{Indirect Prompt:} 
    According to \cite{owasp_llm_top10}, Indirect Prompt refers to another form of sending data to an LLM, but instead of supplying it directly, via a prompt, an external source is used, such as websites, files, or emails. 
    For instance, consider an LLM system responsible for managing user's email.
    An adversary targeting a particular user using such system could hide a prompt injection in the content of an email and send it to the victim.
    When the LLM process the email (e.g., to summarize it for the user), the content (and malicious prompt) are processed and the malicious action may take effect.
    This strategy is used to jailbreak or perform service disruption to an LLM system, as well as steal user's data.

    \item \textbf{Supply Chain:}
    By exploiting third-party LLM system dependencies or pre-trained LLMs (e.g., changing their parameters), adversaries can manipulate (tamper) a system component to create biased outcomes or introduce security breaches to steal sensitive data or perform system disruptions~\cite{owasptop10llm032025}.
    Another threat example is embedding malware to the model (e.g., when it is loaded, a malicious code is executed~\cite{owasptop10llm042025, hitaj2022maleficnet, hitaj2025you}).

    \item \textbf{Poisoning:} 
    An adversary manages to have access to and manipulate the data used in the processes of pre-training, fine-tuning, or RAG of the victim LLM to introduce vulnerabilities, backdoor, or biased behavior~\cite{owasptop10llm042025}), affecting the system performance and its reliability.
    The user feedback can also be a form of poisoning the system~\cite{chen2024dark, altun2024securing}.

    \item \textbf{Insider:} 
    Another threat to LLM systems originate from inside an organization (by its employees) or partners, with the adversary having special privileges to or knowledge about the training data, repository source code, deployment infrastructure, the LLM itself, RAG context data, etc.
    With this special access and knowledge, the adversary can execute actions to cause harm to the system or steal sensitive information (including Intellectual Property).
    The motivation for this type of adversary goes from retaliation to monetary gain, but also includes incompetence, with employees introducing vulnerabilities in the system due to a lack of expertise or by mistake.
    
    \item \textbf{Known Vulnerabilities:}
    An adversary can attack an LLM system by exploiting known software vulnerabilities present in the system, its dependencies, from the infrastructure where the system is deployed, and also from the communication channel between user and infra (when the LLM is deployed in an external server). Examples include vulnerabilities in the hosting operating system, web server, TLS communication (use of an outdated version, such as the v1.0), among others.
    Note that in Supply Chain vulnerabilities, adversaries exploit system dependencies to introduce vulnerabilities, while here, the adversary discovers an existing vulnerability (e.g., originated from a bug) and leverages it to cause harm.

    \item \textbf{Credential Stealing:} 
    Adversaries aim to obtain access to unauthorized data and system.
    To this end, they may apply traditional attacks such as \textit{phishing} or use malware to steal users or developers credentials, so they can have access to the LLM system, training data, code repositories, or other important resources.
    Besides access to sensitive data, they can also cause financial loss to the victim and restrict their access to the system.

    \item \textbf{Exploiting Security Mechanisms:}
    By exploiting the lack of security mechanisms or their proper configuration, adversaries can bypass such defenses and infiltrate in the system to obtain sensitive data, install malware, or cause denial of service.
    Some security mechanisms that can be exploited are the system authentication process or communication channel.
    
    \item \textbf{Reverse Engineering:} 
    LLM systems deployed directly on user's device (i.e., the LLM is running locally on the device rather than a cloud infrastructure) come with the risk of reverse engineering \cite{chen2024llm}.
    This practice allows adversaries to have access to the model, its structure, and metadata (weights and other internal parameters).  
    Adversaries can understand how the model works, modify it to remove jailbreak security mechanisms, apply side-channel attacks easier, and also obtain confidential data, including businesses secrets (which is the LLM itself).

    \item[\circledtwo{\small\color{white}10}] \textbf{Malware:}
    A malicious application is used to cause damage (e.g., service degradation or disruption) or steal confidential LLM-related data (e.g., API keys and user supplied input, which may contain sensitive data or business secretes).
    The malware can be installed in user's device or in the LLM system deployed network.

    \item[\circledtwo{\small\color{white}11}] \textbf{Side-Channel:}
    This attack strategy takes advantage of side effects that occur during the system usage, and by monitoring them, the adversary can extract sensitive information.
    Attacks consist in the analysis of power consumption, acoustic, timing, electromagnetic, system cache, among others.
    The attack usually requires physical access to the victim device and, in many cases, especial equipment to capture the information leakage.
    Insiders can perform side-channel attacks due to their physical access to cloud infrastructures, for instance, but other forms of executing this attack strategy, e.g., remotely by the network, can also be found.

\end{enumerate}

\subsection{Adversary Interaction with the LLM}
\label{sec:att_interaction}

To perform an attack using the strategies previously described, adversaries need to interact with the target LLM system to submit commands and text to try to bypass the security mechanisms or exploit it.
We are considering three forms of interaction with the system, as presented next.
Note that depending on the interaction, attacks may be easier or more complex to perform;
the same applies for defending against them.

\begin{itemize}
    \item \textbf{One-shot:} 
    Adversary only needs a single interaction, via a specifically-crafted prompt, to exploit the model.

    \item \textbf{Multi-turn:} 
    Adversary needs two or more interactions with the target LLM system to exploit it.
    In this form, the adversary takes advantage of the model's ability to retain and process context over multiple prompts \cite{Chen2024news}.

    \item \textbf{Indirectly:} 
    Adversary exploits the LLM by attacking the infrastructure, dependencies, or source of information used by the model when performing tasks.
\end{itemize}

\section{Analysis of Threat Severity Levels}
\label{sec:threats_severity_level}

Attacking LLM systems can be simple or extremely difficult depending on adversary goals and how they choose to execute the attack.
No group, company, or government wishing to deploy LLMs has enough resources to eliminate every possible threat.
Besides, some threats can be so expensive for the adversaries to execute (in terms of computational resources or preparation) that the reward obtained in case of a success is not worth it, or by the time the adversary completes the attack, the data stolen, for instance, has no value anymore.
For such reasons, it is paramount to prioritize the limited resources available to implement defenses against the most significant threats for a particular business or scenario.

In this section, we analyze the threats presented in Section~\ref{sec:attacks_to_LLM} according to the likelihood of each threat occurring and the impact in case of a successful attack.
We employ two methodologies commonly used by the security community to assess the severity of a threat or vulnerability: 
The OWASP Risk Rating\footnote{OWASP Risk Rating Methodology, available at \url{https://owasp.org/www-community/OWASP_Risk_Rating_Methodology} (last accessed at 2025.07.25)} 
and the 
CVSS v3.1\footnote{CVSS v3.1 Specification, available at \url{https://www.first.org/cvss/v3.1/specification-document} (last accessed at 2025.07.25)} (Common Vulnerability Scoring System).
Note that CVSS has many versions, including a more recent one (v4.0).
However, we will use version 3.1 since, by the time of writing this paper, it is still the version most adopted by the security community, including in the CVE database of software vulnerabilities.

To ensure consistency in assessing the severity of all threats, we followed a uniform approach driven by a set of assumptions.
For each threat from Tables~\ref{tab:threats_to_LLM_development} and \ref{tab:threats_to_LLM_operation}, we devised a scenario illustrating how an adversary might carry out the attack to estimate a severity level,
considering a particular attack strategy.
We assumed that the target LLM system is deployed in a manner that meets the threat prerequisites and has a vulnerability (due to a lack of security mechanisms or improper use or configuration of them) that allows an adversary to exploit it via the considered threat.
Whenever possible, we used the same considerations as the authors who presented the threat.
Although methodologies like the OWASP Risk Rating or CVSS offer detailed information on how to compute the severity level of a vulnerability, there is still subjectiveness in the process.
Therefore, these scores reflect our interpretations and are susceptible to discrepancies, serving as references rather than exact measurements.
For example, if new conditions or alternative attack methods are discovered by the time a particular threat is analyzed, this may lead to a different interpretation.
However, with the information available today, we do not expect the analysis to vary significantly.
Another point of attention is that we considered only the technical impact of the threats on CIA aspects, not the business impact, which could vary from one company to another.

\definecolor{myred}{HTML}{FFA1A1}
\definecolor{myorange}{HTML}{FFD59E}
\definecolor{mygreen}{HTML}{B4FF9F}
\definecolor{myyellow}{HTML}{F9FFA4}

\DeclareRobustCommand{\hllow}[1]{{\sethlcolor{mygreen}\hl{#1}}}
\DeclareRobustCommand{\hlmed}[1]{{\sethlcolor{myyellow}\hl{#1}}}
\DeclareRobustCommand{\hlhigh}[1]{{\sethlcolor{myorange}\hl{#1}}}
\DeclareRobustCommand{\hlcrit}[1]{{\sethlcolor{myred}\hl{#1}}}

\begin{table*}
\centering
\scriptsize
\caption{Severity Score of Threats to LLMs using CVSS and OWASP Methodologies.}
\begin{tabular}{cccM{1.4cm}lccM{1.4cm}lc}
\toprule

\multirow{2.6}{*}{\textbf{ID}} & 
\multirow{2.6}{*}{\textbf{\shortstack{Attack \\ Strategy}}} & 
\multicolumn{4}{c}{\textbf{CVSS 3.1}} & 
\multicolumn{4}{c}{\textbf{OWASP Risk Rating}} \\
\cmidrule(lr){3-6}
\cmidrule(lr){7-10}
& &
\textbf{Exploitability} &
\textbf{Tech. Impact} &
\textbf{Overall Score} & 
\textbf{String} & 
\textbf{Likelihood} & 
\textbf{Tech. Impact} &
\textbf{Overall Score} & 
\textbf{String} \\
\midrule

A01 & \aaIP & 1.6 & 4.2 & 5.9 (\hlmed{MEDIUM}) & \href{https://nvd.nist.gov/vuln-metrics/cvss/v3-calculator?vector=AV:N/AC:H/PR:N/UI:R/S:U/C:N/I:L/A:H&version=3.1}{NHNRUNLH} & 5.8 & 5.8 & 33.1 (\hlmed{MEDIUM}) & 917933680599 \\
A02 & \aaIP & 1.6 & 4.2 & 5.9 (\hlmed{MEDIUM}) & \href{https://nvd.nist.gov/vuln-metrics/cvss/v3-calculator?vector=AV:N/AC:H/PR:N/UI:R/S:U/C:N/I:L/A:H&version=3.1}{NHNRUNLH} & 5.8 & 5.8 & 33.1 (\hlmed{MEDIUM}) & 917933680599 \\
A03 & \aaSM & 2.2 & 5.2 & 7.4 (\hlhigh{HIGH}) & \href{https://nvd.nist.gov/vuln-metrics/cvss/v3-calculator?vector=AV:N/AC:H/PR:N/UI:N/S:U/C:N/I:H/A:H&version=3.1}{NHNNUNHH} & 4.8 & 6.8 & 32.1 (\hlhigh{HIGH}) & 914933630999 \\
A04 & \aaIN & 0.2 & 4.2 & 4.4 (\hlmed{MEDIUM}) & \href{https://nvd.nist.gov/vuln-metrics/cvss/v3-calculator?vector=AV:P/AC:H/PR:H/UI:N/S:U/C:N/I:L/A:H&version=3.1}{PHHNUNLH} & 4.1 & 4.3 & 17.5 (\hlmed{MEDIUM}) & 914231490377 \\
A05 & \aaDP & 2.8 & 1.4 & 4.3 (\hlmed{MEDIUM}) & \href{https://nvd.nist.gov/vuln-metrics/cvss/v3-calculator?vector=AV:N/AC:L/PR:L/UI:N/S:U/C:N/I:N/A:L&version=3.1}{NLLNUNNL} & 4.5 & 2.0 & 9.0 (\hllow{LOW}) & 917671410071 \\
A06 & \aaIP & 1.6 & 3.6 & 5.3 (\hlmed{MEDIUM}) & \href{https://nvd.nist.gov/vuln-metrics/cvss/v3-calculator?vector=AV:N/AC:H/PR:N/UI:R/S:U/C:N/I:N/A:H&version=3.1}{NHNRUNNH} & 5.8 & 4.0 & 23.0 (\hlmed{MEDIUM}) & 917933680079 \\
A07 & \aaSM & 1.6 & 4.7 & 6.9 (\hlmed{MEDIUM}) & \href{https://nvd.nist.gov/vuln-metrics/cvss/v3-calculator?vector=AV:N/AC:H/PR:N/UI:R/S:C/C:L/I:N/A:H&version=3.1}{NHNRCLNH} & 6.8 & 6.0 & 40.5 (\hlcrit{CRITICAL}) & 994933986099 \\
C01 & \aaSM & 0.7 & 3.6 & 4.4 (\hlmed{MEDIUM}) & \href{https://nvd.nist.gov/vuln-metrics/cvss/v3-calculator?vector=AV:N/AC:H/PR:H/UI:N/S:U/C:H/I:N/A:N&version=3.1}{NHHNUHNN} & 4.1 & 4.5 & 18.6 (\hlmed{MEDIUM}) & 990233619009 \\
C02 & \aaIN & 0.7 & 3.6 & 4.4 (\hlmed{MEDIUM}) & \href{https://nvd.nist.gov/vuln-metrics/cvss/v3-calculator?vector=AV:N/AC:H/PR:H/UI:N/S:U/C:H/I:N/A:N&version=3.1}{NHHNUHNN} & 3.9 & 4.5 & 17.4 (\hlmed{MEDIUM}) & 990231619009 \\
C03 & \aaSD & 0.4 & 1.4 & 1.8 (\hllow{LOW}) & \href{https://nvd.nist.gov/vuln-metrics/cvss/v3-calculator?vector=AV:P/AC:H/PR:L/UI:N/S:U/C:L/I:N/A:N&version=3.1}{PHLNULNN} & 5.3 & 2.3 & 11.8 (\hllow{LOW}) & 914673482007 \\
C04 & \aaKV & 1.0 & 3.6 & 4.7 (\hlmed{MEDIUM}) & \href{https://nvd.nist.gov/vuln-metrics/cvss/v3-calculator?vector=AV:L/AC:H/PR:N/UI:R/S:U/C:H/I:N/A:N&version=3.1}{LHNRUHNN} & 3.9 & 3.3 & 12.6 (\hlmed{MEDIUM}) & 990431416007 \\
C05 & \aaIN & 0.2 & 3.6 & 3.8 (\hllow{LOW}) & \href{https://nvd.nist.gov/vuln-metrics/cvss/v3-calculator?vector=AV:P/AC:H/PR:H/UI:N/S:U/C:H/I:N/A:N&version=3.1}{PHHNUHNN} & 4.6 & 3.3 & 15.0 (\hlmed{MEDIUM}) & 914271496007 \\
C06 & \aaDP & 1.2 & 1.4 & 2.6 (\hllow{LOW}) & \href{https://nvd.nist.gov/vuln-metrics/cvss/v3-calculator?vector=AV:N/AC:H/PR:L/UI:R/S:U/C:L/I:N/A:N&version=3.1}{NHLRULNN} & 4.1 & 1.8 & 7.2 (\hllow{LOW}) & 914633436001 \\
C07 & \aaIP & 1.6 & 1.4 & 3.1 (\hllow{LOW}) & \href{https://nvd.nist.gov/vuln-metrics/cvss/v3-calculator?vector=AV:N/AC:H/PR:N/UI:R/S:U/C:L/I:N/A:N&version=3.1}{NHNRULNN} & 6.5 & 3.8 & 24.4 (\hlhigh{HIGH}) & 947933986009 \\
C08 & \aaDP & 1.6 & 1.4 & 3.1 (\hllow{LOW}) & \href{https://nvd.nist.gov/vuln-metrics/cvss/v3-calculator?vector=AV:N/AC:H/PR:L/UI:N/S:U/C:L/I:N/A:N&version=3.1}{NHLNULNN} & 4.5 & 1.8 & 7.9 (\hllow{LOW}) & 617631486001 \\
C09 & \aaDP & 1.6 & 1.4 & 3.1 (\hllow{LOW}) & \href{https://nvd.nist.gov/vuln-metrics/cvss/v3-calculator?vector=AV:N/AC:H/PR:L/UI:N/S:U/C:L/I:N/A:N&version=3.1}{NHLNULNN} & 3.9 & 1.8 & 6.8 (\hllow{LOW}) & 614611486001 \\
C10 & \aaDP & 1.6 & 1.4 & 3.1 (\hllow{LOW}) & \href{https://nvd.nist.gov/vuln-metrics/cvss/v3-calculator?vector=AV:N/AC:H/PR:L/UI:N/S:U/C:L/I:N/A:N&version=3.1}{NHLNULNN} & 4.1 & 1.8 & 7.2 (\hllow{LOW}) & 614631486001 \\
C11 & \aaDP & 1.6 & 3.6 & 5.3 (\hlmed{MEDIUM}) & \href{https://nvd.nist.gov/vuln-metrics/cvss/v3-calculator?vector=AV:N/AC:H/PR:L/UI:N/S:U/C:H/I:N/A:N&version=3.1}{NHLNUHNN} & 4.5 & 2.5 & 11.3 (\hllow{LOW}) & 644631489001 \\
C12 & \aaDP & 1.6 & 3.6 & 5.3 (\hlmed{MEDIUM}) & \href{https://nvd.nist.gov/vuln-metrics/cvss/v3-calculator?vector=AV:N/AC:H/PR:L/UI:N/S:U/C:H/I:N/A:N&version=3.1}{NHLNUHNN} & 5.8 & 2.5 & 14.4 (\hllow{LOW}) & 647633989001 \\
C13 & \aaDP & 1.6 & 3.6 & 5.3 (\hlmed{MEDIUM}) & \href{https://nvd.nist.gov/vuln-metrics/cvss/v3-calculator?vector=AV:N/AC:H/PR:L/UI:N/S:U/C:H/I:N/A:N&version=3.1}{NHLNUHNN} & 5.4 & 2.5 & 13.4 (\hllow{LOW}) & 640691989001 \\
C14 & \aaRE & 0.4 & 3.6 & 4.0 (\hlmed{MEDIUM}) & \href{https://nvd.nist.gov/vuln-metrics/cvss/v3-calculator?vector=AV:P/AC:H/PR:L/UI:N/S:U/C:H/I:N/A:N&version=3.1}{PHLNUHNN} & 7.3 & 4.5 & 32.6 (\hlhigh{HIGH}) & 994693999009 \\
C15 & \aaIP & 1.6 & 3.7 & 5.8 (\hlmed{MEDIUM}) & \href{https://nvd.nist.gov/vuln-metrics/cvss/v3-calculator?vector=AV:N/AC:H/PR:N/UI:R/S:C/C:L/I:L/A:L&version=3.1}{NHNRCLLL} & 7.1 & 6.3 & 44.5 (\hlcrit{CRITICAL}) & 997933986379 \\
C16 & \aaIP & 1.6 & 1.4 & 3.1 (\hllow{LOW}) & \href{https://nvd.nist.gov/vuln-metrics/cvss/v3-calculator?vector=AV:N/AC:H/PR:N/UI:R/S:U/C:L/I:N/A:N&version=3.1}{NHNRULNN} & 6.3 & 3.8 & 23.4 (\hlhigh{HIGH}) & 997933916009 \\
C17 & \aaKV & 3.9 & 1.4 & 5.8 (\hlmed{MEDIUM}) & \href{https://nvd.nist.gov/vuln-metrics/cvss/v3-calculator?vector=AV:N/AC:L/PR:N/UI:N/S:C/C:L/I:N/A:N&version=3.1}{NLNNCLNN} & 6.1 & 3.3 & 19.9 (\hlhigh{HIGH}) & 997635916007 \\
I01 & \aaSC & 1.6 & 2.5 & 4.2 (\hlmed{MEDIUM}) & \href{https://nvd.nist.gov/vuln-metrics/cvss/v3-calculator?vector=AV:N/AC:H/PR:N/UI:R/S:U/C:N/I:L/A:L&version=3.1}{NHNRUNLL} & 6.1 & 5.3 & 32.2 (\hlhigh{HIGH}) & 944933980759 \\
I02 & \aaSC & 1.6 & 4.2 & 5.9 (\hlmed{MEDIUM}) & \href{https://nvd.nist.gov/vuln-metrics/cvss/v3-calculator?vector=AV:N/AC:H/PR:N/UI:R/S:U/C:N/I:H/A:L&version=3.1}{NHNRUNHL} & 6.8 & 5.8 & 38.8 (\hlhigh{HIGH}) & 994933980959 \\
I03 & \aaPO & 1.6 & 2.5 & 4.2 (\hlmed{MEDIUM}) & \href{https://nvd.nist.gov/vuln-metrics/cvss/v3-calculator?vector=AV:N/AC:H/PR:L/UI:N/S:U/C:N/I:L/A:L&version=3.1}{NHLNUNLL} & 5.0 & 4.8 & 23.8 (\hlmed{MEDIUM}) & 944691430757 \\
I04 & \aaDP & 1.6 & 2.5 & 4.2 (\hlmed{MEDIUM}) & \href{https://nvd.nist.gov/vuln-metrics/cvss/v3-calculator?vector=AV:N/AC:H/PR:L/UI:N/S:U/C:N/I:L/A:L&version=3.1}{NHLNUNLL} & 5.8 & 5.3 & 30.2 (\hlmed{MEDIUM}) & 947633680759 \\
I05 & \aaSC & 1.6 & 5.3 & 7.5 (\hlhigh{HIGH}) & \href{https://nvd.nist.gov/vuln-metrics/cvss/v3-calculator?vector=AV:N/AC:H/PR:N/UI:R/S:C/C:L/I:H/A:L&version=3.1}{NHNRCLHL} & 6.8 & 7.3 & 48.9 (\hlcrit{CRITICAL}) & 994933986959 \\
I06 & \aaSC & 1.6 & 3.6 & 5.3 (\hlmed{MEDIUM}) & \href{https://nvd.nist.gov/vuln-metrics/cvss/v3-calculator?vector=AV:N/AC:H/PR:N/UI:R/S:U/C:N/I:H/A:N&version=3.1}{NHNRUNHN} & 6.1 & 5.3 & 32.2 (\hlhigh{HIGH}) & 944933982919 \\
I07 & \aaPO & 1.6 & 2.5 & 4.2 (\hlmed{MEDIUM}) & \href{https://nvd.nist.gov/vuln-metrics/cvss/v3-calculator?vector=AV:N/AC:H/PR:L/UI:N/S:U/C:N/I:L/A:L&version=3.1}{NHLNUNLL} & 4.3 & 4.8 & 20.2 (\hlmed{MEDIUM}) & 944631430757 \\
I08 & \aaDP & 1.6 & 2.5 & 4.2 (\hlmed{MEDIUM}) & \href{https://nvd.nist.gov/vuln-metrics/cvss/v3-calculator?vector=AV:N/AC:H/PR:L/UI:N/S:U/C:N/I:L/A:L&version=3.1}{NHLNUNLL} & 4.3 & 4.8 & 20.2 (\hlmed{MEDIUM}) & 944631430757 \\
I09 & \aaDP & 2.8 & 4.7 & 7.6 (\hlhigh{HIGH}) & \href{https://nvd.nist.gov/vuln-metrics/cvss/v3-calculator?vector=AV:N/AC:L/PR:L/UI:N/S:U/C:L/I:H/A:L&version=3.1}{NLLNULHL} & 6.4 & 3.3 & 20.7 (\hlhigh{HIGH}) & 647699912911 \\
I10 & \aaDP & 2.8 & 4.7 & 7.6 (\hlhigh{HIGH}) & \href{https://nvd.nist.gov/vuln-metrics/cvss/v3-calculator?vector=AV:N/AC:L/PR:L/UI:N/S:U/C:L/I:H/A:L&version=3.1}{NLLNULHL} & 6.4 & 3.3 & 20.7 (\hlhigh{HIGH}) & 647699912911 \\
I11 & \aaDP & 2.8 & 4.7 & 7.6 (\hlhigh{HIGH}) & \href{https://nvd.nist.gov/vuln-metrics/cvss/v3-calculator?vector=AV:N/AC:L/PR:L/UI:N/S:U/C:L/I:H/A:L&version=3.1}{NLLNULHL} & 6.4 & 3.3 & 20.7 (\hlhigh{HIGH}) & 647699912911 \\
I12 & \aaDP & 2.8 & 4.7 & 7.6 (\hlhigh{HIGH}) & \href{https://nvd.nist.gov/vuln-metrics/cvss/v3-calculator?vector=AV:N/AC:L/PR:L/UI:N/S:U/C:L/I:H/A:L&version=3.1}{NLLNULHL} & 6.4 & 3.3 & 20.7 (\hlhigh{HIGH}) & 647699912911 \\
I13 & \aaDP & 2.8 & 4.7 & 7.6 (\hlhigh{HIGH}) & \href{https://nvd.nist.gov/vuln-metrics/cvss/v3-calculator?vector=AV:N/AC:L/PR:L/UI:N/S:U/C:L/I:H/A:L&version=3.1}{NLLNULHL} & 6.4 & 3.3 & 20.7 (\hlhigh{HIGH}) & 647699912911 \\
I14 & \aaIP & 1.6 & 6.0 & 8.3 (\hlhigh{HIGH}) & \href{https://nvd.nist.gov/vuln-metrics/cvss/v3-calculator?vector=AV:N/AC:H/PR:N/UI:R/S:C/C:H/I:H/A:H&version=3.1}{NHNRCHHH} & 6.8 & 9.0 & 60.8 (\hlcrit{CRITICAL}) & 997933689999 \\
I15 & \aaDP & 1.8 & 6.0 & 8.5 (\hlhigh{HIGH}) & \href{https://nvd.nist.gov/vuln-metrics/cvss/v3-calculator?vector=AV:N/AC:H/PR:L/UI:N/S:C/C:H/I:H/A:H&version=3.1}{NHLNCHHH} & 5.5 & 7.0 & 38.5 (\hlhigh{HIGH}) & 997633619991 \\
I16 & \aaIP & 1.6 & 4.7 & 6.4 (\hlmed{MEDIUM}) & \href{https://nvd.nist.gov/vuln-metrics/cvss/v3-calculator?vector=AV:N/AC:H/PR:N/UI:R/S:U/C:L/I:H/A:L&version=3.1}{NHNRULHL} & 5.9 & 6.3 & 36.7 (\hlhigh{HIGH}) & 947933482959 \\
I17 & \aaDP & 2.8 & 4.7 & 7.6 (\hlhigh{HIGH}) & \href{https://nvd.nist.gov/vuln-metrics/cvss/v3-calculator?vector=AV:N/AC:L/PR:L/UI:N/S:U/C:L/I:H/A:L&version=3.1}{NLLNULHL} & 6.8 & 3.3 & 21.9 (\hlhigh{HIGH}) & 947699912911 \\
I18 & \aaDP & 2.8 & 4.7 & 7.6 (\hlhigh{HIGH}) & \href{https://nvd.nist.gov/vuln-metrics/cvss/v3-calculator?vector=AV:N/AC:L/PR:L/UI:N/S:U/C:L/I:H/A:L&version=3.1}{NLLNULHL} & 6.4 & 3.3 & 20.7 (\hlhigh{HIGH}) & 647699912911 \\
I19 & \aaDP & 2.8 & 4.7 & 7.6 (\hlhigh{HIGH}) & \href{https://nvd.nist.gov/vuln-metrics/cvss/v3-calculator?vector=AV:N/AC:L/PR:L/UI:N/S:U/C:L/I:H/A:L&version=3.1}{NLLNULHL} & 5.9 & 3.3 & 19.1 (\hlmed{MEDIUM}) & 940699912911 \\
I20 & \aaDP & 2.8 & 4.7 & 7.6 (\hlhigh{HIGH}) & \href{https://nvd.nist.gov/vuln-metrics/cvss/v3-calculator?vector=AV:N/AC:L/PR:L/UI:N/S:U/C:L/I:H/A:L&version=3.1}{NLLNULHL} & 5.9 & 3.3 & 19.1 (\hlmed{MEDIUM}) & 940699912911 \\
I21 & \aaIP & 1.6 & 4.7 & 6.4 (\hlmed{MEDIUM}) & \href{https://nvd.nist.gov/vuln-metrics/cvss/v3-calculator?vector=AV:N/AC:H/PR:L/UI:N/S:U/C:L/I:H/A:L&version=3.1}{NHLNULHL} & 6.6 & 3.3 & 21.5 (\hlhigh{HIGH}) & 944699482911 \\
I22 & \aaIP & 1.6 & 5.3 & 7.5 (\hlhigh{HIGH}) & \href{https://nvd.nist.gov/vuln-metrics/cvss/v3-calculator?vector=AV:N/AC:H/PR:N/UI:R/S:C/C:L/I:H/A:L&version=3.1}{NHNRCLHL} & 6.8 & 7.3 & 48.9 (\hlcrit{CRITICAL}) & 997933686959 \\
I23 & \aaIP & 1.6 & 5.3 & 7.5 (\hlhigh{HIGH}) & \href{https://nvd.nist.gov/vuln-metrics/cvss/v3-calculator?vector=AV:N/AC:H/PR:N/UI:R/S:C/C:L/I:H/A:L&version=3.1}{NHNRCLHL} & 7.1 & 7.3 & 51.7 (\hlcrit{CRITICAL}) & 997933986959 \\
I24 & \aaDP & 1.6 & 4.7 & 6.4 (\hlmed{MEDIUM}) & \href{https://nvd.nist.gov/vuln-metrics/cvss/v3-calculator?vector=AV:N/AC:H/PR:L/UI:N/S:U/C:L/I:H/A:L&version=3.1}{NHLNULHL} & 5.6 & 3.3 & 18.3 (\hlmed{MEDIUM}) & 944639912911 \\
I25 & \aaDP & 1.6 & 4.7 & 6.4 (\hlmed{MEDIUM}) & \href{https://nvd.nist.gov/vuln-metrics/cvss/v3-calculator?vector=AV:N/AC:H/PR:L/UI:N/S:U/C:L/I:H/A:L&version=3.1}{NHLNULHL} & 6.1 & 3.3 & 19.9 (\hlhigh{HIGH}) & 647679912911 \\
I26 & \aaDP & 1.6 & 4.7 & 6.4 (\hlmed{MEDIUM}) & \href{https://nvd.nist.gov/vuln-metrics/cvss/v3-calculator?vector=AV:N/AC:H/PR:L/UI:N/S:U/C:L/I:H/A:L&version=3.1}{NHLNULHL} & 5.6 & 3.3 & 18.3 (\hlmed{MEDIUM}) & 944639912911 \\
I27 & \aaDP & 1.6 & 4.7 & 6.4 (\hlmed{MEDIUM}) & \href{https://nvd.nist.gov/vuln-metrics/cvss/v3-calculator?vector=AV:N/AC:H/PR:L/UI:N/S:U/C:L/I:H/A:L&version=3.1}{NHLNULHL} & 5.1 & 3.3 & 16.7 (\hlmed{MEDIUM}) & 940639912911 \\
I28 & \aaIN & 0.2 & 4.2 & 4.4 (\hlmed{MEDIUM}) & \href{https://nvd.nist.gov/vuln-metrics/cvss/v3-calculator?vector=AV:P/AC:H/PR:H/UI:N/S:U/C:N/I:H/A:L&version=3.1}{PHHNUNHL} & 4.0 & 5.3 & 21.0 (\hlmed{MEDIUM}) & 944231180957 \\

\bottomrule

\end{tabular}
\label{tab:threats_severity_level}
\end{table*}

Table~\ref{tab:threats_severity_level} summarizes the main results. 
We represent each threat of Tables~\ref{tab:threats_to_LLM_development} and \ref{tab:threats_to_LLM_operation} by their ID, and present the attack strategy considered during our estimation. Results are listed for the OWASP Risk Rating and CVSS v3.1 methodologies, including the overall score (and its constituting values - Exploitability/Likelihood and Technical Impact).
We also include a label indicating the severity of the threat (Low, Medium, High, or Critical), and a string vector summarizing the choices made during the estimation for each parameter of the corresponding methodology.

For the OWASP methodology, the string vector is composed of 12 characters, with each string position representing one parameter of the methodology, having specific and predefined values (the 0 is the minimum and 9 the maximum).
The parameters are (in order of appearance): \textit{Skill Level}, \textit{Motive}, \textit{Opportunity}, \textit{Size}, \textit{Ease of Discovery}, \textit{Ease of Exploit}, \textit{Awareness}, \textit{Intrusion Detection}, \textit{Loss of Confidentiality}, \textit{Loss of Integrity}, \textit{Loss of Availability}, and \textit{Loss of Accountability}.

\begin{figure*}
\centering
\subfloat[CVSS 3.1]{\includegraphics[width=0.3\textwidth]{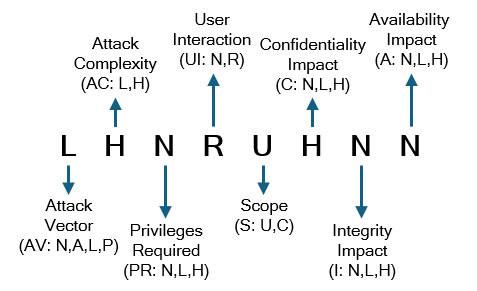}\label{fig:sub1}}
\subfloat[OWASP Risk Rating]{\includegraphics[width=0.39\textwidth]{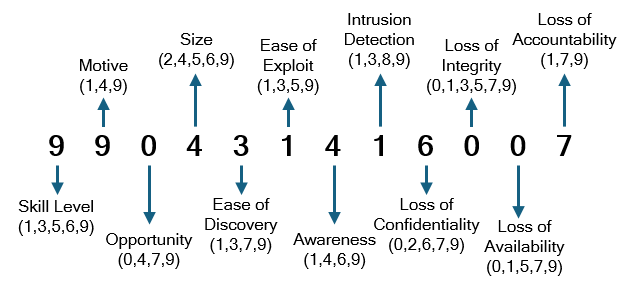}\label{fig:sub2}}
\caption{Example on how to interpret the severity score string values. On part (a) we provide the scoring produced using CVSS 3.1 ratings, whereas on (b), we provide an example of the OWASP risk rating. 
The acceptable values for each character is listed within parentheses.}
\label{fig:string_explanation_cvss_owasp}
\end{figure*}

The CVSS follows a similar way of representing the string vector, but this methodology adopts 8 (eight) different parameters to compute the severity score.
Each string position has some possible values, represented by their initial letter.
The parameters are (in order of appearance): \textit{Attack Vector}, \textit{Attack Complexity}, \textit{Privileges Required}, \textit{User Interaction}, \textit{Scope}, \textit{Confidentiality Impact}, \textit{Integrity Impact}, and \textit{Availability Impact}.
Figure~\ref{fig:string_explanation_cvss_owasp} presents the meaning and possible values of each character of CVSS and OWASP Risk Rating strings of Table~\ref{tab:threats_severity_level} (for a complete description, see their references).
The CVSS also has an official calculator to assist in computing the severity level, and based on the string that is inputted in the URL, it can fills the fields automatically.
We provide the links to the calculator in the string vector representation.

In this analysis, we adopted CVSS v3.1 using only the Base Score Metrics, but it also provides other components to customize the severity score assessment, such as the Temporal (analyzing the existence of exploits, patches, and credibility about the vulnerability) and Environment Score Metrics (allowing a score customization based on asset importance).
However, we will not use these other parameters since some of them are specific for a business case and others only apply to vulnerabilities, not threats.
In the same way, we choose to ignore the Business Impact factor of OWASP methodology.

There are a few differences among the parameters considered in the methodologies.
For instance, OWASP takes into consideration the skill level of adversaries choosing to exploit a vulnerability, while CVSS does not.
One can argue that with many exploits available today and the rise of LLMs, one can easily obtain enough knowledge to execute some attacks, and therefore, attacker knowledge is not an obstacle that needs to be considered anymore. 
However, what is more important to consider may be the resources required to perform an attack.
This aspect is reflected by both methodologies, the \textit{Attack Complexity} in CVSS and \textit{Opportunity} in OWASP.

Another difference between the methodologies is that CVSS considers \textit{User Interaction} in the process, while OWASP does not.
This is important because some threats can be exploited only in the moment the user performs a certain action, while others can be exploited at any time, elevating the risks.
On the other hand, OWASP considers the \textit{Motive} of adversaries in exploring a threat, based on the reward obtained.
This can also increase or decrease the chances for a threat being exploited:
the greater the potential rewards, the more likely adversaries are to invest time and resources in attempting the exploitation.

In the Impact category, OWASP provides a more granular level of choices when estimating the damage of a successful exploitation, allowing values from 0 to 9. 
CVSS provides only 3 choices: No Impact, Low Impact, and High Impact.
Having more values to choose enables a more precise estimation, but also increases the chances for disparities related to subjectiveness.
Another difference is that OWASP includes Loss of Accountability in the process, which considers the impact of tracing back adversary actions on the affected system to individuals. CVSS does not include this category, but does include Scope, which analyzes the impact of an exploit in other resources beyond the affected one.

OWASP methodology proposes using only a label to represent the severity level of a vulnerability, whereas CVSS assigns a value from 0.1 to 10, from which a label is derived.
In Table~\ref{tab:threats_severity_level}, in addition to the label obtained following OWASP methodology, we also presented a value obtained by multiplying the Likelihood and Technical Impact averages to understand how the values were distributed among the labels.
Note that by the results, we had values, such as for A02 threat of 33.06 with a MEDIUM label, but lower values were attributed a HIGH label, such as for the C07 case, with a value of 24.37.
By considering the possible values as 0 to 9, the minimum and maximum values for OWASP would be 0 and 81, respectively.

We can also see by the results that, for the same threat, different severity levels were obtained in some cases.
In most cases, the severity level of OWASP was higher than that of CVSS.
Another difference in the results is that six threats were assigned a CRITICAL label in OWASP, while no case was defined in CVSS.

Note that many of the threats presented the same severity level.
Such examples are mostly related to Jailbreak attacks, a common threat to LLMs.
We have considered different forms of performing a jailbreak; however, when computing their severity level, most of them have the same impact and exploitability/likelihood parameters.
For this type of threat, we have considered high impact levels in integrity and low levels in confidentiality and availability, since after a successful exploit of the LLM, adversaries can have access to some sensitive data and make it behave as they wish (i.e., completely control of the LLM).

Based on the results, one can prioritize the most severe threats according to the overall score or focus on those that have a higher impact or higher chances of occurring.
\section{Existing Mitigation Strategies}
\label{sec:mitigation_strategies}

Based on the threats discussed so far, it is evident that the use of LLMs demands specialized security measures that traditional controls fail to address. 
To secure systems using this technology, additional countermeasures must be integrated throughout the entire software and LLM development life cycles.
This section presents techniques to secure the development, deployment, and use of LLM systems that can mitigate threats and potential risks associated with LLMs.

In this section, we describe how we divided the mitigation techniques into eight categories, each dealing with one part of the LLM system.
Note that some techniques may be applicable to more than one category, and for a single threat, more than one technique can be applied.
Next, we explain the categories and provide examples of techniques for each one of them.
Table~\ref{tab:Mitigation_Techniques} presents a compilation of some techniques proposed for mitigating the threats presented in Section~\ref{sec:attacks_to_LLM}, highlighting the life cycle phase and use case scenario in which it should be implemented (or impacts).

We highlight that this is not an exhaustive literature revision, nor does it presents all possible defensive techniques used by security professionals in real-world scenarios.
It is a compilation of techniques referenced in the literature and proposed as mitigation to threats, as well as those considered most effective by security specialists.
The eight mitigation categories described below were obtained from our literature review in Section~\ref{sec:related-work} and sources such as the Department of Homeland Security \cite{dhlsecurity2024} and NIST \cite{nist_adversarial_ml_attacks_2023}.

\newcommand*\circledw[1]{\tikz[baseline=(char.base)]{\node[shape=circle,fill=white,draw,inner sep=1.5pt] (char) {#1};}}

\begin{enumerate}[label=\protect\circledwone{\color{black}\small \arabic*}]

    \item \textbf{Data Management:}
    An LLM, as with any IA technique, is as good as the data it was trained on. The difference is that LLMs require huge amounts of data compared with other techniques. However, one must be cautious about the data it uses to train the model, since many problems can arise, ranging from privacy issues to poisoning. 
    Some techniques that should be used in any LLM pipeline are data cleaning (including the application of anonymization on sensitive data \cite{yan2024protecting, fed24, johnson2020deidentification}); sanitization (to detect and remove adversarial data) \cite{owasp_llm_top10, shang2024intentobfuscator, nist_adversarial_ml_attacks_2023}; encryption (at rest, on training data stored in the LLM infrastructure to avoid compromise) \cite{dhlsecurity2024}; and a strict access control policy to limit the access to and manipulation of the training data \cite{fed24, dhlsecurity2024}.
    Other important techniques are related to Data Provenance Analysis \cite{owasp_llm_top10, fed24}, by recording the obtained data metadata (e.g., source, modification date, etc), and Multiple Training Data Providers \cite{fed24}, by adopting a source credibility analysis or getting data from multiple sources in different time intervals, all with the aim of mitigating poisoning and supply chain threats.

    \item \textbf{Infrastructure - Development Environment:}
    Besides protecting the data, it is also paramount to protect the LLM development infrastructure, to counter any supply chain, insider, or other threat that may compromise the system during its development.
    Some important techniques that can be applied here are: a strict access control policy to limit the access to data and resources to the minimum level, and only for those who require it \cite{fed24, dhlsecurity2024}; encryption of data (in transit) \cite{dhlsecurity2024}; strong authentication \cite{owasp_llm_top10}; and secure session management to provide and maintain access to developers; restriction to only trusted development tools \cite{owasp_llm_top10, dhlsecurity2024} with integrity verification processes \cite{owasp_llm_top10, nist_adversarial_ml_attacks_2023} to avoid supply chain threats; VPN for remote access to the infrastructure \cite{tidjon2022threat}; and many others presented in Table~\ref{tab:Mitigation_Techniques}.
    For this category, it is also important
    to adopt Security and Privacy Best Practices, i.e., \textit{DevSecOps/LLMOps/SSDLC/Security by Design} \cite{dhlsecurity2024} and \textit{Privacy by Design} \cite{cavoukian2021privacy}, which will aid in mitigating many threats and compliance issues.
    
    \item \textbf{Infrastructure - Deployment Environment:}
    When LLM systems are deployed and ready for use, the security of the infrastructure relies on how the system was developed and whether security best practices were used. 
    Infrastructure security also relies on the techniques that can be applied to protect the system, data, and users from threats, since despite the adoption of security practices during development, failures can still occur and vulnerabilities can be discovered. 
    Some of the defense techniques include creating a segmented network and isolated environment to deploy the LLM system \cite{liu2024demystifying}, so that it does not compromise other resources in case of a breach.
    Also important are logging and audit trails \cite{fed24, owasp_llm_top10, dhlsecurity2024, chernyshev2024short}; routine vulnerability scanning on infrastructure \cite{dhlsecurity2024}; software update and patching \cite{owasp_llm_top10}; and environment monitoring, including solutions to detect threats \cite{owasp_llm_top10, cui2024risk, tidjon2022threat, virustotal, dhlsecurity2024} and monitor resource utilization \cite{owasp_llm_top10, cui2024risk}. 
    Finally, restricting resources usage can mitigate DoS attacks, with some techniques limiting the control over the LLM context window \cite{owasp_llm_top10, cui2024risk}; energy consumption \cite{shumailov2021sponge}; resources per request \cite{fed24, owasp_llm_top10}; and also defining a rate limiting and throttling \cite{fed24, owasp_llm_top10} on requests to LLMs.
    Another line of defense can be the offense, with solutions exploring vulnerabilities in automated attacks to disrupt operations \cite{pasquini2024hacking}.

    \item \textbf{LLM / App Robustness and Protection:}
    The core of an LLM system is the model.
    Many attacks target the model, trying to cause a jailbreak, manipulate data, or obtain sensitive data.
    The techniques to protect the model and make it more robust against attacks can be further classified into four distinct groups:
    \begin{itemize}
        \item \textbf{Model Adversarial Protection:} Techniques ranging from adversarial training \cite{fed24, cui2024risk} and backdoor detection \cite{qi2020onion, cui2024risk} to model retraining (or fine-tuning) \cite{shang2024intentobfuscator, liu2023jailbreaking, kandpal2023backdoor}, distillation \cite{cui2024risk, papernot2016distillation}, and compression \cite{tanveer2024pruning}.
        There are also techniques based on the moving target defense \cite{chen2024flexllm} approach.

        \item \textbf{Model Privacy Protection:} Techniques adding noise or new data in the model training process to protect sensitive data (differential privacy \cite{garg2024task, fed24, yan2024protecting, cui2024risk}, watermarking \cite{mozes2023use, cui2024risk, pagnotta2024tattooed}, regularization \cite{cui2024risk}, gradient-based defenses \cite{mozes2023use}),
        training the models in a specialized way such that sensitive data is protected or avoided during the model conception (task-specific knowledge distillation \cite{garg2024task}, federated learning methods \cite{yan2024protecting})
        or removing sensitive data from already trained models (unlearning \cite{chen2023unlearn, eldan2023s, fed24, struppek2024finding, sun2024psy, ren2024unveiling} and privacy with backdoors \cite{hintersdorf2023defending, fed24}).
        Another type of technique consists in protecting models from side-channel attacks caused by
        timing and power leakage.
        In the LLM context, many works show that graphical processing units (GPUs) can leak sensitive data and be exploited.
        Defenses can be the use of software-based techniques that create some sort of obfuscation during processing, such as shuffling, random workload execution, or using constant-time algorithms \cite{zhan2022graphics, horvathbarracuda, yaman2023agent}.
        Other approaches, based on hardware, aim to reduce the leakage by using an electromagnetic shielding or via noise generation (with a radio-frequency device) \cite{zhan2022graphics, horvathbarracuda}.

        \item \textbf{Model Supply Chain Protection:} Techniques for selecting foundation open-source models \cite{fed24}, defining and analyzing LLM system software bill of materials and machine learning bill of materials (SBOM / MLBOM) \cite{owasp_llm_top10}, and also performing software dependencies verification and selection \cite{dhlsecurity2024}.

        \item \textbf{Model General Protection:} Model defense techniques against memory manipulation \cite{coalson2024prisonbreak, nazari2024forget, wang2023aegis, liu2023neuropots} or exploitation of model deserialization vulnerabilities, using safe persistence formats to deploy models \cite{nist_adversarial_ml_attacks_2023}, since many common libraries have vulnerabilities that allow code execution in the hosting environment, e.g., via I47 attack, exploiting CVE-2022-29216 (for tensorfow), or CVE-2019-6446 (for pickle in neural network tools).
        This group also considers the execution of red teaming practices on LLM (including LLM vulnerability scanners) \cite{fed24, owasp_llm_top10, mozes2023use, shang2024intentobfuscator, bullwinkellessons, tidjon2022threat, brokman2024insights, lee2025exploring} to search for risks and vulnerabilities in the LLM system.
        Although such practices are constantly encouraged by the industry, \cite{feffer2024red} argues that red teaming is often divergent regarding its purpose, settings, and target, and brings several concerns about this practice, characterizing its scope, structure, and criteria.
        Finally, using techniques such as reinforcement learning from human feedback (RLHF) \cite{nist_adversarial_ml_attacks_2023, su2024mission} or supervised fine-tuning~\cite{ge2023supervised} can help align the model with the system goals, preventing unwanted behaviors and leaking of sensitive content. 
        In particular for the I50 attack, \cite{huang2024harmful} discusses different techniques to defend against such a threat depending on the fine-tuning technique used.
        Defensive prompts \cite{wu2024the} can be another alternative when additional LLM training is not possible.
    \end{itemize}

    \item \textbf{Input Preprocessing:}
    Analyzing and processing inputs to the LLM system should also be another concern.
    It is paramount to consider any entity supplying data to the system as unreliable, and apply different techniques before data is ready for use within the system. 
    Techniques exist for validating \cite{fed24, owasp_llm_top10, shang2024intentobfuscator, nist_adversarial_ml_attacks_2023, cui2024risk}, sanitizing \cite{fed24, owasp_llm_top10, shang2024intentobfuscator, nist_adversarial_ml_attacks_2023}, and formatting \cite{fed24, nist_adversarial_ml_attacks_2023, cui2024risk, yi2023benchmarking, feffer2024signedprompt, lee2024prompt} the input, for instance, so that the model can distinguish between user and injected instructions.
    Another class of techniques encompasses the detection of malicious content within the input, using techniques such as poisoning protection~\cite{qi2020onion, cui2024risk, qi2021hidden}, context-aware filtering \cite{shang2024intentobfuscator}, adversarial examples detection~\cite{cui2024risk}, segmentation \cite{shang2024intentobfuscator}, warning~\cite{juuti2019prada, cui2024risk}, latent-space monitoring \cite{bailey2025obfuscatedactivationsbypassllm, ball2024understandingjailbreaksuccessstudy}, Key-Value Caches Optimization~\cite{jiang2024robustkv}, classifiers~\cite{chen2024defending, galinkin2024improved}, and also using another LLM to verify the input for malicious content \cite{schulhoff2023ignore, nist_adversarial_ml_attacks_2023, liu2023prompt}.
    Other works analyze the inputted prompt using mechanisms such as the attention distribution~\cite{liu2024feint, hung2024attention}, Gradient Cuff~\cite{hu2024gradient}, or the LLM activations during inference that are outside a defined safety boundary \cite{gao2024shaping} to detect malicious content (a jailbreak attempting).

    \item \textbf{Output Processing:}
    It is also important to verify the output of an LLM, since sensitive content may be returned to users and applications.
    Some techniques to this purpose include validation \cite{fed24, owasp_llm_top10, shang2024intentobfuscator, greshake2023not}, sanitization \cite{fed24}, and encoding \cite{owasp_llm_top10, shang2024intentobfuscator}, with some of them aimed to mitigate undesired code execution or to make automatic processing of outputs more difficult.
    Techniques for malicious content detection can also be used, including the use of another LLM to verify the content to detect offensive or otherwise undesired content \cite{schulhoff2023ignore, du2024detecting} or provide model theft protection \cite{dubinski2024bucks, dziedzic2022increasing}.

    \item \textbf{User's Device:}
    When an LLM runs on a user device, specific techniques are required to provide security in an unsafe and uncontrolled (by the LLM provider) environment.
    To protect sensitive data or company intellectual property information from leaks, techniques such as a trusted execution environment technology \cite{chen2024llm, gim2024confidential} can be adopted.
    Besides, all communication between a user's device and the LLM provider should be encrypted \cite{dhlsecurity2024} and have strong authentication \cite{owasp_llm_top10} and secure session management. 
    Antivirus or control-flow integrity techniques can provide device endpoint security \cite{cui2024risk}, and device authentication and attestation can allow a trust evaluation between the parties.

    \item \textbf{User Awareness:}
    The last category is about people, with some solutions requiring the user to approve or confirm (declare consent) about the execution of privileged operations executed by the LLM system, e.g., human-in-the-loop \cite{fed24, owasp_llm_top10}, especially in scenarios where LLM-integrated applications are used and can perform operations in the device.
    Another important aspect involving users, including end users and developers in general, is training \cite{fed24, dhlsecurity2024}.
    It is important to inform those responsible for operating the system or developing it about LLM threats and usage risks, as well as Security by Design / Secure Software Development Lifecycle (SSDLC) principles \cite{dhlsecurity2024} (including Defense-in-Depth, Least Privilege, Separation of Duties, Weakest Link, among other security principles), and Privacy by Design \cite{fed24, cavoukian2021privacy}.
    Having knowledge about these topics can be effective in counter LLM attacks.

\end{enumerate}

\begin{table*}
\setlength\tabcolsep{5pt}
\setlength{\fboxsep}{0.6pt} 
\setlength{\fboxrule}{0pt} 
\renewcommand{\citepunct}{]\,\allowbreak[}
\centering
\scriptsize
\caption{List of mitigation techniques, their applicability to LLM life cycle phases, and the affected design choices per use case.}
\begin{tabular}{cp{8cm}cp{0.1cm}p{0.1cm}p{0.1cm}p{0.1cm}p{0.1cm}p{0.1cm}C{0.2cm}C{0.2cm}C{0.2cm}C{0.2cm}C{0.2cm}C{0.2cm}}
\toprule
& & & \multicolumn{6}{c}{\textbf{LLM System Life Cycle}} & & & & & & \\
\cmidrule(lr){4-9}
& & & \multicolumn{5}{c}{\textbf{D}} & \textbf{O} & \multicolumn{6}{c}{\textbf{}} \\
\cmidrule(lr{0.8ex}){4-8}  \cmidrule(l{1ex}r{2.3ex}){9-9}
& & & & & & & & & \multicolumn{6}{c}{\textbf{Design Choices per Use Case}} \\
\cmidrule(lr){10-15}
& & & & & & & & &  \multicolumn{3}{c}{\textbf{D}} & \multicolumn{3}{c}{\textbf{O}} \\
\cmidrule(lr{0.8ex}){10-12} \cmidrule(lr){13-15}
\multirow{-7.2}{*}{\textbf{ID}} & \centering \multirow{-7.2}{*}{\textbf{Mitigation Technique}} &  \multirow{-7.2}{*}{\textbf{\shortstack{Mitigation \\ Category}}} & \multirow{-3.5}{*}{\rotatebox[origin=l]{90}{\textbf{Planning}}} & \multirow{-3.6}{*}{\rotatebox[origin=l]{90}{\textbf{Sys. Dev.}}} & \multirow{-4}{*}{\rotatebox[origin=l]{90}{\textbf{Data Eng.}}} & \multirow{-4.1}{*}{\rotatebox[origin=l]{90}{\textbf{LLM Dev.}}} & \multirow{-3.9}{*}{\rotatebox[origin=l]{90}{\textbf{LLM Int.}}} & \multirow{-4}{*}{\rotatebox[origin=l]{90}{\textbf{Operation}}} & \textbf{FM} & \textbf{FT} & \textbf{RG} & \textbf{CB} & \textbf{AP} & \textbf{AG} \\
\midrule
M01 & Data Cleaning \cite{yan2024protecting, fed24, johnson2020deidentification} & \circledwonesmall{1} & - & - & \Spt & - & - & - & \multicolumn{3}{c}{\mybox{1.3cm}{ DP, CL}} & - & - & - \\
M02 & Data Provenance Analysis \cite{owasp_llm_top10, fed24} & \circledwonesmall{1} & - & - & \Spt & - & - & - & \multicolumn{3}{c}{\mybox{1.3cm}{ DP, DI, SI, CL}} & - & - & - \\
M03 & Multiple Training Data Providers \cite{fed24} & \circledwonesmall{1} & - & - & \Spt & - & - & - & \multicolumn{3}{c}{\mybox{1.3cm}{ DP, CL}} & - & - & - \\
M04 & Training Data Encryption (at rest) \cite{dhlsecurity2024} & \circledwonesmall{1} & - & - & \Spt & - & - & - & \multicolumn{3}{c}{\mybox{1.3cm}{ DI, SI}} & - & - & - \\
M05 & Training Data Sanitization \cite{owasp_llm_top10, shang2024intentobfuscator, nist_adversarial_ml_attacks_2023} & \circledwonesmall{1} & - & - & \Spt & - & - & - & \multicolumn{3}{c}{\mybox{1.3cm}{ DP, CL}} & - & - & - \\
M06 & Strict Access Control Policy \cite{fed24, dhlsecurity2024,owasp_llm_top10} & \circledwonesmall{1} \circledwonesmall{2} \circledwonesmall{3} & \Spt & \Spt & \Spt & \Spt & \Spt & \Spt & \multicolumn{6}{c}{\mybox{3cm}{ DI, SI, AR}} \\
M07 & Development Environment Monitoring \cite{owasp_llm_top10} & \circledwonesmall{2} & \Spt & \Spt & \Spt & \Spt & \Spt & - & \multicolumn{6}{c}{\mybox{3cm}{ DI, SL, SI}} \\
M08 & Restriction to Trusted Development Tools \cite{owasp_llm_top10, dhlsecurity2024} & \circledwonesmall{2} & \Spt & \Spt & \Spt & \Spt & \Spt & - & \multicolumn{6}{c}{\mybox{3cm}{ DI, SL}} \\
M09 & Security and Privacy Best Practices Adoption \cite{dhlsecurity2024, cavoukian2021privacy} & \circledwonesmall{2} & \Spt & \Spt & \Spt & \Spt & \Spt & \Spt & \multicolumn{6}{c}{\mybox{3cm}{ DP, DI, SL, SI, IO, AR, CL}} \\
M10 & Signature and Integrity Verification \cite{owasp_llm_top10, nist_adversarial_ml_attacks_2023} & \circledwonesmall{2} & \Spt & \Spt & \Spt & \Spt & \Spt & - & \multicolumn{6}{c}{\mybox{3cm}{ SL}} \\
M11 & VPN for Remote Access \cite{tidjon2022threat} & \circledwonesmall{2} & \Spt & \Spt & \Spt & \Spt & \Spt & - & \multicolumn{6}{c}{\mybox{3cm}{ DI}} \\
M12 & Incident Response Plan \cite{dhlsecurity2024} & \circledwonesmall{2} \circledwonesmall{3} & \Spt & - & - & - & - & - & \multicolumn{6}{c}{\mybox{3cm}{ DI}} \\
M13 & Logging and Audit Trails \cite{fed24, owasp_llm_top10, dhlsecurity2024, chernyshev2024short} & \circledwonesmall{2} \circledwonesmall{3} & \Spt & \Spt & \Spt & \Spt & \Spt & \Spt & \multicolumn{6}{c}{\mybox{3cm}{ DI, SI, AR}} \\
M14 & Network Segmentation \cite{nistspcontrols} & \circledwonesmall{2} \circledwonesmall{3} & \Spt & \Spt & \Spt & \Spt & \Spt & \Spt & \multicolumn{6}{c}{\mybox{3cm}{ DI, SI}} \\
M15 & Network Traffic Analyzer \cite{tidjon2022threat} & \circledwonesmall{2} \circledwonesmall{3} & \Spt & \Spt & \Spt & \Spt & \Spt & \Spt & \multicolumn{6}{c}{\mybox{3cm}{ DI}} \\
M16 & Routine Software Update and Patching \cite{owasp_llm_top10} & \circledwonesmall{2} \circledwonesmall{3} & \Spt & \Spt & \Spt & \Spt & \Spt & \Spt & \multicolumn{6}{c}{\mybox{3cm}{ SL}} \\
M17 & Routine Vulnerability Scanning on Infrastructure \cite{dhlsecurity2024} & \circledwonesmall{2} \circledwonesmall{3} & \Spt & \Spt & \Spt & \Spt & \Spt & \Spt & \multicolumn{6}{c}{\mybox{3cm}{ DI}} \\
M18 & Secure Environment Configuration \cite{dhlsecurity2024} & \circledwonesmall{2} \circledwonesmall{3} & \Spt & \Spt & \Spt & \Spt & \Spt & \Spt & \multicolumn{6}{c}{\mybox{3cm}{ DI, SI, AR}} \\
M19 & Communication Data Encryption (in transit) \cite{dhlsecurity2024} & \circledwonesmall{2} \circledwonesmall{3} \circledwonesmall{7} & \Spt & \Spt & \Spt & \Spt & \Spt & \Spt & \multicolumn{6}{c}{\mybox{3cm}{ DI, IO, AR, CL}} \\
M20 & Secure and Strong Authentication \cite{owasp_llm_top10} & \circledwonesmall{2} \circledwonesmall{3} \circledwonesmall{7} & \Spt & \Spt & \Spt & \Spt & \Spt & \Spt & \multicolumn{6}{c}{\mybox{3cm}{ DI}} \\
M21 & Secure Session Management \cite{owaspcheatsheet, rfc2109} & \circledwonesmall{2} \circledwonesmall{3} \circledwonesmall{7} & \Spt & \Spt & \Spt & \Spt & \Spt & \Spt & \multicolumn{6}{c}{\mybox{3cm}{ DI}} \\
M22 & Data Processing Encryption (in use) \cite{costan2026intelsgx} & \circledwonesmall{3} & - & - & - & - & - & \Spt &  - & - & - & \multicolumn{3}{c}{\mybox{1.34cm}{ DI, SI}} \\
M23 & Deployment Environment Monitoring \cite{owasp_llm_top10, cui2024risk, tidjon2022threat, virustotal, dhlsecurity2024} & \circledwonesmall{3} & - & - & - & - & - & \Spt &  - & - & - & \multicolumn{3}{c}{\mybox{1.34cm}{ DI, SI, AR}} \\
M24 & Execution Environment Isolation \cite{liu2024demystifying} & \circledwonesmall{3} & - & - & - & - & - & \Spt &  - & - & - & \multicolumn{3}{c}{\mybox{1.34cm}{ DI, SI, AR}} \\
M25 & Limiting Infrastructure Resources Use \cite{owasp_llm_top10, cui2024risk, shumailov2021sponge, fed24} & \circledwonesmall{3} & - & - & - & - & - & \Spt &  - & - & - & \multicolumn{3}{c}{\mybox{1.34cm}{ DI}} \\
M26 & Load Balancing Adoption \cite{abdulkareem2022loadbalance} & \circledwonesmall{3} & - & - & - & - & - & \Spt &  - & - & - & \multicolumn{3}{c}{\mybox{1.34cm}{ DI}} \\
M27 & Minimum Software Permissions and Execution Rights \cite{liu2024demystifying, fed24, owasp_llm_top10} & \circledwonesmall{3} \circledwonesmall{7} & - & - & - & - & - & \Spt &  - & - & - & \multicolumn{3}{c}{\mybox{1.34cm}{ DI, SI, AR}} \\
M28 & Trusted Execution Environment Technology Adoption \cite{chen2024llm, gim2024confidential} & \circledwonesmall{3} \circledwonesmall{7} & - & - & - & - & - & \Spt &  - & - & - & \multicolumn{3}{c}{\mybox{1.34cm}{ DI, SI}} \\
M29 & Model Adversarial Protection \cite{fed24, cui2024risk, qi2020onion, shang2024intentobfuscator, liu2023jailbreaking, papernot2016distillation, kandpal2023backdoor, hajipour2024codelmsec, hsu2024safe, kim2024mitigating, liu2024mitigating-1, peng2024rapid, tao2024robustness, varshney2023the, wang2024adversarial, zhang2025shortlength, chen2024flexllm, tanveer2024pruning} & \circledwonesmall{4} & - & - & - & \Spt & - & - & \multicolumn{6}{c}{\mybox{3cm}{ DP, IO}} \\
M30 & Model General Protection \cite{coalson2024prisonbreak, nazari2024forget, wang2023aegis, liu2023neuropots, fed24, owasp_llm_top10, mozes2023use, shang2024intentobfuscator, bullwinkellessons, tidjon2022threat, brokman2024insights, nist_adversarial_ml_attacks_2023, yi2023benchmarking, bhatt2024alert, chen2024agentpoison, deng2025adversaflow, feffer2024red, ganguli2022red, lee2024learning, perez2022red, raheja2024recent, traykov2024a, verma2024operationalizing, xu2024redagent, xue2023explore, derczynski2024garak, su2024mission, jiang2024wildteaming, lee2025exploring, cao2023red, holanda2024a, der2024threatfinderai, wu2024the, ge2023supervised} & \circledwonesmall{4} & - & - & - & \Spt & - & \Spt & \multicolumn{6}{c}{\mybox{3cm}{ DI, SL, SI, IO, AR}} \\
M31 & Model Privacy Protection \cite{garg2024task, fed24, yan2024protecting, cui2024risk, mozes2023use, hintersdorf2023defending, chen2023unlearn, eldan2023s, struppek2024finding, sun2024psy, ren2024unveiling, du2024privacy, feretzakis2024trustworthy, fulvio2024data, jang2023knowledge, kim2023propile, liu2024learning, mai2024splitanddenoise, luo2024privacy, patil2024can, pan2020privacy} & \circledwonesmall{1} \circledwonesmall{4} & - & - & - & \Spt & - & \Spt & \multicolumn{6}{c}{\mybox{3cm}{ DP, IO}} \\
M32 & Model Supply Chain Protection \cite{fed24, owasp_llm_top10, dhlsecurity2024} & \circledwonesmall{4} & \Spt & - & - & - & - & - & \multicolumn{6}{c}{\mybox{3cm}{ DI, SL}} \\
M33 & Malicious Input Content Detection \cite{juuti2019prada, cui2024risk, qi2020onion, qi2021hidden, schulhoff2023ignore, nist_adversarial_ml_attacks_2023, liu2023prompt, greshake2023not, shang2024intentobfuscator, bailey2025obfuscatedactivationsbypassllm, ball2024understandingjailbreaksuccessstudy, hu2024token, liu2024feint, gao2024shaping, chen2024defending, galinkin2024improved, hu2024gradient, hung2024attention, cao2024guide, ji2024defending, jiang2024robustkv, li2024chainofscrutiny, muliarevych2024enhancing, wang2024defensive, xi2023defending, tan2024knowledge} & \circledwonesmall{5} & - & - & - & - & - & \Spt &  - & - & - & \multicolumn{3}{c}{\mybox{1.34cm}{ IO}} \\
M34 & Prompt Instruction and Formatting \cite{fed24, nist_adversarial_ml_attacks_2023, cui2024risk, yi2023benchmarking, feffer2024signedprompt, lee2024prompt} & \circledwonesmall{5} & - & - & - & - & - & \Spt &  - & - & - & \multicolumn{3}{c}{\mybox{1.34cm}{ IO}} \\
M35 & Sanitization \cite{fed24, owasp_llm_top10, shang2024intentobfuscator, nist_adversarial_ml_attacks_2023} & \circledwonesmall{5} & - & - & - & - & - & \Spt &  - & - & - & \multicolumn{3}{c}{\mybox{1.34cm}{ IO}} \\
M36 & Validation \cite{fed24, owasp_llm_top10, shang2024intentobfuscator, nist_adversarial_ml_attacks_2023, cui2024risk} & \circledwonesmall{5} & - & - & - & - & - & \Spt &  - & - & - & \multicolumn{3}{c}{\mybox{1.34cm}{ IO}} \\
M37 & Encoding \cite{owasp_llm_top10, shang2024intentobfuscator} & \circledwonesmall{6} & - & - & - & - & - & \Spt &  - & - & - & \multicolumn{3}{c}{\mybox{1.34cm}{ IO}} \\
M38 & Malicious Output Content Detection \cite{schulhoff2023ignore, dubinski2024bucks, dziedzic2022increasing, du2024detecting} & \circledwonesmall{6} & - & - & - & - & - & \Spt &  - & - & - & \multicolumn{3}{c}{\mybox{1.34cm}{ DP}} \\
M39 & Sanitization \cite{fed24} & \circledwonesmall{6} & - & - & - & - & - & \Spt &  - & - & - & \multicolumn{3}{c}{\mybox{1.34cm}{ DP}} \\
M40 & Validation \cite{fed24, owasp_llm_top10, shang2024intentobfuscator, greshake2023not} & \circledwonesmall{6} & - & - & - & - & - & \Spt &  - & - & - & \multicolumn{3}{c}{\mybox{1.34cm}{ DP}} \\
M41 & Device Authentication and Attestation \cite{rfc9711, tcg2023tpm} & \circledwonesmall{7} & \Spt & \Spt & \Spt & \Spt & \Spt & \Spt & \multicolumn{6}{c}{\mybox{3cm}{ DI, SL, SI, IO, AR, CL}} \\
M42 & Device Endpoint Security Solution Adoption \cite{cui2024risk} & \circledwonesmall{7} & \Spt & \Spt & \Spt & \Spt & \Spt & \Spt & \multicolumn{6}{c}{\mybox{3cm}{ DI, SI, IO, AR, CL}} \\
M43 & Human-in-the-loop \cite{fed24, owasp_llm_top10} & \circledwonesmall{8} & - & - & - & - & - & \Spt &  - & - & - & - & \multicolumn{2}{l}{\mybox{0.84cm}{DI, IO, AR}} \\
M44 & Training Developers \cite{fed24, cavoukian2021privacy, dhlsecurity2024} & \circledwonesmall{8} & \Spt & - & - & - & - & - & \multicolumn{6}{c}{\mybox{3cm}{ DP, DI, SL, SI, IO, AR, CL}} \\
M45 & Processing Obfuscation \cite{zhan2022graphics, yaman2023agent, horvathbarracuda} & \circledwonesmall{4} & - & - & - & - & - & \Spt &  - & - & - & \multicolumn{3}{c}{\mybox{1.34cm}{ DI, SI}} \\
M46 & Device Shielding \cite{zhan2022graphics, horvathbarracuda} & \circledwonesmall{3}, \circledwonesmall{4} & - & - & - & - & - & \Spt &  - & - & - & \multicolumn{3}{c}{\mybox{1.34cm}{ DI, SI}} \\
M47 & Hack-back \cite{pasquini2024hacking} & \circledwonesmall{3} & - & - & - & - & - & \Spt &  - & - & - & \multicolumn{3}{c}{\mybox{1.34cm}{ DI}} \\

\bottomrule

\end{tabular}
\label{tab:Mitigation_Techniques}
\end{table*}

In Table~\ref{tab:Mitigation_Techniques_and_threats}, we present the attack strategies attenuated by the mitigation techniques above. 
All marked cells indicate that a particular attacker strategy has had its effects reduced or completely mitigated by the adoption of the technique.
Note that many techniques may reduce the risk of a particular threat, and an effective mitigation should employ multiple techniques concurrently as a Defense-in-Depth approach.

\begin{table}
\centering
\scriptsize
\caption{Mapping attack strategies to Corresponding Mitigation Techniques.}

\renewcommand*{\fullcirc}[1][1ex]{~\tikz\fill (0,0) circle (#1);}
\begin{tabular}{cp{0.1cm}M{0.1cm}M{0.1cm}M{0.1cm}M{0.1cm}M{0.1cm}M{0.1cm}M{0.1cm}M{0.1cm}M{0.1cm}M{0.1cm}}
\toprule
\textbf{Mitigation} & 
\multicolumn{11}{c}{\textbf{Attack Strategies Attenuated}}
\\
\textbf{Technique (ID)} & \circledonesmall{1} & \circledonesmall{2} & \circledonesmall{3} & \circledonesmall{4} & \circledonesmall{5} & \circledonesmall{6} & \circledonesmall{7} & \circledonesmall{8} & \circledonesmall{9} & \circledtwosmall{10} & \circledtwosmall{11} \\

\midrule

M01 & \fullcirc[0.6ex] & \fullcirc[0.6ex] & \fullcirc[0.6ex] & & \fullcirc[0.6ex] & \fullcirc[0.6ex] & & \fullcirc[0.6ex] & \fullcirc[0.6ex] & \fullcirc[0.6ex] & \fullcirc[0.6ex]\\
M02 &  & & \fullcirc[0.6ex] & \fullcirc[0.6ex] & & & & & & & \\
M03 &  & & \fullcirc[0.6ex] & \fullcirc[0.6ex] & & & & & & & \\
M04 &  & & \fullcirc[0.6ex] & & \fullcirc[0.6ex] & \fullcirc[0.6ex] & & \fullcirc[0.6ex] & & \fullcirc[0.6ex] & \\
M05 &  & & \fullcirc[0.6ex] & \fullcirc[0.6ex] & & & & & & & \\
M06 &  & \fullcirc[0.6ex] & & & \fullcirc[0.6ex] & \fullcirc[0.6ex] & \fullcirc[0.6ex] & \fullcirc[0.6ex] & & \fullcirc[0.6ex] & \fullcirc[0.6ex]\\
M07 &  & & \fullcirc[0.6ex] & & \fullcirc[0.6ex] & \fullcirc[0.6ex] & \fullcirc[0.6ex] & \fullcirc[0.6ex] & & \fullcirc[0.6ex] & \\
M08 &  & & \fullcirc[0.6ex] & & & & & & & & \\
M09 &  \fullcirc[0.6ex] & \fullcirc[0.6ex] & \fullcirc[0.6ex] & \fullcirc[0.6ex] & \fullcirc[0.6ex] & \fullcirc[0.6ex] & \fullcirc[0.6ex] & \fullcirc[0.6ex] & \fullcirc[0.6ex] & \fullcirc[0.6ex] & \fullcirc[0.6ex]\\
M10 &  & & \fullcirc[0.6ex] & & & & & & & & \\
M11 &  & & & & & \fullcirc[0.6ex] & \fullcirc[0.6ex] & \fullcirc[0.6ex] & & \fullcirc[0.6ex] & \fullcirc[0.6ex] \\
M12 &  & \fullcirc[0.6ex] & \fullcirc[0.6ex] & \fullcirc[0.6ex] & \fullcirc[0.6ex] & \fullcirc[0.6ex] & \fullcirc[0.6ex] & \fullcirc[0.6ex] & & \fullcirc[0.6ex] & \\
M13 &  & \fullcirc[0.6ex] & & & \fullcirc[0.6ex] & \fullcirc[0.6ex] & \fullcirc[0.6ex] & \fullcirc[0.6ex] & & \fullcirc[0.6ex] & \\
M14 &  & \fullcirc[0.6ex] & & & & \fullcirc[0.6ex] & & \fullcirc[0.6ex] & & \fullcirc[0.6ex] & \fullcirc[0.6ex]\\
M15 &  & \fullcirc[0.6ex] & \fullcirc[0.6ex] & & & \fullcirc[0.6ex] & \fullcirc[0.6ex] & \fullcirc[0.6ex] & & \fullcirc[0.6ex] & \\
M16 &  & & \fullcirc[0.6ex] & & & \fullcirc[0.6ex] & & & & \fullcirc[0.6ex] & \\
M17 &  & & \fullcirc[0.6ex] & & & \fullcirc[0.6ex] & & \fullcirc[0.6ex] & & & \\
M18 &  & & & & \fullcirc[0.6ex] & \fullcirc[0.6ex] & & \fullcirc[0.6ex] & & & \fullcirc[0.6ex] \\
M19 &  & & & & \fullcirc[0.6ex] & \fullcirc[0.6ex] & \fullcirc[0.6ex] & \fullcirc[0.6ex] & & & \\
M20 &  \fullcirc[0.6ex] & \fullcirc[0.6ex] & & & \fullcirc[0.6ex] & & \fullcirc[0.6ex] & \fullcirc[0.6ex] & & & \\
M21 &  \fullcirc[0.6ex] & \fullcirc[0.6ex] & & & & \fullcirc[0.6ex] & \fullcirc[0.6ex] & \fullcirc[0.6ex] & & & \\
M22 &  & & & & \fullcirc[0.6ex] & & & & & \fullcirc[0.6ex] & \fullcirc[0.6ex]\\
M23 &  \fullcirc[0.6ex] & \fullcirc[0.6ex] & & & \fullcirc[0.6ex] & \fullcirc[0.6ex] & \fullcirc[0.6ex] & \fullcirc[0.6ex] & & \fullcirc[0.6ex] & \\
M24 &  & \fullcirc[0.6ex] & & & & \fullcirc[0.6ex] & & \fullcirc[0.6ex] & & \fullcirc[0.6ex] & \fullcirc[0.6ex]\\
M25 &  \fullcirc[0.6ex] & \fullcirc[0.6ex] & & & & \fullcirc[0.6ex] & & \fullcirc[0.6ex] & & \fullcirc[0.6ex] & \fullcirc[0.6ex] \\
M26 &  \fullcirc[0.6ex] & \fullcirc[0.6ex] & & & & \fullcirc[0.6ex] & & \fullcirc[0.6ex] & & & \\
M27 &  & \fullcirc[0.6ex] & & & \fullcirc[0.6ex] & \fullcirc[0.6ex] & & \fullcirc[0.6ex] & & \fullcirc[0.6ex] & \\
M28 &  & & & & & & & & \fullcirc[0.6ex] & \fullcirc[0.6ex] & \fullcirc[0.6ex]\\
M29 &  \fullcirc[0.6ex] & \fullcirc[0.6ex] & & \fullcirc[0.6ex] & & & & & & & \\
M30 &  \fullcirc[0.6ex] & \fullcirc[0.6ex] & \fullcirc[0.6ex] & & & \fullcirc[0.6ex] & & \fullcirc[0.6ex] & & & \\
M31 &  \fullcirc[0.6ex] & \fullcirc[0.6ex] & \fullcirc[0.6ex] & & \fullcirc[0.6ex] & \fullcirc[0.6ex] & \fullcirc[0.6ex] & \fullcirc[0.6ex] & \fullcirc[0.6ex] & \fullcirc[0.6ex] & \fullcirc[0.6ex]\\
M32 &  & & \fullcirc[0.6ex] & \fullcirc[0.6ex] & & & & & & & \\
M33 &  \fullcirc[0.6ex] & \fullcirc[0.6ex] & & \fullcirc[0.6ex] & & & & & & & \\
M34 &  \fullcirc[0.6ex] & \fullcirc[0.6ex] & & & & & & & & & \\
M35 &  \fullcirc[0.6ex] & \fullcirc[0.6ex] & & & & & & & & & \\
M36 &  \fullcirc[0.6ex] & \fullcirc[0.6ex] & & & & & & & & & \\
M37 &  \fullcirc[0.6ex] & \fullcirc[0.6ex] & & & & & & & & & \\
M38 &  \fullcirc[0.6ex] & \fullcirc[0.6ex] & & & & & & & & & \\
M39 &  \fullcirc[0.6ex] & \fullcirc[0.6ex] & & & & & & & & & \\
M40 &  \fullcirc[0.6ex] & \fullcirc[0.6ex] & & & & & & & & & \\
M41 &  \fullcirc[0.6ex] & \fullcirc[0.6ex] & & & & & \fullcirc[0.6ex] & & & \fullcirc[0.6ex] & \\
M42 &  & & & & & & & & & \fullcirc[0.6ex] & \\
M43 &  & \fullcirc[0.6ex] & & & & & & & & & \\
M44 &  \fullcirc[0.6ex] & \fullcirc[0.6ex] & \fullcirc[0.6ex] & \fullcirc[0.6ex] & \fullcirc[0.6ex] & \fullcirc[0.6ex] & \fullcirc[0.6ex] & \fullcirc[0.6ex] & \fullcirc[0.6ex] & \fullcirc[0.6ex] & \fullcirc[0.6ex] \\
M45 &  & & & & \fullcirc[0.6ex] & & & & & & \fullcirc[0.6ex]\\
M46 &  & & & & \fullcirc[0.6ex] & & & & & & \fullcirc[0.6ex]\\
M47 &  & & & & & \fullcirc[0.6ex] & & \fullcirc[0.6ex] & & &\\

\bottomrule
\end{tabular}
\label{tab:Mitigation_Techniques_and_threats}
\end{table}
\section{Analysis of LLM Deployment Scenarios}
\label{sec:threats_LLM_use_cases}

Based on the scenarios presented in Section~\ref{sec:llm_scenarios} and the threats presented in Section~\ref{sec:attacks_to_LLM}, we can see that not every threat applies to all LLM deployment scenarios.
For instance, threat C14 (Model Reverse Engineering) only affects models deployed on users' devices, not in external servers (e.g., the cloud); for those, threat A06 (Time-consuming Background Tasks) is applicable, with adversaries trying to cause instability in the system for legitimate users.
For this reason, a deeper analysis of a specific LLM scenario is necessary to identify the relevant threats and best defense strategies to put in place.

To assess the full impact of the surveyed threats, we conduct structured threat modeling across the four representative LLM scenarios described in Section~\ref{sec:ex_llm_scenarios}. This process allows for a consistent evaluation of how literature threats associates to practical threat evaluations, regarding severity, attack strategies, and mitigations, enabling future researchers to understand to what extent each threat applies to a given deployment. While not exhaustive, the provided modeling bridges theoretical insights from literature with practical guidelines, such as from OWASP Top 10 LLM~\cite{owasp_llm_top10}, supporting more grounded and actionable security recommendations.

\subsection{Application of STRIDE for LLM Scenarios}

For the threat modeling, we apply the STRIDE framework \cite{kohnfelder1999threats}.
We present potential and general risks or threats associated with one or more STRIDE mnemonic(s) 
(and the OWASP Top 10 LLM category, whenever available), 
a brief remark about the adversary goal and capabilities, the possible strategy adopted to perform the attack (from Section~\ref{sec:att_methods}), 
and the severity score considering the scenario under analysis and any relevant consideration described in the remark.
We associate these general risks/threats with the threats mapped in Section~\ref{sec:attacks_to_LLM}, and provide potential mitigation strategies from Section~\ref{sec:mitigation_strategies}.
For the severity score methodology, we consider only the CVSS v3.1 due to the advantages and objectiveness noticed during its application in Section~\ref{sec:threats_severity_level}, but we highlight that OWASP methodology would fit as well.
Notice, however, that the severity score of a particular risk may have a different value compared to the associated threats mapped, since the score in the STRIDE analysis is adapted to the scenario under consideration.
For instance, in Table~\ref{tab:TM_STRIDE_uc_on_device_chatobot_no_internet}, the \textit{Token Wasting} risk has a different CVSS value than the A07 threat presented in Table~\ref{tab:threats_severity_level}, since the conditions for the calculus changed (the privileges required changed from \textit{High} to \textit{None}, increasing the chances of a successful attack, thus elevating the score).

\begin{table*}
\centering
\footnotesize
\caption{Application of STRIDE for the LLM Context}
\begin{tabular}{p{3cm}p{13cm}}
\toprule
\multicolumn{1}{c}{\textbf{Category}}                   & \multicolumn{1}{c}{\textbf{Description}} \\
\midrule

\textbf{(S)poofing}                   & Adversary impersonates a legitimate user or system developer / administrator to gain access and manipulate or misuse the model's capabilities.\\

\textbf{(T)ampering}                  & Adversary manipulates data (training, fine-tuning, or RAG context data), model inputs/outputs or the environment where the model runs (configuration files, access control permissions, etc.). \\

\textbf{(R)epudiation}                & Adversary manipulates inputs to the LLM or performs adversarial attacks without being discovered, because there is no logging / authentication feature or it is insufficient.\\

\textbf{(I)nformation Disclosure}     & Adversary obtains sensitive information from the LLM, memorized by the model during the training, fine-tuning, or RAG data, by means of prompts or reverse engineering it. \\

\textbf{(D)enial of Service}          & Adversary makes the LLM system unavailable through resource exhaustion or API flooding/wasting.\\

\textbf{(E)levation of Privilege}     & Adversary gains unauthorized control of model or system resources.\\

\bottomrule

\end{tabular}
\label{tab:stride_for_llm}
\end{table*}

For each STRIDE acronym, we consider the interpretation adapted to the LLM context presented in Table~\ref{tab:stride_for_llm}.
Tables~\ref{tab:TM_STRIDE_uc_on_premises_chatobot_model_creation} to~\ref{tab:TM_STRIDE_uc_on_cloud_device_llm_agent} present our analysis results.

\begin{table*}
\centering
\setlength\tabcolsep{4pt}
\footnotesize
\caption{Threat Modeling of LLM Scenario 1 - Development of an LLM for Chat-bot Application.}
\label{tab:TM_STRIDE_uc_on_premises_chatobot_model_creation}
\begin{tabular}{P{1.2cm}P{2.7cm}p{8cm}P{0.8cm}P{1.05cm}P{1.2cm}P{1.2cm}}
\toprule

\textbf{STRIDE Acronym} &
\textbf{Risk \newline (Owasp Top 10 LLM)} & 
\centering \multirow{2}{*}{\textbf{Remark}} & 
\textbf{CVSS Score} & 
\textbf{Attack Strategy} &
\textbf{Related Threats} &
\textbf{Possible Defenses}\\ 
\midrule

\textbf{S T D} & 
Impersonating the User (LLM04:2025) & 
Adversary impersonates a legitimate user and manipulates information (obtained from public sources) used during LLM training. 
Adversary injects corrupted, malformed, irrelevant, or adversarial data into the training set to overwhelm and make the learning process more difficult or to cause performance degradation.
& 
\href{https://nvd.nist.gov/vuln-metrics/cvss/v3-calculator?vector=AV:N/AC:H/PR:N/UI:N/S:U/C:N/I:H/A:N&version=3.1}{4.8} &
\circledone{\color{white}4} & I01, I03 & M02,M06, M13,M20 \\

\textbf{S T D E} & 
Impersonating Vendor (LLM03:2025) & 
Adversary impersonates a legitimate vendor or open-source contributor, responsible for providing software used during the model development pipeline.
& 
\href{https://nvd.nist.gov/vuln-metrics/cvss/v3-calculator?vector=AV:N/AC:H/PR:N/UI:N/S:C/C:L/I:H/A:H&version=3.1}{8.9} &
\circledone{\color{white}3} & I05-06 & M06-10, M32 \\

\textbf{T} & 
Model Manipulation (LLM03:2025) & 
Adversary manipulates the weights, hyper-parameters or other metadata that affects the LLM training. & 
\href{https://nvd.nist.gov/vuln-metrics/cvss/v3-calculator?vector=AV:N/AC:H/PR:H/UI:N/S:U/C:N/I:H/A:L&version=3.1}{5.0} &
\circledone{\color{white}3} & I02 & M06-07, M10,M13 \\

\textbf{T, I, D} & 
Data Manipulation (LLM04:2025) & 
Adversary (insider or one compromising developer's credentials or security mechanisms) manipulates the data used during LLM training, stored on company's premises. & 
\href{https://nvd.nist.gov/vuln-metrics/cvss/v3-calculator?vector=AV:N/AC:H/PR:H/UI:N/S:U/C:H/I:H/A:L&version=3.1}{6.2}/\href{https://nvd.nist.gov/vuln-metrics/cvss/v3-calculator?vector=AV:L/AC:H/PR:H/UI:N/S:U/C:H/I:H/A:L&version=3.1}{6.0} &
\circledone{\color{white}4} \circledone{\color{white}5} & I01 & M02,M04, M06-07, M13,M20\\

\textbf{R} & 
Lack of Logging (-) & 
Adversary performs malicious and unauthorized actions du- ring LLM development without being discovered (there is no mechanism to track who added or changed data). & 
\href{https://nvd.nist.gov/vuln-metrics/cvss/v3-calculator?vector=AV:N/AC:H/PR:L/UI:N/S:U/C:L/I:L/A:L&version=3.1}{5.0/5.0} &
\circledone{\color{white}7} \circledone{\color{white}8} & - & M13\\

\textbf{I} & 
Sensitive Data Leakage (LLM02:2025) & 
LLM may expose sensitive information during its operation if trained on such data, due to its data memorization capability. & 
\href{https://nvd.nist.gov/vuln-metrics/cvss/v3-calculator?vector=AV:N/AC:L/PR:L/UI:N/S:U/C:H/I:N/A:N&version=3.1}{6.5} &
\circledone{\color{white}1} & C12-13 & M01,M31 \\

\bottomrule
\end{tabular}
\end{table*}


\begin{table*}
\centering
\setlength\tabcolsep{4pt}
\footnotesize
\caption{Threat Modeling of LLM Scenario 2 - Chat-bot Application on User's device.}
\label{tab:TM_STRIDE_uc_on_device_chatobot_no_internet}
\begin{tabular}{P{1.2cm}P{2.7cm}p{8cm}P{0.8cm}P{1.05cm}P{1.2cm}P{1.2cm}}
\toprule

\textbf{STRIDE Acronym} &
\textbf{Risk \newline (Owasp Top 10 LLM)} & 
\centering \multirow{2}{*}{\textbf{Remark}} & 
\textbf{CVSS Score} & 
\textbf{Attack Strategy} &
\textbf{Related Threats} &
\textbf{Possible Defenses}\\ 
\midrule

\textbf{S T I} & 
Impersonating the User (LLM03:2025) & 
Adversary (with physical access to the physical device or using malware) makes requests as if they were the legitimate user to obtain sensitive information or influence model behavior.
Adversary may also manipulates the weights, hyper-parameters, or other metadata (including RAG data) that affects the LLM. & 
\href{https://nvd.nist.gov/vuln-metrics/cvss/v3-calculator?vector=AV:P/AC:H/PR:L/UI:R/S:U/C:H/I:L/A:L&version=3.1}{5.0/5.0} &
\circledone{\color{white}1} \circledone{\color{white}7} & C15, I09-27, I02 & M20,M29, M31, M33-36, M42\\

\textbf{R} &
Lack of Logging (-) & 
Adversary performs malicious and unauthorized actions when using the model without being discovered (there is no mechanism to track who added or changed data). &
\href{https://nvd.nist.gov/vuln-metrics/cvss/v3-calculator?vector=AV:P/AC:H/PR:L/UI:N/S:U/C:L/I:L/A:L&version=3.1}{3.8/3.8} &
\circledone{\color{white}7} \circledone{\color{white}8} & - & M13 \\

\textbf{R} &
Denial of Malicious Inputs (-) &
Adversary performs an impersonation attack on a particular user. 
That user cannot deny malicious actions performed by the adversary.
& 
\href{https://nvd.nist.gov/vuln-metrics/cvss/v3-calculator?vector=AV:P/AC:H/PR:L/UI:N/S:U/C:H/I:H/A:N&version=3.1}{5.6} &
\circledone{\color{white}1} & - & M13,M20, M21,M41 \\

\textbf{I} &
Local Data Leakage (LLM02:2025) &
Adversary exploits vulnerabilities in the device, a compromised LLM-related software component (supply chain attack), or makes the user to install malware.
The goal is to steal sensitive data handled by or provided to the LLM (conversations, search history, and documents). & 
\href{https://nvd.nist.gov/vuln-metrics/cvss/v3-calculator?vector=AV:P/AC:H/PR:N/UI:N/S:U/C:H/I:L/A:L&version=3.1}{5.3}/\href{https://nvd.nist.gov/vuln-metrics/cvss/v3-calculator?vector=AV:N/AC:H/PR:N/UI:R/S:U/C:H/I:L/A:L&version=3.1}{6.4}&
\circledone{\color{white}3} \circledtwo{\color{white}10} & C15-16, I05-06 & M16,M27, M32, M41-42, M44 \\

\textbf{I} & 
Side-channel Leak (-) &
Adversary performs side-channel attacks in the device to obtain sensitive information from the LLM (internals, user queries, etc.). & 
\href{https://nvd.nist.gov/vuln-metrics/cvss/v3-calculator?vector=AV:P/AC:H/PR:L/UI:N/S:U/C:L/I:N/A:N&version=3.1}{1.8}&
\circledone{\color{white}1} & C03 & M28,M31, M45 \\

\textbf{I} &
Reverse Engineering (RE) (-) &
Adversary reverses engineering the LLM to obtain sensitive data by analyzing the binary file available in user's device or to create a shadow model (via prompts). The goal is to steal IP data (for business competitiveness) or attack other devices using the same model. & 
\href{https://nvd.nist.gov/vuln-metrics/cvss/v3-calculator?vector=AV:P/AC:H/PR:L/UI:N/S:U/C:H/I:H/A:N&version=3.1}{5.6}/\href{https://nvd.nist.gov/vuln-metrics/cvss/v3-calculator?vector=AV:P/AC:H/PR:L/UI:N/S:U/C:H/I:N/A:N&version=3.1}{4.0}&
\circledone{\color{white}9} \circledone{\color{white}1} & C13-14 & M04,M19, M28 \\

\textbf{D} &
Device Resource Exhaustion (LLM10:2025) &
Adversary overwhelms the device by instructing the model to perform time-consuming tasks before answering requests, affecting user experience (service degradation); this is done in the background (not visible to user) 
and can be achieved using malware. & 
\href{https://nvd.nist.gov/vuln-metrics/cvss/v3-calculator?vector=AV:N/AC:H/PR:H/UI:R/S:U/C:N/I:N/A:H&version=3.1}{4.2} &
\circledtwo{\color{white}10} & A06 & M25,M29, M33-36 \\

\textbf{D} &
Token wasting (-) &
Adversary steals user's API key (e.g., using malware) and perform several operations with it, incurring in financial loss or running out resource usage. & 
\href{https://nvd.nist.gov/vuln-metrics/cvss/v3-calculator?vector=AV:N/AC:H/PR:H/UI:R/S:U/C:L/I:N/A:H&version=3.1}{4.8} &
\circledtwo{\color{white}10} & A07 & M20,M25, M41-42 \\

\textbf{T I D E} & 
Privilege Escalation through Exploits (-) &
Adversary exploits vulnerabilities in the system (OS or application) to escalate privileges (usually using malware), having access to sensitive data and the ability to change the application behavior. & 
\href{https://nvd.nist.gov/vuln-metrics/cvss/v3-calculator?vector=AV:N/AC:H/PR:H/UI:R/S:U/C:H/I:H/A:H&version=3.1}{6.4} &
\circledtwo{\color{white}10} & I05-06 & M24,M27, M42\\
\bottomrule

\end{tabular}
\end{table*}


\begin{table*}
\setlength\tabcolsep{4pt}
\centering
\footnotesize
\caption{Threat Modeling of LLM Scenario 3 - LLM-Integrated App. on-Cloud.}
\label{tab:TM_STRIDE_uc_on_cloud_llm_integrated_app}
\begin{tabular}{P{1.2cm}P{2.7cm}p{8cm}P{0.8cm}P{1.05cm}P{1.2cm}P{1.2cm}}
\toprule

\textbf{STRIDE Acronym} &
\textbf{Risk \newline (Owasp Top 10 LLM)} & 
\centering \multirow{2}{*}{\textbf{Remark}} & 
\textbf{CVSS Score} & 
\textbf{Attack Strategy} &
\textbf{Related Threats} &
\textbf{Possible Defenses}\\ 
\midrule

\textbf{S I T D} & 
Impersonating the User (-) & 
Adversary impersonates a legitimate user (by exploiting failures in the authentication process, stealing an API key or session tokens, MITM, etc.) making requests as if they were the user to obtain sensitive information, influencing the model behavior (e.g., via jailbreak) or causing a denial of service (Token wasting / Device Resource Exhaustion). & 
\href{https://nvd.nist.gov/vuln-metrics/cvss/v3-calculator?vector=AV:N/AC:H/PR:L/UI:N/S:U/C:H/I:L/A:L&version=3.1}{6.4} &
\circledone{\color{white}8} & I09-11, I13, I15, I17-I18, I20, C15, A06-07 & M20,M23, M25,M31, M33-36\\

\textbf{S I T D} & 
Impersonating the LLM-integrated App (LLM03:2025) & 
Adversary impersonates the App (e.g., via supply chain attack) by spoofing its requests to the LLM system (on-cloud), thus manipulating the behavior of the model, getting sensitive information, or causing service degradation (device resource exhaustion). & 
\href{https://nvd.nist.gov/vuln-metrics/cvss/v3-calculator?vector=AV:N/AC:H/PR:L/UI:R/S:U/C:L/I:L/A:L&version=3.1}{4.6} &
\circledone{\color{white}3} & C16, I06, A03 & M06-10, M32 \\

\textbf{S I T D} & 
Server Spoofing (-) & 
Adversary spoofs the CSP by exploiting communication vulnerabilities and performs MITM, redirecting users to a malicious server controlled by them, reading sensitive content, and changing data sent to the LLM-integrated app. & 
\href{https://nvd.nist.gov/vuln-metrics/cvss/v3-calculator?vector=AV:N/AC:H/PR:H/UI:R/S:C/C:L/I:L/A:L&version=3.1}{5.1} &
\circledone{\color{white}6} & C15-16 & M16-17, M19,M41 \\

\textbf{T, I, D} & 
Insider (-) & 
Adversary is someone with privilege access to the LLM and its infrastructure (someone inside the CSP or a company developer), changing configuration files, model parameters, the model itself, or reading sensitive information. & 
\href{https://nvd.nist.gov/vuln-metrics/cvss/v3-calculator?vector=AV:L/AC:L/PR:H/UI:N/S:U/C:H/I:H/A:L&version=3.1}{6.3} &
\circledone{\color{white}5} & I02, C15-16 & M04, M06-07, M13,M20\\

\textbf{T} &
Feedback Poisoning (LLM04:2025) & 
Adversary manipulates the LLM evaluation option, used to take users' feedback into consideration for improving (retraining) the model. & 
\href{https://nvd.nist.gov/vuln-metrics/cvss/v3-calculator?vector=AV:N/AC:L/PR:L/UI:N/S:U/C:N/I:L/A:L&version=3.1}{5.4} &
\circledone{\color{white}1} & I08 & M02,M05, M33, M35-36\\

\textbf{T, I, D} & 
Data Manipulation (LLM04:2025) & 
Adversary (insider or one compromising developer's credentials or security mechanisms) manipulates private data used during LLM fine-tuning or RAG, stored on-cloud. & 
\href{https://nvd.nist.gov/vuln-metrics/cvss/v3-calculator?vector=AV:L/AC:H/PR:H/UI:N/S:U/C:H/I:H/A:L&version=3.1}{6.0}/\href{https://nvd.nist.gov/vuln-metrics/cvss/v3-calculator?vector=AV:N/AC:H/PR:H/UI:N/S:U/C:H/I:H/A:L&version=3.1}{6.2} &
\circledone{\color{white}5} \circledone{\color{white}8} & I01 & M02, M04-07, M13,M20 \\

\textbf{T R} &
Lack of Logging Mechanism (-) & 
Adversary performs malicious and unauthorized actions to the LLM without being discovered.
There is no mechanism to track who performed a particular action (misconfiguration) or tampered with logs. &
\href{https://nvd.nist.gov/vuln-metrics/cvss/v3-calculator?vector=AV:N/AC:H/PR:L/UI:N/S:U/C:L/I:L/A:L&version=3.1}{5.0}/\href{https://nvd.nist.gov/vuln-metrics/cvss/v3-calculator?vector=AV:N/AC:H/PR:H/UI:N/S:U/C:L/I:H/A:L&version=3.1}{5.5}
&
\circledone{\color{white}7} \circledone{\color{white}8} & - & M13\\

\textbf{I} & 
Sensitive Data Leakage (LLM02:2025) & 
LLM may expose proprietary, personal, or other sensitive information during its operation if trained on such data, due to its data memorization capability. & 
\href{https://nvd.nist.gov/vuln-metrics/cvss/v3-calculator?vector=AV:N/AC:L/PR:L/UI:N/S:U/C:H/I:N/A:N&version=3.1}{6.5} &
\circledone{\color{white}1} & C12-17 & M01,M31, M38,M40 \\

\textbf{I} &
Reverse Engineering (RE) (-) &
Adversary reverses engineering the application (available in user's device) responsible for communicating and accessing the LLM system
to obtain API keys or any other sensitive data. 
The adversary can also use malware to this end. & 
\href{https://nvd.nist.gov/vuln-metrics/cvss/v3-calculator?vector=AV:P/AC:H/PR:L/UI:N/S:U/C:H/I:N/A:L&version=3.1}{4.6}/\href{https://nvd.nist.gov/vuln-metrics/cvss/v3-calculator?vector=AV:N/AC:H/PR:H/UI:R/S:U/C:H/I:N/A:L&version=3.1}{4.8}&
\circledone{\color{white}9} \circledtwo{\color{white}10} & C14-15 & M04,M19, M28 \\

\textbf{TIDE} & 
Remote Code Execution (RCE) (LLM05:2025) & 
Adversary sends a problem-solving malicious request to the app, which creates a prompt to the LLM. The LLM generates code to solve the problem and returns it to the app. 
The app executes the command (triggering a malicious activity, e.g., steal confidential data) and returns the result to adversary. & 
\href{https://nvd.nist.gov/vuln-metrics/cvss/v3-calculator?vector=AV:N/AC:H/PR:L/UI:R/S:U/C:H/I:H/A:H&version=3.1}{7.1} &
\circledone{\color{white}2} & I14 & M24,M27, M33-36 \\

\textbf{D} & 
Service disruption (LLM10:2025) & 
Adversary exploits a vulnerability or forces a failure in the application to crush the system infrastructure, denying responses to other users. & 
\href{https://nvd.nist.gov/vuln-metrics/cvss/v3-calculator?vector=AV:N/AC:H/PR:L/UI:N/S:U/C:N/I:N/A:H&version=3.1}{5.3} &
\circledone{\color{white}6} & - & M16-17, M25, M33-36\\

\textbf{TIDE} & 
Privilege Escalation through Exploits (-) &
Adversary exploits vulnerabilities in the system (OS or application) to escalate privileges, having access to sensitive data and the ability to change the application behavior. & 
\href{https://nvd.nist.gov/vuln-metrics/cvss/v3-calculator?vector=AV:N/AC:H/PR:H/UI:N/S:U/C:H/I:H/A:H&version=3.1}{6.6} &
\circledone{\color{white}8} & I05-06 & M23-24, M27\\

\textbf{TID} & 
LLM Behavior Manipulation (LLM1:2025) & 
Adversary modifies the behavior of an LLM by indirectly making it retrieve malicious content from the internet (e.g., from a web page controlled by him/her). & 
\href{https://nvd.nist.gov/vuln-metrics/cvss/v3-calculator?vector=AV:N/AC:H/PR:L/UI:R/S:U/C:H/I:L/A:L&version=3.1}{5.9} &
\circledone{\color{white}2} & I14-15, I22 & M29, M33-36\\

\textbf{TIDE} & 
Abusing Shared Infrastructure (LLM3:2025) & 
Adversary abuses the CSP for not having a proper isolated infrastructure, allowing access to unauthorized data and resources available on that infrastructure (e.g., \textit{Cloud Jacking} attack). & 
\href{https://nvd.nist.gov/vuln-metrics/cvss/v3-calculator?vector=AV:N/AC:H/PR:L/UI:N/S:U/C:H/I:L/A:L&version=3.1}{6.4/6.4} &
\circledone{\color{white}6} \circledtwo{\color{white}11} & C06, A04 & M16-17, M24-25, M27,M28, M45-46\\
\bottomrule

\end{tabular}
\end{table*}


\begin{table*}
\centering
\setlength\tabcolsep{4pt}
\footnotesize
\caption{Threat Modeling of LLM Scenario 4 - LLM-based Agent for User Assistance.}
\label{tab:TM_STRIDE_uc_on_cloud_device_llm_agent}
\begin{tabular}{P{1.2cm}P{2.7cm}p{8cm}P{0.8cm}P{1.05cm}P{1.2cm}P{1.2cm}}
\toprule

\textbf{STRIDE Acronym} &
\textbf{Risk \newline (Owasp Top 10 LLM)} & 
\centering \multirow{2}{*}{\textbf{Remark}} & 
\textbf{CVSS Score} & 
\textbf{Attack Strategy} &
\textbf{Related Threats} &
\textbf{Possible Defenses}\\ 
\midrule


\textbf{S I T D} & 
Impersonating the LLM-Agent (LLM03:2025) & 
Adversary impersonates an agent (e.g., via supply chain) by spoofing its requests to the LLM system (on-cloud), thus manipulating the behavior of the model, getting sensitive information, or causing service degradation (device resource exhaustion). & 
\href{https://nvd.nist.gov/vuln-metrics/cvss/v3-calculator?vector=AV:N/AC:H/PR:L/UI:N/S:U/C:L/I:L/A:L&version=3.1}{5.0} &
\circledone{\color{white}3} & C16, I06, A03 & M06-10, M32 \\

\textbf{S I T D} & 
Server Spoofing (-) & 
Adversary spoofs the CSP by exploiting communication vulnerabilities and performs MITM, redirecting users to a malicious server, 
reading sensitive content, and changing data sent to the LLM system. & 
\href{https://nvd.nist.gov/vuln-metrics/cvss/v3-calculator?vector=AV:N/AC:H/PR:H/UI:R/S:C/C:L/I:L/A:L&version=3.1}{5.1} &
\circledone{\color{white}6} & C15-16 & M16-17, M19,M41 \\

\textbf{T, I, D} & 
Insider (-) & 
Adversary is someone with privilege access to the LLM and its infrastructure (someone inside the CSP or a company developer), changing configuration files, model parameters, the model itself, or reading sensitive information. & 
\href{https://nvd.nist.gov/vuln-metrics/cvss/v3-calculator?vector=AV:L/AC:L/PR:H/UI:N/S:U/C:H/I:H/A:L&version=3.1}{6.3} &
\circledone{\color{white}5} & I02, C15-16 & M04, M06-07, M13,M20\\

\textbf{T} &
Federated Learning Poisoning (LLM04:2025) & 
Adversary, controlling multiple clients or a collusion of multiple adversaries, send manipulated data to the LLM for model improvement (taking advantage of the federated learning process). & 
\href{https://nvd.nist.gov/vuln-metrics/cvss/v3-calculator?vector=AV:N/AC:H/PR:L/UI:N/S:U/C:N/I:H/A:L&version=3.1}{5.9} &
\circledone{\color{white}8} & I07 & M02,M05, M33,M41\\

\textbf{T, I, D} & 
Data Manipulation (LLM04:2025) & 
Adversary (insider or one compromising developer's credentials or security mechanisms) manipulates private data (RAG) stored on-cloud. & 
\href{https://nvd.nist.gov/vuln-metrics/cvss/v3-calculator?vector=AV:L/AC:H/PR:H/UI:N/S:U/C:H/I:H/A:L&version=3.1}{6.0}/\href{https://nvd.nist.gov/vuln-metrics/cvss/v3-calculator?vector=AV:N/AC:H/PR:H/UI:N/S:U/C:H/I:H/A:L&version=3.1}{6.2} &
\circledone{\color{white}5} \circledone{\color{white}8} & I01 & M02, M04-07, M13,M20 \\

\textbf{T R} &
Lack of Logging Mechanism (-) & 
Adversary performs malicious and unauthorized actions to the LLM without being discovered.
There is no mechanism to track who performed a particular action (misconfiguration) or tampered with logs. &
\href{https://nvd.nist.gov/vuln-metrics/cvss/v3-calculator?vector=AV:N/AC:H/PR:L/UI:N/S:U/C:L/I:L/A:L&version=3.1}{5.0}/\href{https://nvd.nist.gov/vuln-metrics/cvss/v3-calculator?vector=AV:N/AC:H/PR:H/UI:N/S:U/C:L/I:H/A:L&version=3.1}{5.5}
&
\circledone{\color{white}7} \circledone{\color{white}8} & - & M13\\

\textbf{I} &
Sensitive Data Leakage (LLM02:2025) & 
LLM agents manipulate sensitive data, obtained from hardware and software sensors (e.g., GPS location, running apps on the device, etc.), that is sent to the LLM central system for processing and analysis.
Adversary obtains such data during transmission, compromising the cloud, or as an insider. & 
\href{https://nvd.nist.gov/vuln-metrics/cvss/v3-calculator?vector=AV:L/AC:L/PR:H/UI:N/S:U/C:H/I:H/A:L&version=3.1}{6.3}/\href{https://nvd.nist.gov/vuln-metrics/cvss/v3-calculator?vector=AV:N/AC:H/PR:H/UI:R/S:C/C:H/I:H/A:L&version=3.1}{7.5} &
\circledone{\color{white}5} \circledone{\color{white}6} & C15 & M01,M04, M06,M13, M16,M19, M41 \\

\textbf{I} &
Reverse Engineering (RE) (-) &
Adversary reverses engineering the application (available in user's device) responsible for communicating and accessing the LLM system 
to obtain API keys or any other sensitive data. 
The adversary can also use malware to this end. & 
\href{https://nvd.nist.gov/vuln-metrics/cvss/v3-calculator?vector=AV:P/AC:H/PR:L/UI:N/S:U/C:H/I:N/A:L&version=3.1}{4.6}/\href{https://nvd.nist.gov/vuln-metrics/cvss/v3-calculator?vector=AV:N/AC:H/PR:H/UI:R/S:U/C:H/I:N/A:L&version=3.1}{4.8}&
\circledone{\color{white}9} \circledtwo{\color{white}10} & C14-15 & M04,M19, M28 \\

\textbf{TIDE} & 
Remote Code Execution (RCE) (LLM01:2025) & 
Adversary sends a malicious email to the victim.
The LLM agent processes the email, containing a malicious hidden instruction, executing it and triggering the malicious activity. & 
\href{https://nvd.nist.gov/vuln-metrics/cvss/v3-calculator?vector=AV:N/AC:H/PR:L/UI:R/S:U/C:H/I:H/A:H&version=3.1}{7.1} &
\circledone{\color{white}2} & I14 & M24,M27, M33-36 \\

\textbf{STID} &
Automatic Speech Recognition (ASR) system Compromise&
LLMs with voice command feature are susceptive to adversarial attacks on the ASR system, responsible to convert an acoustic waveform (user voice) into text.
Attacks on these systems, such as the DolphinAttack, add particular noise into the audio (imperceptible to humans) that when interpreted by the ASR, result in malicious instructions that are executed into the system. &
\href{https://nvd.nist.gov/vuln-metrics/cvss/v3-calculator?vector=AV:N/AC:H/PR:N/UI:N/S:U/C:H/I:H/A:N&version=3.1}{7.4} &
\circledone{\color{white}8} & \cite{zhang2017dolphinattack, bhanushali2024adversarial, gong2018overview} & \cite{zhang2017dolphinattack, bhanushali2024adversarial, gong2018overview} \\

\textbf{TIDE} & 
Abusing Shared Infrastructure (LLM3:2025) & 
Adversary abuses the CSP for not having a proper isolated infrastructure, allowing access to unauthorized data and resources available on that infrastructure (e.g., \textit{Cloud Jacking} attack). & 
\href{https://nvd.nist.gov/vuln-metrics/cvss/v3-calculator?vector=AV:N/AC:H/PR:L/UI:N/S:U/C:H/I:L/A:L&version=3.1}{6.4/6.4} &
\circledone{\color{white}6} \circledtwo{\color{white}11} & C06,A04 & M16-17, M24-25, M27-28, M45-46\\

\bottomrule

\end{tabular}
\end{table*}

\subsection{Analysis of Threats Vectors across Different Scenarios}

Based on the threat model analysis, we examine the security implications of the main architectural differences and design choices across LLM deployment scenarios. We then identify key security concerns that emerge from the differing threat vectors inherent to each scenario and provide recommendations aiming to support future researchers and practitioners in developing robust security controls for LLM systems.

\subsubsection{On-device vs. Off-device Model Deployment}

The first design choice when deploying LLMs is to define where the model will be hosted, on a user's device (Section~\ref{sec:llm_scenario_2}, Table~\ref{tab:TM_STRIDE_uc_on_device_chatobot_no_internet}) or on an external server (Section~\ref{sec:llm_scenario_3}, Table~\ref{tab:TM_STRIDE_uc_on_cloud_llm_integrated_app} or Section~\ref{sec:llm_scenario_4}, Table~\ref{tab:TM_STRIDE_uc_on_cloud_device_llm_agent}).
Here, we are not concerned about the processing power or other resource limitations of running LLMs on users' devices, only in the threats to which the model will be exposed.

The first major threat of on-device deployed models is the possibility of reverse engineering (attack strategy \circledone{\color{white}\small 9}).
Having the model deployed on the device allows adversaries to learn about how the model works, obtain internal information, submit any amount of queries to the model without rate limiting (a good security mechanism for external deployed models), and create a shadow model easier.

Another threat to models deployed on user devices is malware (attack strategy \circledtwo{\color{white}\small 10}).
The LLM provider has no control over what tools and applications are running on a user's device, where the model is being executed.
Malicious applications can steal users' data (including chat history and keys to access the model - C15-16), perform unauthorized actions on the model (I09-27), and also cause denial of service (A01-03, A06-07).
Protecting the model in a user's device is a difficult task, requiring additional hardware to provide some level of security (M28) and endpoint software solutions (M42).

In addition to these two major threats,
side-channel attacks are facilitated when LLMs run on user devices.
Although this type of attack can also occur when models are deployed in external services (C04, C06), by having access to the model, adversaries have more techniques at their disposal to perform the attack. 
In this way, they can learn a model's sensitive information and internals (C03, C05), and are not limited to only performing queries to the target model and trying to learn from the results.
In addition, adversaries can perform jailbreak by hardware manipulation (I28) to abuse the LLM.

However, some threats are attenuated when models are deployed on a device, such as the \textit{Impersonating the User threat}. 
Note that the CVSS score decreased from 6.4 (Table~\ref{tab:TM_STRIDE_uc_on_cloud_llm_integrated_app}) to 5.0 (Table~\ref{tab:TM_STRIDE_uc_on_device_chatobot_no_internet}), since an attack would require the adversary to obtain physical access to the device compared to deployments in external servers, in which adversaries can access the LLM remotely.
Besides, on-device models can better preserve user privacy, since no data is required to leave the device.

Although deploying LLMs in external servers has advantages (including not depending on device processing power or worrying about its energy consumption), 
it also exposes the model to more threats, facilitating attacks strategies such as \circledone{\color{white}\small 5}, \circledone{\color{white}\small 6}, \circledone{\color{white}\small 7}, and \circledone{\color{white}\small 8}.
A common issue to services deployed in external servers is proper attestation.
Adversaries may perform man-in-the-middle (MITM) attacks to intercept user communications, read sensitive information, and/or change it, or launch spoofing attacks to redirect users to malicious services.
In addition, Insiders (attack strategy \circledone{\color{white}\small 5}) become a threat in such a configuration when it comes to the LLM operation, because people with special access or knowledge about the infrastructure and solution become an adversary.

Lack or improper use of logging is not a threat itself, but it is a significant risk due to the inability to identify some attacks aimed at stealing data, as well as silent attacks that do not interrupt a service or cause perceptual damage.
Performing logging for models deployed on a user device can also be a protection choice, but it is less effective because it can be easily disabled or changed (by malware, for instance).

Another risk that comes with the deployment of LLMs is allowing different users (or applications) to access the same model, hosted in a shared infrastructure, with each one having access to an instance of the running service.
This exposes users to additional threats, especially because the model needs to maintain context information for each user/app invocation (containing sensitive data) (C16, I14), shared resources (A04), databases (I15), etc.
Isolation of environments and LLM instances (M24) or even the use of trusted execution environments (M28) are some possible mitigation strategies.

External servers require additional security tools to limit the access to the LLM (M25), such as the use of encryption (in transit - M19), firewall (M15), VPNs (M11), access control (M06), authentication technologies (with strong recommendations of two factor authentication - M20), and proper configuration of all of these tools (M18).

In summary, for LLMs deployed on user devices, the adversary needs either physical access to the device or to convince the user to install malware. For LLMs deployed on external servers that adversaries can access from anywhere, the adversary only needs to know the address to make requests, resulting in a higher number of threats and the need to put even more security mechanisms in place. 
For hybrid deployments, both threat scenarios must be taken into consideration.

\subsubsection{On-cloud vs. On-premises Infrastructure}

Another design choice when deploying LLMs, in this case, in external servers with users making queries to the model via APIs or websites, is to use the companies own infrastructure (on-premise) or a third-party Cloud Service Provider (CSP).
Again, here we make the analysis from a security perspective and not considering financial costs or other aspects (e.g., physical space, maintenance).

Using company premises to deploy LLMs (Section~\ref{sec:llm_scenario_1}, Table~\ref{tab:TM_STRIDE_uc_on_premises_chatobot_model_creation}) requires the purchase (for some technologies), installation, and configuration of many tools to provide security to the model and infrastructure, such as firewall (M15), VPN (M11), authentication (M20), access control (M06), and rate limiting and throttling (M25), among other techniques and technologies.
This is important to ensure a Defense-in-Depth approach \cite{Schneier2006defenseindepth}.

By using a third-party CSP, we can have all these tools already in place, requiring one-click to activate them.
However, it is imperative to configure tools properly (M18) and to avoid (or pay special attention to) default configurations, which could be inefficient and even make software vulnerable to attacks.

External servers are vulnerable to Insiders (attack strategy \circledone{\color{white}\small 5}).
Both on-premise and on-cloud options share this threat, with the latter being amplified due the use of an infrastructure owned by a third party, in which trust is paramount due to the LLM developer not knowing (or being able to control) who has access to the infrastructure and data stored.
Note, however, that the CVSS (and other methodologies) do not present a way to highlight this difference regarding the insider threat, thus not reflecting it in the severity score.
More details are discussed in Section~\ref{sec:CVSS_Score_Limitations}.

The lack or improper implementation (and use) of logging to track users' actions every time they perform a query, log in to the system, or perform modifications in the model or infrastructure, can also be a major risk for both deployment scenarios.
This practice (M13) can help detect attacks and anomalies in the system, but requires regular review, using manual or automated methods~\cite{lindqvist1997systematically,lindqvist1999detecting}.

In summary, deploying models on premises is more complex, but it does not require trusting a third party.
The level of trust regarding highly sensitive information can be a turning point in deciding which approach to choose.
Although it is easier to obtain top-notch security tools when using a CSP, the lack or improper configuration of the environment and tools can be dangerous, and the use of default configurations is not always the best option.

\subsubsection{LLM Development vs. Operation Phases}

Different threats occur to the phases related to the LLM development
(Section~\ref{sec:llm_scenario_1}, Table~\ref{tab:TM_STRIDE_uc_on_premises_chatobot_model_creation}) and operation (Section~\ref{sec:llm_scenario_2}, Table~\ref{tab:TM_STRIDE_uc_on_device_chatobot_no_internet}; Section~\ref{sec:llm_scenario_3}, Table~\ref{tab:TM_STRIDE_uc_on_cloud_llm_integrated_app}; and Section~\ref{sec:llm_scenario_4}, Table~\ref{tab:TM_STRIDE_uc_on_cloud_device_llm_agent}), 
requiring the adoption of different defense strategies.

During the LLM Development Phase, major threats are poisoning \circledone{\color{white}\small 4} and supply chain \circledone{\color{white}\small 3}.
The first threat targets the data used during LLM training (I01) or the model itself (I02, I08).
Since LLMs require a huge amount of data for their creation, it is difficult to carefully analyze all the data obtained and used during the training, taking care of not introducing any data that may cause bias or unexpected behavior from the model during its operation (e.g., introducing backdoor - I02).
In practice, we have seen LLM providers not being careful enough during model training \cite{CatoReport2025, hornbyCopilotPirateWindows2025}, and instead, they spend significant resources to review the model outputs when in production.
The problem is that testing every possible scenario is infeasible, and vulnerabilities may remain in production models, as constantly reported in the news~\cite{Cuthbertson-news-aiworm, Chen2024news, Russinovich-jailbreakattack, news-Wickens-shadowlogic2024, news-Kassianik-TAP2023, Kovacs-news-2025-2, cohen-aiworm, Shimony-news-jailbreak2025, Wickens-shadowlogic2024}.
Poisoning can also target model fine-tuning or RAG processes when external content is used to specialize an LLM, especially when these data come from users without proper validation (I03) or are manipulated by insiders or adversaries \cite{wang2024poisoned} (I01).

Supply-chain threats are also present during model development, although they are not restricted to this scenario.
Adversaries may manipulate pre-trained LLMs (foundation models) that may be used by applications (I02), since creating an LLM from scratch is an expensive task.
Adversaries might also create fake plugins (I06) and dependencies (I05) that developers may adopt during the development process, introducing vulnerabilities in the model or development environment and putting at risk the LLM integrity and associated sensitive data.

When models become ready to users (Operation phase), threats related to user interaction may arise.
LLMs allow users and applications to submit prompts having any possible content;
although this is an interesting characteristic and maybe the reason for making this technology so popular, since the model can interact with users simulating human-like conversations, it can also be a major threat (attack strategies \circledone{\color{white}\small 1} \circledone{\color{white}\small 2} - threats C08-09, C12, I09-27, A01-02, etc).
To protect LLMs, developers create security mechanisms and guardrails to restrict and avoid sensitive and forbidden information to be disclosed by the LLM, but as presented in Table~\ref{tab:threats_to_LLM_operation}, there are many variations of jailbreak attacks to bypass the defenses in place, and it is difficult to prepare for every possible word combination that could compromise the model.
There are many defense alternatives to this sort of attack (M29-31, M33-40), but we continue to see new attacks to prevail.

Supply Chain and poisoning can also be a threat to LLMs during operation.
The supply chain threat can be seen in the tools used to make the model available for users.
Many open-source tools are available nowadays that can make the processes of deploying LLMs easier and faster, but some of these tools could be compromised and be a threat to users and companies (I05).
When it comes to poisoning, the feedback feature about users experience with the LLM system can be a way of adversaries manipulate a particular behavior of the model (I08).
Another form of feedback to contribute with model's training process is through federated learning.
Poisoning (I02) and Byzantine (I07) attacks are another threat in this scenario, with adversaries sending erroneous updates to the central aggregation system to manipulate the LLM \cite{wen2023survey, lee2025exploring}.
Changing data that will be used for model re-training (fine-tuning) or consumption (RAG) can be a serious threat to users, that can be manipulated by insiders or when the infrastructure security mechanisms are compromised (I01).
In the federated learning context, data exfiltration \cite{hitaj2023fedcomm} or leakage \cite{hitaj2017deep} are also relevant threats that should be considered.

The lack or improper implementation (and use) of logging affects both phases, the development and operation.
During LLM development, especially during training data processing and collection in Data Engineering Phase, logging and other security mechanisms implemented in the infrastructure storing the data is paramount (M02, M13), especially with a strict access control policy (M06) and strong authentication methods (M20) to avoid unauthorized people to access and manipulate data. 
Log has an important role of helping identifying any manipulation on data or other abuses.
For LLM Operation Phase, logging is of great importance to help detecting attacks and abuses of the model and infrastructure.

In essence, LLM development is mostly affected by poisoning \circledone{\color{white}\small 4} and supply chain \circledone{\color{white}\small 3}, while LLM operation, still being at risk for this two attack strategies, has other serious threats related to users interaction, via direct \circledone{\color{white}\small 1} or indirect \circledone{\color{white}\small 2} prompts submitted to the system.

\subsubsection{General-purpose vs. Goal-oriented LLM Interaction}

Different LLM use cases allow different forms of user interaction.
LLMs become widely known by people via general-purpose chat-bot applications (Section~\ref{sec:llm_scenario_2}, Table~\ref{tab:TM_STRIDE_uc_on_device_chatobot_no_internet}), mostly due to their ability to recognize text from users (via a text prompt) and respond accordingly, simulating real-life conversations.
However, this is not the only use case of or interaction form with LLMs.
We can also have goal-oriented LLMs, with applications interacting with users and, based on some conditions, prompt the LLM to solve a particular problem (Section~\ref{sec:llm_scenario_3}, Table~\ref{tab:TM_STRIDE_uc_on_cloud_llm_integrated_app}), 
or even having LLMs with predefined goals acting like an agent, possibly executing or controlling tools and other LLMs (Section~\ref{sec:llm_scenario_4}, Table~\ref{tab:TM_STRIDE_uc_on_cloud_device_llm_agent}).
Each of these use cases have their own limitations and associated threats based on the way they interact with users, presented in the following.

General-purpose chat-bots allow users to send any input (usually via a text prompt) to the LLM interpret and respond, amplifying threats related to prompt injection (attack strategy \circledone{\color{white}\small 1}).
Adversaries targeting this LLM use case usually execute jailbreak attacks, trying to bypass the security mechanisms and guardrails to obtain sensitive information from the LLM, e.g., \textit{"how to build a bomb''} (I09-27, I16, I28).
However, adversaries can also target users API keys (C15), credentials \circledone{\color{white}\small 7}, or chat history (C16), since this will allow them to obtain sensitive information (or even business data).
Some attack strategies used to this end are malware \circledtwo{\color{white}\small 10}, MITM attack \circledone{\color{white}\small 6}, or users' credentials compromise \circledone{\color{white}\small 7}, all targeting user interaction with the LLM system.

Although chat-bots use text as the main form of receiving user input, some solutions are also adopting the voice as another form of interaction \cite{llmvoicebot, gpt4o2024}.
This brings new threats to the system, as another element will be responsible for converting the voice into text before the LLM can make inferences on it.
One can use a multimodal model to receive the voice directly \cite{shen2024voice} (not in the scope of this paper) or use an LLM with an additional tool (usually an ASR - Automated Speech Recognition - system) to make the voice conversion into text, which will feed the LLM. 
For the second case, adversaries may attack the ASR, 
for instance, using adversarial attacks such as the Dolphin \cite{zhang2017dolphinattack} or any other ones \cite{bhanushali2024adversarial, gong2018overview} to force a wrong transcription or the embedding of malicious text into the results.

On the other hand, goal-oriented LLMs, such as LLM-integrated applications, are exposed to other types of threats due to the way they interact with users (attack strategy \circledone{\color{white}\small 2}).
In this scenario, the LLM receives data from the integrated application, which will create a prompt (or use a predefined one) with a particular goal and send it to the LLM to execute; the results will be interpreted by the application and presented to users.
Note that user input to the application may be appended to the prompt, which could represent a threat.
Adversaries can attack the LLM via inputs to the application or find a way to change the standard prompt used by the application.
Another way of attacking LLM-integrated applications is making the LLM fetches and processes external content created by the adversary with hidden malicious instructions (I16) that will be activated when processed by the LLM system.

Another type of goal-oriented LLM is the LLM agent, which can interact with users via text (using a chat-bot interface),
voice commands (using an ASR system) or with an LLM-integrated application able to receive or send agent requests.
In such cases, the LLM agent will inherit the threats from the chosen user interaction method, as described before.

With goal-oriented LLMs, where integrated applications or agents can execute predefined actions or commands on the host system, remote code execution (RCE) vulnerabilities (I14) represent a major threat. Compared to traditional chat-bots, susceptibility to denial-of-service attacks (A01-A02, A06) is significantly higher, and the risk of leaking user sensitive information (C15-16) also increases.

In summary,
all forms of LLM interaction require a strict input (M33-36) and output (M37-40) analysis, as well as other mitigation strategies to harden the model (M29-32) and other resources to withstand adversarial attacks.
In addition, there is a significant risk in LLM-integrated applications or LLM agents for their ability to execute code, which can be attenuated by defining minimum permissions and execution rights to applications (M27), perform environment isolation (M14, M24), implement logging (M13), and also having constant monitoring of the environment for detecting attacks and leakage of sensitive data (M15, M23).

\subsubsection{Enabled vs. Disabled Access to Resources}

The last design choice of an LLM system discussed here is about allowing access to external content and tools to the LLM 
(Section~\ref{sec:llm_scenario_3}, Table~\ref{tab:TM_STRIDE_uc_on_cloud_llm_integrated_app} and Section~\ref{sec:llm_scenario_4}, Table~\ref{tab:TM_STRIDE_uc_on_cloud_device_llm_agent})
or not 
(Section~\ref{sec:llm_scenario_1}, Table~\ref{tab:TM_STRIDE_uc_on_premises_chatobot_model_creation} and Section~\ref{sec:llm_scenario_2}, Table~\ref{tab:TM_STRIDE_uc_on_device_chatobot_no_internet}), such as access to content on the internet, tools, databases, or hardware/software device sensors.

Searching for content online can improve LLM capabilities, but also expose the model to new threats, such as indirect prompt injection \circledone{\color{white}\small 2}.
For instance, an LLM with access to user' emails can be target of adversaries that craft email messages containing hidden instructions within its content, and executed by the model during the message processing.
Another example refers to LLMs that process content from a website (I22), similar to the case just presented, but with adversaries hiding the malicious instructions within a page that will be accessed and processed by an LLM when responding to users' queries.

LLM-integrated applications or agents may have the ability to execute commands on the hosting system, having even control of tools to perform specific tasks.
Adversaries targeting such use cases seek to obtain remote code execution attacks to take control of the whole system, via indirect prompt injection \circledone{\color{white}\small 2} (I14), supply chain \circledone{\color{white}\small 4}, or exploiting known vulnerabilities \circledone{\color{white}\small 6} (I05) and security mechanisms \circledone{\color{white}\small 8}.
Function calling is also another possibility to allow LLMs to use external tools, but they can also be manipulated via indirect prompt injection \circledone{\color{white}\small 2}, even for simpler attacking methods, since models using such feature may not be prepared with the same safety alignment training as chat-bot LLMs \cite{wu2024the}.

Access to databases containing sensitive information may also be another feature of LLM-integrated applications or agents.
LLMs can query databases to obtain additional information about a particular topic (e.g., sales information during a particular period) and execute tasks based on some condition.
However, adversaries may try to perform SQL injection attacks (I15) to steal these data, using direct \circledone{\color{white}\small 1} or indirect \circledone{\color{white}\small 2} prompts.

Some LLM agents may have the ability to read from hardware or software sensors available in a device to act upon such information.
Note that sensitive data may be obtained in this case, such as the GPS data or a list of running applications in the device, for instance.
Allowing the LLM to access such data puts it as a high target for adversaries.
Besides, there are threats when the collected data is sent to a cloud for further processing.
Attacks such as indirect prompt injection \circledone{\color{white}\small 2}, exploit of security mechanisms \circledone{\color{white}\small 8} and use of malware \circledtwo{\color{white}\small 10} are some of the alternatives used by adversaries to obtain this highly sensitive data.

Even when LLMs have no access to external content directly or the ability to execute commands, risks still exist, such as the \textit{Copied Prompt Injection} attack (I16) \circledone{\color{white}\small 2}.
In this attack, users are lured to copy content from a malicious website with hidden instructions and provide to the LLM, via a paste operation in the text field of the application interface they use to communicate with the LLM.
Without noticing the hidden instructions (obfuscated by the adversaries using style features of the interface), users send the copied content to be processed by the LLM, along with the hidden instructions.
Note that such attack targets chat-bots applications and can be a combination of threats, with I16 and another one to jailbreak the LLM (I09-I29) or cause denial of service (A01-03, A06).

To summarize, by enabling or not LLMs to access external content, when processing any user or application supplied content requires a deeper inspection, using a combination of the defense techniques to handle the inputted data (M33-36), review the output (M37-40), and protect the LLM from jailbreaks (M29-31).
During LLM development, consider using security best practices (M09), which encompasses a zero-trust over any data supplied to the system.
To further limit the damage of attacks that bypass the defenses, it is recommended to set minimum permissions and execution rights to applications (M27), keep all software up-to-date (M16), perform isolation of execution environment (M14, M24), implement correctly logging (M13), and keep monitoring the environment for attacks and leakage of sensitive data (M15, M23).
\section{Discussions, Open Challenges, and Limitations}
\label{sec:discussion}

In this section, we answer the research questions that led us to conduct this research, discuss some open challenges identified in the field according to our perspective, and present the limitations of this paper.

\subsection{Research Questions}

The research questions introduced at the beginning of this work are discussed below.

\begin{itemize}

    \item[RQ1] \textbf{What are the main use cases and design choices of LLM systems from a security perspective?}

    Not every threat applies to every possible scenario.
    Understanding the scenario where the LLM will be used is the first step towards building a secure system.
    In Section~\ref{sec:llm_scenarios}, we listed possible LLM use cases, from development to operation, considering different ways of using LLMs. We also discussed the characteristics and design choices of a LLM system that may directly affect security in Section~\ref{sec:llm_scenarios_design_choices}.
    Then, we proposed a structured definition of an LLM deployment scenario, grounded in a set of distinct configurations and a canonical vector string. Using this framework, we derive four representative real-world scenarios from a security perspective in Section~\ref{sec:ex_llm_scenarios}, in which we examined in detail, 
    showing that the design choices are critical engineering decisions that shape the threat landscape of real-world LLM-based systems and influence the selection of corresponding mitigation strategies.

    \item[RQ2] \textbf{What are the major threats to real-world LLM-based systems and how can they be mitigated?}
    After proposing a structured definition for a LLM scenario, we scrutinized in Section~\ref{sec:attacks_to_LLM} both academic literature and industry reports to uncover and characterize known threats and recent attacks to LLM systems, ranging from supply chain to jailbreak attacks.
    It is important to note that many of these threats can be executed through multiple attack strategies, in which we outline in Section~\ref{sec:att_methods}. For instance, an adversary can steal a user's API key (C15) through an indirect prompt injection strategy \aaaIP, credential stealing \aaaCS, or malware infection \aaaMW.
    The ease or difficulty of executing an attack depends on the strategy employed, which in turn affects both the likelihood of the attack and the choice of appropriate defense strategies.
    To evaluate the practical impact of the threats surveyed in the literature, we assessed the risk associated with each threat in the context of LLM real-world scenarios. 
    We found that the most significant risks stem from attacks targeting LLM prompting and system integrity, particularly the strategies manipulating prompts directly \aaaDP~or indirectly \aaaIP~or exploiting security mechanisms \aaaSM. Moreover, when examining through the lens of the four real-world LLM deployment scenarios, insider threats (\aaaIN) and the exploitation of known vulnerabilities (\aaaKV) also emerge as critical concerns in practical LLM deployments.
    Another important conclusion is that there is no ``silver bullet'' to mitigate all attacks or even an entire class of attacks.
    We have seen that the traditional ``Defense-in-Depth'' approach is the best solution to protect systems, meaning that 
    it is important to use 
    one or more mitigation strategies (from Section~\ref{sec:mitigation_strategies}) to implement in the LLM system, depending on the use case and design choices adopted.

    \item[RQ3] \textbf{How do different use cases and design choices affect the security and privacy of LLM-based systems?}

    Finally, in Section~\ref{sec:threats_LLM_use_cases} we presented an in-depth security analysis of the four LLM scenarios defined in this work (from Section~\ref{sec:ex_llm_scenarios}), applying the STRIDE framework to model threats. 
    The analysis identified specific threats relevant to each scenario and outlined possible defense mechanisms.
    We also highlighted how different design choices and use cases affect security, showing the threats that should be considered for each setting and corresponding mitigation strategies.
    Notably, changing the scenario settings can increase or decrease the severity score level for a particular attack, and in some cases, if appropriate design choices are selected, they are completely mitigated.

\end{itemize}

\subsection{Open Challenges}

After completing this research, we identified some challenges that, from our perspective are worthy of further exploration.
We separated these challenges by topic, as discussed below.
Although some research has already been developed on some of these topics, solutions to satisfactorily address them are still lacking.

\begin{itemize}

    \item \textbf{The never ending Jailbreak of LLMs.}
    Although much research has been conducted to mitigate jailbreak attacks, we continue to see novel methods of performing attacks capable of bypassing the security mechanisms and guardrails in place.
    We suggest that further research focus on developing effective and alternative strategies for detecting malicious input content.

    \item \textbf{Testbed of attacks to LLM systems.}
    Some research has focused on creating or gathering many jailbreak attack methods into a single dataset of malicious prompts \cite{cui2024risk, liu2024formalizing, chao2024jailbreakbench}.
    However, a more comprehensive testbed to evaluate additional attacks can ease the process of finding other vulnerabilities, including privacy-related and side-channel attacks, among others. Testing the integration of both the application and the LLM is also necessary, since many software vulnerabilities can be exploited to take control of the system and steal valuable data.

    \item \textbf{Reproducibility.}
    Many researchers have explored various methods of compromising an LLM, but some of the proposed attacks lack straightforward ways to apply and test them on these models.
    Further research is needed to test these attacks on real-world deployments, in order to refine our analytical approach with empirical results. The community should also strengthen efforts to standardize and provide reproducibility mechanisms to make such comparisons more feasible.

    \item \textbf{Finding and using good quality data.}
    LLMs require vast amounts of training data to achieve good performance.
    More often than not, the data is scrapped from the web with little to no oversight, which may inadvertently compromise the LLM training and consequently the produced results.
    This may introduce many problems to the LLM, including backdoors via data poisoning, toxicity, misinformation, bias, among others.
    Another emerging challenge is the widespread scraping of AI-generated content, which, after successive generations of reuse in LLM training, can degrade data quality over time and ultimately lead to model collapse~\cite{peterson2025aicollapse}.
    More research is required on developing effective methods for analyzing large datasets to detect malicious or unwanted content.

    \item \textbf{Respecting privacy laws and avoiding misinformation.}
    Copyright is a concern due to LLMs being trained on data that is protected and should not be used without it being referenced.
    Research is needed on how to attribute the source of information when LLMs produce their results.
    This can help
    to mitigate misinformation or hallucination problems, since the source will be available in the content, allowing others to verify the information.

    \item \textbf{Measuring efficacy of mitigation strategies.}
    We presented different mitigation strategies that apply for various configurations of LLM use cases and design choices, and showed, through risk and threat modeling assessments, that employing multiple strategies concurrently is crucial to protect the LLM system.
    However, we have not determined how to assess the efficacy of combining two or more strategies, including the implantation complexity and costs, as well as the extent of protection they offer.
    What is needed is a framework or methodology to perform such an analysis that takes into consideration the insights brought by this work.
\end{itemize}

\subsection{Limitations}
\label{sec:CVSS_Score_Limitations}

Some of the limitations of this paper, presented below, are inherent to a field or particular technique, while others resulted from our choices during this research.

\begin{itemize}
    
    \item \textbf{Peer review is important}.
    Much research encompassing LLMs and LLM security is currently being published in repositories with no peer review process, such as ArXiv. 
    Although this makes the research results available faster to the public, no peers are analyzing and filtering this material (including technical reports from companies or blog presentations) to ensure that high-quality papers and documents are being created and shared.

    \item \textbf{Same attack, different name}.
    In this paper, we choose to specify different ways of performing jailbreak attacks on LLMs, since this category is particular to the field 
    and has generated considerable amounts of research.
    Although there are many variations of prompts an adversary can create to bypass security mechanisms and guardrails, as detailed in Section~\ref{sec:attacks_to_LLM}, one particular form of jailbreak attack may be referred to by different names.
    It is common to observe the same idea to craft a prompt (or another one with strong similarity) being referred by different researchers to different names.   

    \item \textbf{Focus on security and privacy threats.}
    This paper focuses on security and privacy threats and does not cover alignment-related threats such as LLMs that perform hate speech or other unethical actions.

    \item \textbf{Limitations of CVSS scoring system.}
    Although CVSS is a standard methodology to assign severity levels to vulnerabilities, when it comes to the threats we addressed in Section~\ref{sec:threats_severity_level}, 
    it provides a limited view of the problem, and does not highlight some of the differences existing between two related threats.
    One example, the Insiders threat \circledone{\color{white}\small 5},
    will have the same severity level in two distinct scenarios, such as when models are deployed in a company's own infrastructure and when models are deployed in a third-party infrastructure.
    Note that they are similar, but in the latter case, the threat is more likely to occur because the number of people having access or privileged knowledge about the system may be larger (the same company employees of the first case, plus additional ones responsible for the third-party infrastructure).

\end{itemize}
\section{Conclusions}
\label{sec:conclusion}

The integration of Generative AI, in particular LLMs, to software applications for distinct purposes demands a proactive and vigilant approach to secure all elements within a system. 
Not only must common software vulnerabilities be identified and mitigated, but also intrinsic threats and vulnerabilities from LLMs that are discovered as this technology becomes more mature.
In this paper, we conducted an extensive systematic literature review of threats and mitigation strategies to LLM systems, defining and exploring different use cases and settings that such technology may be used and evaluate their impact on the security of a system.
We explored not only chat-bot use cases, but also LLMs integrated to applications, including in forms of agents, as well as cases related to the LLM development process.
We performed a threat characterization and presented existing mitigation strategies to deal with different attack strategies, analyzed the severity score of threats, and performed threat modeling considering examples of real-world scenarios.

Although we are far from having systems without security or privacy concerns, we identified possible countermeasures and mapped them according to the LLM use case and settings. These countermeasures can be applied, from the development to operation, including data management techniques, hardening of device and development/deployment infrastructures, LLM and application robustness and protection, input pre-processing, output processing, and user awareness.
Each category encompasses the use of different techniques, and we concluded that the best defense is still via a \textit{Defense-in-Depth} approach, mixing different techniques at different layers to protect the whole system.



\bibliographystyle{IEEEtranS}
\bibliography{all_references_final}

\newpage
\newpage

 




\begin{IEEEbiography}[{\includegraphics[width=1in,height=1.25in,clip,keepaspectratio]{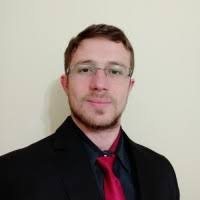}}]{Vitor Hugo Galhardo Moia} is a Cybersecurity researcher and consultant at Instituto de Pesquisas Eldorado (Brazil) since 2022. He received his M.Sc. (2016) and Ph.D. (2020) degrees in Computer Engineering from the School of Electrical and Computer Engineering (FEEC), University of Campinas (UNICAMP), Brazil. During his master’s studies, he focused on the security and privacy on cloud data storage. His Ph.D. research centered on digital forensics. He also worked for about three years at Samsung R\&D Institute as a Security Researcher, where he led projects and conducted applied research in mobile and network security. Since 2021, Vitor is part of SBSeg program committee. His research interests include Red teaming activities, secure software development practices, malware detection and analysis, and the application of Artificial Intelligence to security problems.
\end{IEEEbiography}

\begin{IEEEbiography}[{\includegraphics[width=1in,height=1.25in,clip,keepaspectratio]{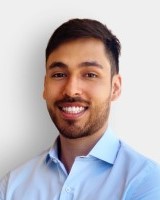}}]{Igor Jochem Sanz} is a researcher and technology consultant at Instituto de Pesquisas Eldorado since 2022, where he coordinates research groups and lead R\&D projects. His main specialties includes software security, network security and artificial intelligence. He also worked as a security researcher from Samsung R\&D Institute Brazil (SRBR) between 2018 and 2022. He received a master's degree in Electrical Engineering from the Federal University of Rio de Janeiro (COPPE/UFRJ) in 2018 and a bachelor's degree in Electronic and Computer Engineering from the same institution in 2017, with sandwich graduation at Bangor University, UK. He has published scientific texts, including two patents, among several topics of computer and security fields, such as intrusion detection, malware analysis, wireless security, mobile computing, machine-learning applied to security, network function virtualization, software-defined networks, cryptography, and blockchain.
\end{IEEEbiography}

\begin{IEEEbiography}[{\includegraphics[width=1in,height=1.25in,clip,keepaspectratio]{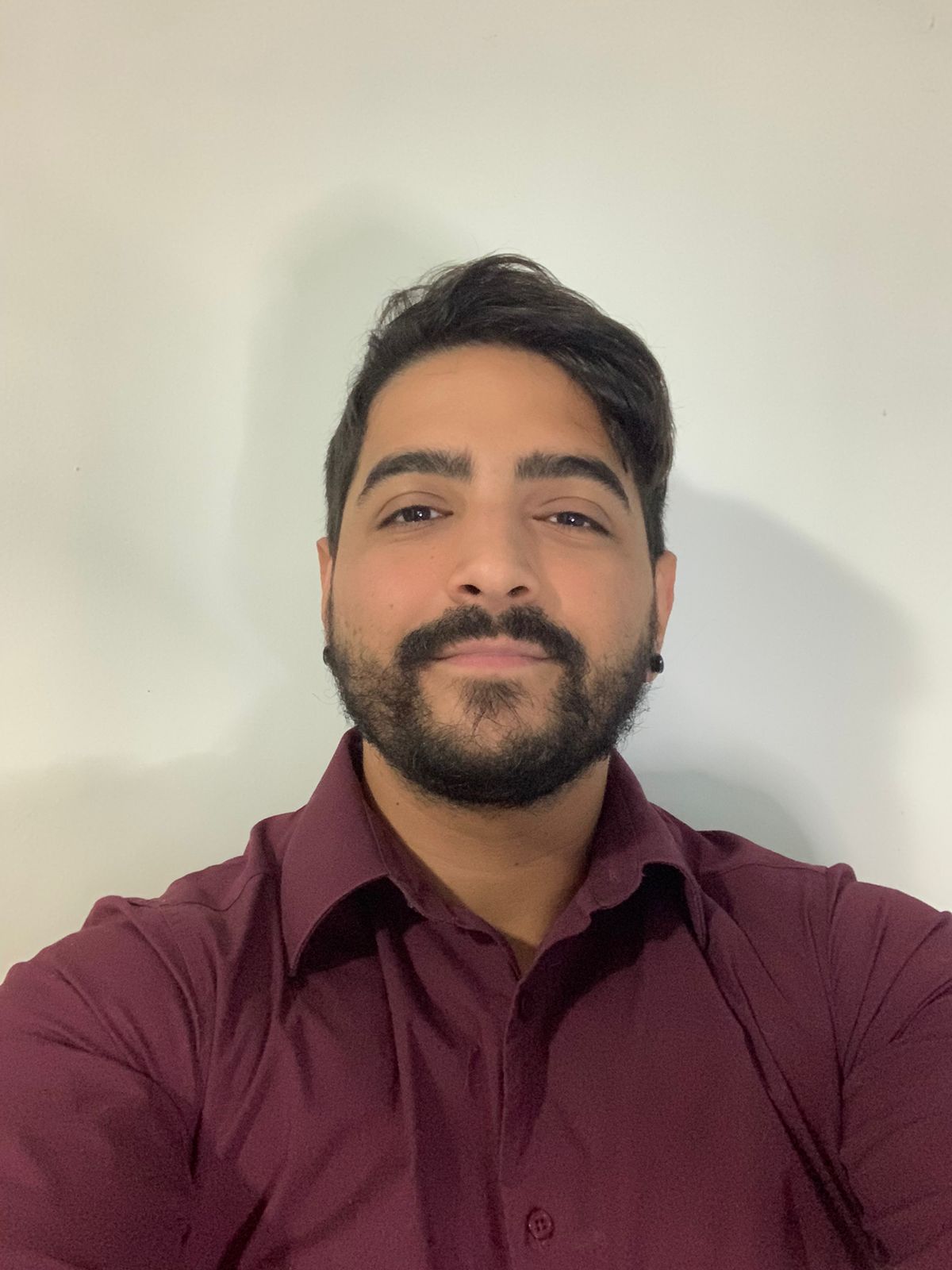}}]{Gabriel Antonio Fontes Rebello} is a cybersecurity researcher and consultant at Instituto de Pesquisas Eldorado (Brazil).  He earned a Ph.D. degree in Computer Science from Sorbonne Université in 2023, a master's degree in Electrical Engineering from Universidade Federal do Rio de Janeiro (UFRJ) in 2019, and a \textit{cum laude} B.Eng. degree in Computer Engineering from UFRJ in 2019. His areas of expertise lie in AI for cybersecurity, cloud security, network security, and blockchains, having published several papers in IEEE conferences and journals such as INFOCOM, COMST, TNSM, and others. 
\end{IEEEbiography}

\begin{IEEEbiography}[{\includegraphics[width=1in,height=1.25in,clip,keepaspectratio]{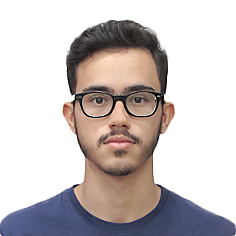}}]{Rodrigo Duarte de Meneses} is a cybersecurity intern at Instituto de Pesquisas Eldorado (Brazil) since 2024. He is currently an undergraduate student in Electrical and Electronics Engineering at the School of Electrical and Computer Engineering (FEEC) from the University of Campinas (UNICAMP). To date, he has published various papers at the Brazilian Symposium on Information Security and Computer Systems (SBSeg) and the International Symposium on the Internet of Things (SIoT), spanning several topics such as post-quantum cryptography, zero-knowledge proofs, homomorphic encryption and embedded systems security.
\end{IEEEbiography}

\begin{IEEEbiography}[{\includegraphics[width=1in,height=1.25in,clip,keepaspectratio]{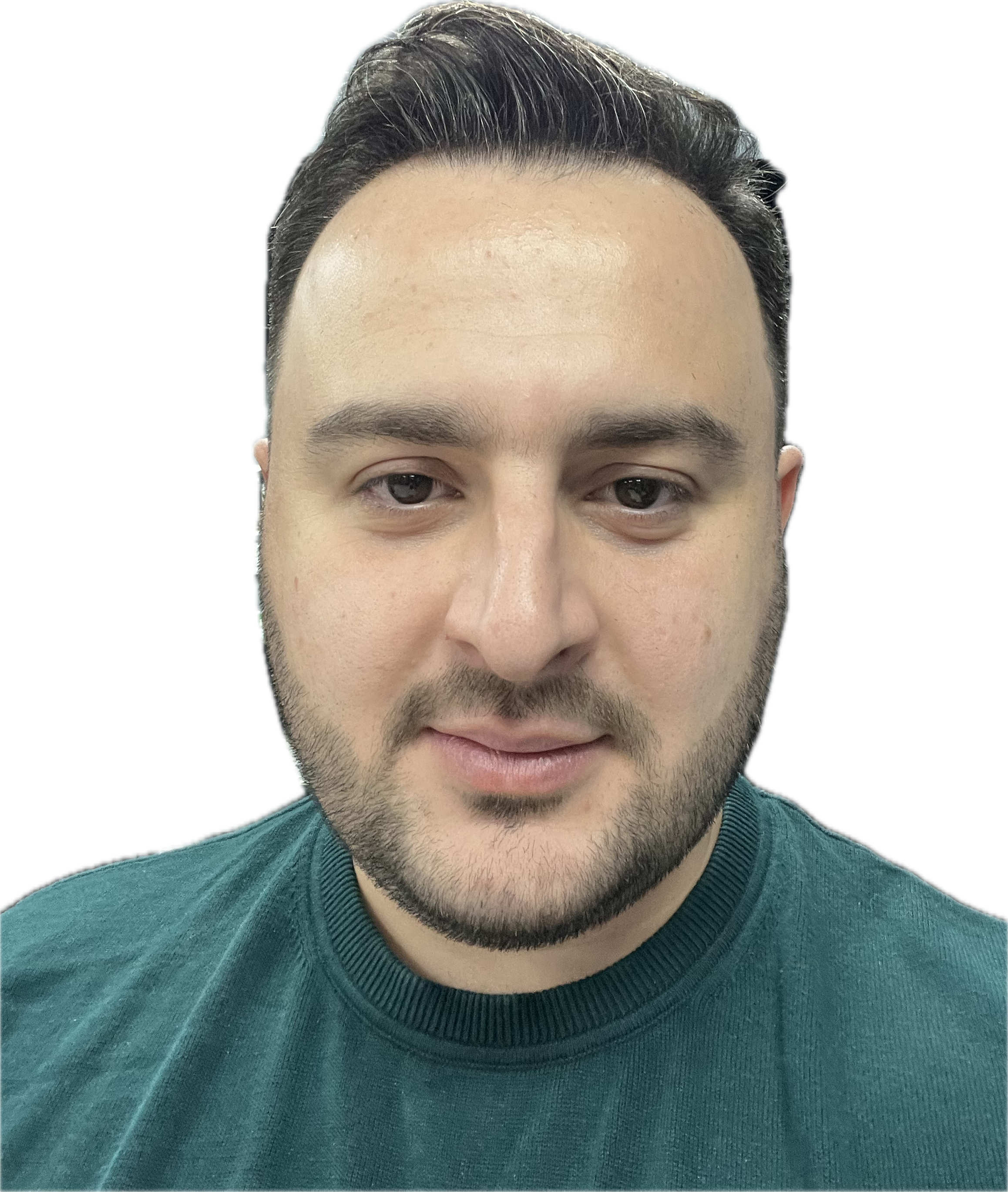}}]{Briland Hitaj}
 is an Advanced Computer Scientist at SRI's Computer Science Laboratory, bringing extensive expertise on security and privacy of machine learning models, with a particular focus on applications of generative models as AI Red-teaming mechanisms. Dr. Hitaj has studied the privacy implications of collaborative (federated) learning, resulting in the demonstration of the first data reconstruction attack in the field based on generative adversarial networks (GANs). More recently, Dr. Hitaj's research focus has been on studying the security and privacy risks related to the use of machine learning models hosted on third-party, often unvetted repositories. Dr. Hitaj has demonstrated that an adversary could embed malicious payloads within the weights of a neural network without hindering the performance on the model, paving the way to a new class of attacks on end-user devices and infrastructure. His work also spans other critical topics in AI security such as adversarial samples, covert communication on top of federated learning, deep neural network watermarking, and password security. Dr. Hitaj received his Ph.D. in Computer Science from Sapienza University of Rome in 2018.
\end{IEEEbiography}
\vskip 0pt plus -1fil
\begin{IEEEbiography}
[{\includegraphics[width=1in,height=1.25in,clip,keepaspectratio]{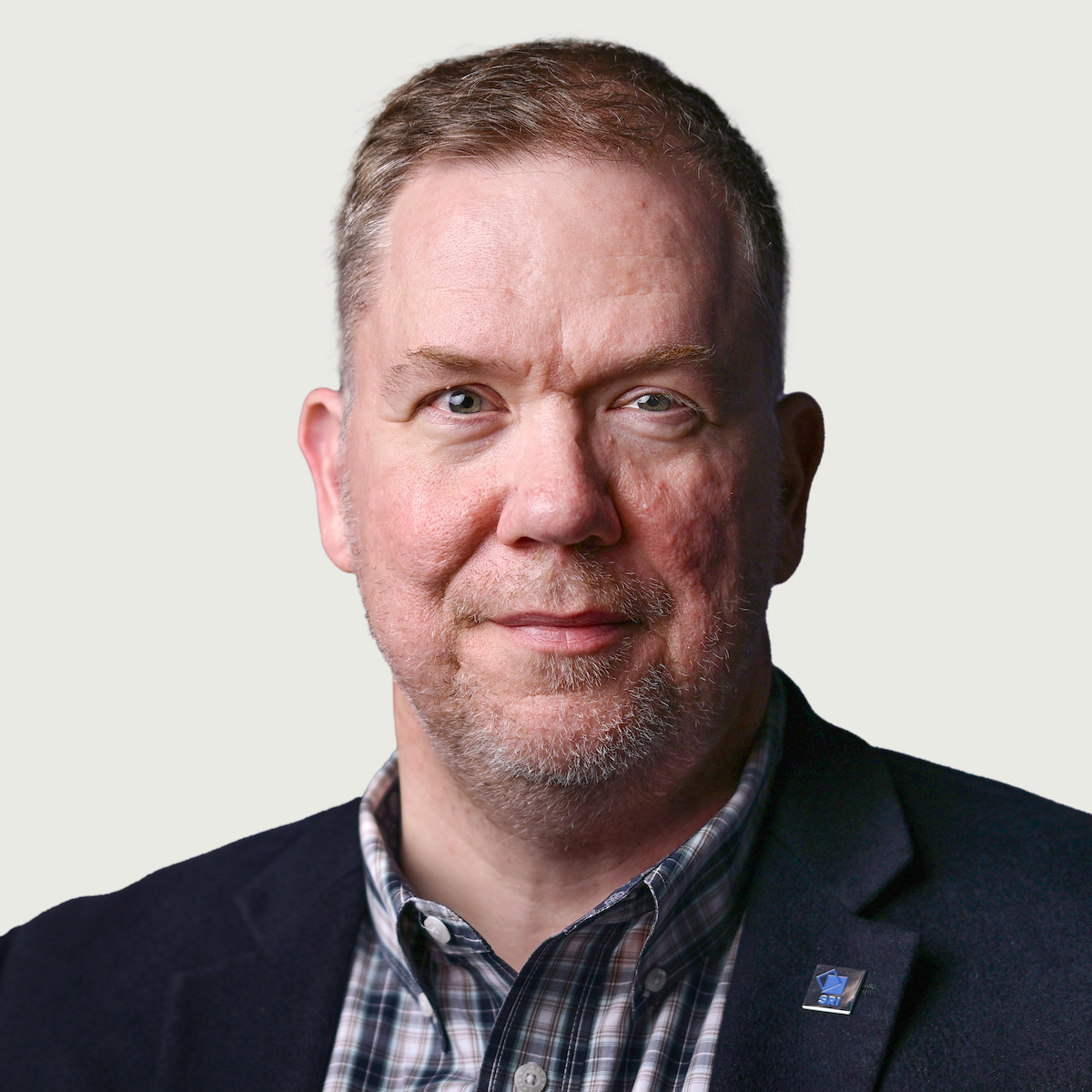}}]{Ulf Lindqvist} is a Senior Technical Director at SRI's Computer Science Laboratory where he manages research and development programs. Dr.~Lindqvist established and leads SRI’s program in infrastructure security research, which is focused on cybersecurity for critical infrastructure systems, including specialized systems in the Internet of Things, electric power, oil and gas, telecommunications, finance, automotive, aviation, and space sectors.  His expertise and interests are focused on cyber security, infrastructure systems, intrusion detection in computer systems, and security for systems that interact with the physical world. He has more than 40 publications in the computer security area, many of which are bridging the gap between theoretical and applied research, and he holds several patents. He served as the 2016-2017 Chair of the IEEE Computer Society’s Technical Committee on Security and Privacy and also served as the Vice Chair of the IEEE Cybersecurity Initiative. He holds a Ph.D. in computer engineering from Chalmers University of Technology in Sweden and was named an SRI Fellow in 2016.
\end{IEEEbiography}
\vfill

\end{document}